\documentclass[preprint,tightenlines,eqsecnum,floats,aps,amsmath,amssymb,nofootinbib,prd,showpacs]{revtex4}

\usepackage{amsmath,amssymb,amsfonts}
\usepackage{graphicx}
\usepackage{enumerate} 
\usepackage{colordvi} 

\def\be{\begin{equation}}
\def\ee{\end{equation}}
\def\ba{\begin{eqnarray}}
\def\ea{\end{eqnarray}}

\def\H{{\cal H}}
\def\Hkg{\H_{\rm kin}^{\rm grav}}
\def\Hk{\H_{\rm kin}^{\rm total}}
\def\Hkwdw{\H_{\rm kin}^{\rm wdw}}
\def\Ha{\H_{\rm aux}}
\def\Hawdw{\H_{\rm aux}^{\rm wdw}}
\def\Hp{\H_{\rm phy}}

\def\Hpwdw{\H_{\rm phy}^{\rm wdw}}

\def\V{{\cal V}}
\def\h{\hat }

\def\Cyl{{\rm Cyl}}
\def\cyl{\Cyl}

\def\SU{{\rm SU}}

\def\Tr{{\rm Tr\,}}

\def\g{\gamma}
\def\lp{{\ell}_{\rm Pl}}
\def\Comp{\mathbb{C}}

\def\R{\mathbb{R}}
\def\Z{\mathbb{Z}}
\def\M{M}

\def\q{{}^o\!q}
\def\e{{}^o\!e}
\def\w{{}^o\!\omega}

\newcommand{\rcr}{\rho_{\mathrm{crit}}}
\newcommand{\snf}{\sin(\mu_o c)}
\newcommand{\snfs}{\sin^2(\mu_o c)}
\newcommand{\csf}{\cos(\mu_o c)}

\newcommand{\heff}{{\cal H}_{\mathrm{eff}}}

\newcommand{\p}{\partial}

\newcommand{\f}{\frac}

\newcommand{\ep}{\epsilon}

\def\M{M}
\newcommand{\integer}{\mathbb{Z}}
\newcommand{\re}{\mathbb{R}}

\usepackage{enumerate}
\DeclareMathOperator*{\sgn}{sgn}

\usepackage{colordvi}
\newcounter{mnotecount}[section]

\newcommand{\comment}[1]{}

\def\b{$\bullet\,\, $}

\def\f{\frac}

\def\La{\mathcal{L}}
\def\a{\mathfrak{a}}

\def\dd{\textrm{d}}
\def\ub{\underbar}
\def\ul{\underline}
\def\WDW{WDW\,\,}
\def\t{\tilde}
\def\epsilon{\varepsilon}

\begin{document}
\preprint{\vbox{\baselineskip=12pt \rightline{IGPG-06/03-2}
}}

\title{Quantum Nature of the Big Bang:\\ An Analytical and
Numerical Investigation}

\author{Abhay Ashtekar${}^{1,2,3}$}
\email{ashtekar@gravity.psu.edu}
\author{Tomasz Pawlowski${}^{1}$}
     \email{pawlowsk@gravity.psu.edu}
    \author{Parampreet Singh${}^{1,2}$}
    \email{singh@gravity.psu.edu}
\affiliation{${}^{1}$Institute for Gravitational Physics and Geometry\\
Physics Department, Penn State, University Park, PA 16802,
U.S.A.\\
${}^2$Inter-University Centre for Astronomy and Astrophysics\\
post bag 4, Ganeshkhind, Pune 411 017, India\\
${}^3$Isaac Newton Institute for Mathematical Sciences,\\
20 Clarkson Road, Cambridge CB3 0EH, UK}


\begin{abstract}

Analytical and numerical methods are developed to analyze the
quantum nature of the big bang in the setting of loop quantum
cosmology. They enable one to explore the effects of quantum
geometry both on the gravitational and matter sectors and
significantly extend the known results on the resolution of the
big bang singularity. Specifically, the following results are
established for the homogeneous isotropic model with a massless
scalar field: i) the scalar field is shown to serve as an internal
clock, thereby providing a detailed realization of the `emergent
time' idea; ii) the physical Hilbert space, Dirac observables and
semi-classical states are constructed rigorously; iii) the
Hamiltonian constraint is solved numerically to show that
\emph{the big bang is replaced by a big bounce}. Thanks to the
non-perturbative, background independent methods, unlike in other
approaches the quantum evolution is \emph{deterministic across the
deep Planck regime}. Our constructions also provide a conceptual
framework and technical tools which can be used in more general
models. In this sense, they provide foundations for analyzing
physical issues associated with the Planck regime of loop quantum
cosmology as a whole.

\end{abstract}


\maketitle

\section{Introduction}
\label{s1}

Loop quantum gravity (LQG) is a background independent,
non-perturbative approach to quantum gravity
\cite{alrev,crbook,ttbook}. It is therefore well-suited for the
analysis of certain long standing questions on the quantum nature
of the big bang. Examples of such questions are:

\medskip
{\narrower{\noindent$\bullet$ How close to the Big Bang does a
smooth space-time of general relativity make sense? In particular,
can one show from first principles that this approximation is
valid at the onset of inflation?\\
$\bullet$ Is the Big-Bang singularity naturally resolved by
quantum gravity? Or, is some external input such as a new
principle or a boundary condition at the Big Bang essential?\\
$\bullet$ Is the quantum evolution across the `singularity'
deterministic? Since one needs a fully non-perturbative framework
to answer this question in the affirmative, in the Pre-Big-Bang
\cite{pbb1} and Ekpyrotic/Cyclic \cite{ekp1,ekp2} scenarios, for
example, so far the answer has been in the negative
\cite{pbb2,kklt}.\\
$\bullet$ If the singularity is resolved, what is on the `other
side'? Is there just a `quantum foam', far removed from any
classical space-time (as suggested, e.g., in
\cite{paddy-narlikar}), or, is there another large, classical
universe (as suggested, e.g., in \cite{pbb1,ekp1,ekp2})?}}

\medskip
\noindent Over the years, these and related issues had been
generally relegated to the `wish list' of what one would like the
future, satisfactory quantum gravity theory to eventually address.
However, over the past five years, thanks to the seminal ideas
introduced by Bojowald and others, notable progress was made on
such questions in the context of symmetry reduced, minisuperspaces
\cite{mbrev}. In particular, it was found that Riemannian quantum
geometry, which comes on its own in the Planck regime, has just
the right features to resolve the big-bang singularity \cite{mb1}
in a precise manner \cite{abl}. However, the physical
ramifications of this resolution ---in particular the answer to
what is `on the other side'--- have not been worked out. It is
therefore natural to ask whether one can complete that analysis
and systematically address the questions listed above, at least in
the limited context of simple cosmological models. It turns out
that the answer is in fact in the affirmative. A brief summary of
arguments leading to this conclusion appeared in \cite{aps1}. The
purpose of this paper is to provide the detailed constructions and
numerical simulations that underlie those results.

Let us begin with a brief summary of the main results of loop
quantum cosmology (LQC). (For a comprehensive survey, see, e.g.,
\cite{mbrev}.) They can be divided in two broad classes:\\ i)
singularity resolution based on exact quantum equations (see e.g.
\cite{mb1,bh,hm1,abl,bdv,mb3,ab1,lm}), and,\\ ii) phenomenological
predictions based on `effective' equations (see e.g.
\cite{mb2,bd1,tsm,jw,dh1,dh2,bms,hw,jel,sd,gh,ps1,sv}).

Results in the first category make a crucial use of the effects of
quantum geometry on the \emph{gravitational part} of the
Hamiltonian constraint. Because of these effects, the quantum
`evolution' is now dictated by a second order \emph{difference
equation} rather than the second order differential equation of
the Wheeler-DeWitt (WDW) theory. Nonetheless, the intuitive idea
of regarding the scale factor $a$ as `internal time' was
maintained. The difference is that the `evolution' now occurs in
discrete steps. As explained in detail in Sec. \ref{s2.2}, this
discreteness descends directly from the quantum nature of geometry
in LQG; in particular, the step size is dictated by the lowest
non-zero eigenvalue of the area operator. When the universe is
large, the \WDW differential equation is an excellent
approximation to the `more fundamental' difference equation.
However, in the Planck regime, there are major deviations. In
particular, while the classical singularity generically persists
in the \WDW theory without additional inputs, this is not the case
in LQG. This difference does \emph{not} arise simply because the
discrete `evolution' enables one to `jump over' the classical
singularity. Indeed, even when the discrete evolution passes
\emph{through} the point $a=0$, the difference equation remains
well-defined and enables one to `evolve' \emph{any} initial data
across this classically singular point. It is in this sense that
the singularity was said to be resolved.

In the second category of results, by and large the focus was on
quantum geometry modifications of the \emph{matter part} of the
Hamiltonian constraint. The main idea is to work with
approximation schemes which encode the idea of semi-classicality
and to incorporate quantum geometry effects on the matter
Hamiltonian by adding suitable `effective' terms to the classical
Hamiltonian constraint. The hope is that dynamics generated by the
`quantum-corrected' classical Hamiltonian would have a significant
domain of validity and provide a physical understanding of certain
aspects of the full quantum evolution. This strategy has led to a
number of interesting insights. For example, using effective
equations, it was argued that the singularity resolution is
generic for all homogeneous models; that there is an early phase
of inflation driven by quantum gravity effects; that this phase
leads to a reduction of power spectrum in CMB at large angular
scales; and that quantum geometry effects suppress the classical
chaotic behavior of Bianchi IX models near classical singularity.

Attractive as these results are, important limitations have
persisted. Let us begin with the results on singularity
resolution. As in most non-trivially constrained systems, solutions
to the quantum constraint fail to be normalizable in the
kinematical Hilbert space on which the constraint operators are
well-defined. Therefore one has to endow physical states (i.e.
solutions to the Hamiltonian constraint) with a new,
\emph{physical inner product}. Systematic strategies ---e.g., the
powerful `group averaging procedure' \cite{dm,almmt}--- typically
require that the constraint be represented by a self-adjoint
operator on the kinematic Hilbert space while, so far, most of
the detailed discussions of singularity resolution are based on
Hamiltonian constraints which do not have this property.
Consequently, the space of solutions was not endowed with a
Hilbert space structure. This in turn meant that one could not
introduce Dirac observables nor \emph{physical} semi-classical
states. Therefore, as pointed out, e.g. in \cite{bt}, the physical
meaning of singularity resolution remained somewhat obscure. In
particular, even in simple models, there was no clear-cut answer
as to what the universe did `before' the big-bang. Was there a
genuine `quantum foam' or was the quantum state peaked at a large
classical universe on the `other side'? In absence of a physical
Hilbert space, the second and the third questions posed in the
beginning of this section could only be answered partially and the
first and the fourth remained unanswered.

The phenomenological predictions have physically interesting
features and serve as valuable guides for future research.
However, the current form of this analysis also faces some
important limitations. Many of these discussions focus only on the
quantum geometry modifications of the matter Hamiltonian. This
strategy has provided new insights and does serve as a useful
starting point. However, there is no a priori reason to believe
that it is consistent to ignore modifications of the gravitational
part of the Hamiltonian and retain only the matter modifications.
Conclusions drawn from such analysis can be taken as attractive
suggestions, calling for more careful investigations, rather than
firm predictions of LQC, to be compared with observations. On the
conceptual side, a number of semi-classical approximations are
made while deriving the effective equations. Many of them are
largely violated in the Planck regime. Therefore it is difficult
to regard conclusions drawn from effective equations on
singularity avoidance, and on the fate of the universe beyond, as
reliable. It does happen surprisingly often in physics that
approximation schemes turn out to work even outside the regimes
for which they were originally intended. But there is no a priori
reason to think that this \emph{must} happen. To develop intuition
on the validity of approximations, it is essential to make, in at
least a few simple models, detailed comparisons of predictions of
effective equations with those of quantum equations that are being
approximated. To summarize, while LQC has led to significant
progress by opening new avenues and by indicating how
qualitatively new and physically desirable results can arise, the
program has remained incomplete even within the realm of symmetry
reduced, mini-superspace models.

The goal of this paper is to complete the program in the simplest
of models by using a combination of analytical and numerical
methods. The resulting theory will enable us to answer, in the
context of these models, the questions raised in the beginning of
this section. Specifically, we will show from first principles
that: i) a classical space-time continuum is an excellent
approximation till very early times; 
ii) the singularity is resolved in the sense that a
complete set of \emph{Dirac observables} on the \emph{physical}
Hilbert space remains well-defined throughout the evolution; iii)
the big-bang is replaced by a big-bounce in the \emph{quantum}
theory; iv) there is a large classical universe on the `other
side',  and,  v) the evolution bridging the two classical branches
is deterministic, thanks to the background independence and
non-perturbative methods. While the paper is primarily concerned
with basic conceptual and computational issues, our constructions
also provide some tools for a more systematic analysis of
phenomenological questions. Finally, our approach can be used in
more general models. In particular, our constructions can be used
also for anisotropic models and for models in which the scalar
field has a potential, although certain conceptual subtleties have
to be handled carefully and, more importantly, the subsequent
numerical analysis is likely to be significantly more complicated.
Nonetheless, in a rather well defined sense, these constructions
provide a foundation from which one can systematically analyze the
Planck regime in LQC well beyond the specific model discussed in
detail.

The main ideas of our analysis can be summarized as follows.
First, our Hamiltonian constraint is self-adjoint on the
kinematical (more precisely, auxiliary) Hilbert space. Second, we
use the scalar field as `internal time'. In the classical theory
of $k\!=\! 0$ models with a massless scalar field $\phi$, the
scale factor $a$ as well as $\phi$ are monotonic functions of time
in any given solution. In the $k\! =\! 1$ case, on the other hand,
since the universe recollapses, only $\phi$ has this property.
Therefore, it seems more natural to use $\phi$ as `internal time',
which does not refer to space-time coordinates or any other auxiliary
structure. It turns out that $\phi$ is well-suited to be
\emph{`emergent time'} also in the quantum theory. Indeed, our
self-adjoint Hamiltonian constraint is of the form
\be \f{\partial^2\Psi}{\partial \phi^2} = - {\Theta} \Psi\ee
where ${\Theta}$ does not involve $\phi$ and is a positive,
self-adjoint, difference operator on the auxiliary Hilbert space
of quantum geometry. Hence, the quantum Hamiltonian constraint can
be readily regarded as a means to `evolve' the wave function with
respect to $\phi$. Moreover, this interpretation makes the group
averaging procedure similar to that used in the quantization of a
`free' particle in a static space-time, and therefore conceptually
more transparent. Third, we can carry out the group averaging and
arrive at the physical inner product. Fourth, we identify complete
sets of Dirac observables on the physical Hilbert space. One such
observable is provided by the momentum $\h{p}_\phi$ conjugate to
the scalar field $\phi$ and a set of them by $\hat{a}|_{\phi_o}$,
the scale factor at the `instants of emergent time' $\phi_o$.
\footnote{As we will see in section \ref{s2}, to construct a
Hamiltonian framework in the open model, one has to fix a fiducial
cell. The scale factor $a$ (and the momentum $p$ conjugate
to the gravitational connection introduced later) refers to the volume of this
cell. Alternatively, one can avoid the reference to the fiducial
cell by fixing  a $\phi_o$ and considering the ratios
$a|_\phi/a_{\phi_o}$ as Dirac observables. However, for simplicity
of presentation we will not follow this route.}
Fifth, we construct semi-classical states which are peaked at
values of these observables at late times, when the universe is
large. 
Finally  we `evolve' them backwards using the Hamiltonian
constraint using the adaptive step, 4th order Runge-Kutta method.
The numerical tools are adequate to keep track not only of how the
peak of the wave function `evolves' but also of  fluctuations in
the Dirac observables in the course of `evolution'. A variety of
numerical simulations have been performed and the existence of the
bounce is robust.

The paper is organized as follows. In section \ref{s2} we
summarize the framework of LQC in the homogeneous, isotropic
setting, keeping the matter field generic. We first summarize the
kinematical structure \cite{abl}, highlighting the origin of the
qualitative difference between the \WDW theory and LQC already at
this level. We then provide a self-contained and systematic
derivation of the quantum Hamiltonian constraint, spelling out the
underlying assumptions. 
In section \ref{s3} we restrict the matter field to be a massless
scalar field, present the detailed \WDW theory and show that the
singularity is not resolved. The choice of a massless scalar field
in the detailed analysis was motivated by two considerations.
First, it makes the basic constructions easier to present and the
numerical simulations are substantially simpler. More importantly,
in the massless case \emph{every} classical solution is singular,
whence the singularity resolution by LQC is perhaps the clearest
illustration of the effects of quantum geometry. In section
\ref{s4} we discuss this model in detail within the LQC framework,
emphasizing the dynamical differences from the \WDW theory.
Specifically, we present the general solution to the quantum
constraint, construct the physical Hilbert space and introduce
operators corresponding to Dirac observables. In section \ref{s5}
we present the results of our numerical simulations in some
detail. Solutions to the Hamiltonian constraint are constructed
using two different methods, one using a Fast Fourier Transform
and another by `evolving' initial data using the adaptive step,
4th order Runge-Kutta method. To further ensure the robustness of
conclusions, the initial data (at late `times' when the universe
is large) is specified in three different ways, reflecting three
natural choices in the construction of semi-classical states. In
all cases, the classical big-bang is replaced by a quantum
big-bounce and the two `classical branches' are joined by a
deterministic quantum evolution. Section \ref{s6} compares and
contrasts the main results with those in the literature.

Issues which are closely related to (but are not an integral part
of) the main results are discussed in three appendices. Appendix
\ref{a1} is devoted to certain heuristics on effective equations
and uncertainty relations which provide a physical intuition for
`mechanisms' underlying certain constructions and results. In
Appendix \ref{a2} we discuss technical aspects of numerical
simulations which are important but whose inclusion in the main
text would have broken the flow of the argument. Finally in
Appendix \ref{a3} we present an alternate physical Hilbert space
which can be constructed by exploiting certain special features of
LQC which are not found in the general setting of constrained
systems. This space is more closely related to the \WDW theory and
could exist also in more general contexts. Its existence
illustrates an alternate way to address semi-classical issues
which may well be useful in full LQG.

\section{LQC in the $k\!=\! 0$ homogeneous, isotropic setting}
\label{s2}

This section is divided into three parts. To make the paper
self-contained, in the first we provide a brief summary of the
kinematical framework, emphasizing the conceptual structure that
distinguishes LQC from the \WDW theory. In the second, we
introduce the self-adjoint Hamiltonian constraints and their \WDW
limits. By spelling out the underlying assumptions clearly, we
also pave the way for the construction of a more satisfactory
Hamiltonian constraint in \cite{aps3}.%
\footnote{Our conventions are somewhat different from those in the
literature, especially \cite{abl}. First, we follow the standard
quantum gravity convention and set $\lp^2 = G\hbar$ (rather than
$8\pi G\hbar$). Second, we follow the general convention in
geometry and set the volume element $e$ on $\M$ to be $e :=
\sqrt{|\det E|}$ (rather than $e := \sqrt{|\det E|}\, {\rm sgn}
(\det E)$). This gives rise to some differences in factors of
${\rm sgn} \, p$ in various terms in the expression of the
Hamiltonian constraint. Finally, the role of the the minimum
non-zero eigenvalue of area is spelled out in detail, and the
typographical error in the expression of $\mu_o$ that features in
the Hamiltonian constraint is corrected. }
In the third part, we first list issues that must be addressed to
extract physics from this general program and then spell out the
model considered in the rest of the paper.

\subsection{Kinematics}
\label{s2.1}
\subsubsection{Classical phase space}
\label{s2.1.1}

In the $k\! = \! 0$ case, the spatial 3-manifold $\M$ is
topologically $\R^3$, endowed with the action of the Euclidean
group. This action can be used to introduce on $\M$ a fiducial,
flat metric $\q_{ab}$ and an associated constant orthonormal triad
$\e^a_i$ and co-triad $\w_a^i$. In full general relativity, the
gravitational phase space consists of pairs $(A_a^i,\, E^a_i)$ of
fields on $\M$, where $A_a^i$ is an $\SU(2)$ connection and
$E^a_i$ an orthonormal triad (or moving frame) with density
weight 1 \cite{alrev}. As one would expect, in the present
homogeneous, isotropic context, one can fix the gauge and spatial
diffeomorphism freedom in such a way that $A_a^i$ has only one
independent component, $\tilde{c}$, and $E^a_i$, only one
independent component, $\tilde{p}$:
\be\label{ss} A\ =\ \tilde{c}\,\, \w^{i} \tau_i,\quad {E}\ =\
\tilde{p}\, \sqrt{\q}\,\,\, \e_{i}\tau^i \, ,\ee
where $\tilde{c}$ and $\tilde{p}$ are constants and the density
weight of ${E}$ has been absorbed in the determinant of the
fiducial metric.%
\footnote{Our conventions are such that $\tau_i\, \tau_j =
\f{1}{2}\epsilon_{ijk}\tau^k - \f{1}{4} \delta_{ij}$. Thus, $2i
\tau_k = \sigma_k$, where $\sigma_i$ are the Pauli matrices.}
Denote by $\Gamma^S_{\rm grav}$ the subspace of the gravitational
phase space ${\Gamma_{\rm grav}}$ defined by (\ref{ss}). Because
$\M$ is non-compact and our fields are spatially homogeneous,
various integrals featuring in the Hamiltonian framework of the
full theory diverge. However, the presence of spatial homogeneity
enables us to bypass this problem in a natural fashion: Fix a
`cell' $\V$ adapted to the fiducial triad and restrict all
integrations to this cell. (For simplicity, we will assume that
this cell is cubical with respect to $\q_{ab}$.) Then the
gravitational symplectic structure $\Omega_{\rm grav}$ on
$\Gamma_{\rm grav}$ and fundamental Poisson brackets are given by
\cite{abl}:
\be \Omega^S_{\rm grav}\, =\, \frac{3V_o}{8\pi \gamma G} \,\,
\dd\tilde{c} \wedge \dd\tilde{p}, \quad {\rm and}\quad
\{\tilde{c},\, \tilde{p}\} = \f{8\pi\g G}{3V_o} \ee
where $V_o$ is the volume of $\V$ with respect to the auxiliary
metric $\q_{ab}$. Finally, there is a freedom to rescale the
fiducial metric $\q_{ab}$ by a constant and the canonical
variables $\tilde{c}, \tilde{p}$ fail to be invariant under this
rescaling. But one can exploit the availability of the elementary
cell $\V$ to eliminate this additional `gauge' freedom. For,
\be  c = V_o^{\f{1}{3}} \tilde{c} \quad {\rm and} \quad
     p = V_o^{\f{2}{3}} \tilde{p} \ee
are independent of the choice of the fiducial metric $\q_{ab}$.
Using $(c,\, p)$, the symplectic structure and the fundamental
Poisson bracket can be expressed as
\be \label{qcsym} \Omega^S_{\rm grav}\, =\, \frac{3}{8\pi \gamma
G} \,\, \dd c \wedge \dd p \quad {\rm and} \quad \{c,\, {p}\} =
\f{8\pi\g G}{3}\, . \ee
Since these expressions are now independent of the volume $V_o$ of
the cell ${\cal V}$ and make no reference to the fiducial metric
$\q_{ab}$, it is natural to regard the pair $(c,p)$ \textit{as the
basic canonical variables} on $\Gamma^S_{\rm grav}$. In terms of
$p$, the physical triad and co-triad are given by:
\be \label{e1} e^a_i = ({\rm sgn}\,p) |p|^{-\frac{1}{2}}\,\,\,
(V_o^{\frac{1}{3}}\,  \e^a_i), \quad {\rm and} \quad \omega_a^i =
({\rm sgn}\, {p}) |p|^{\frac{1}{2}}\,\,\, (V_o^{- \frac{1}{3}}\,
\w_a^i) \ee
The function ${\rm sgn}\, p$ arises because in connection-dynamics
the phase space contains triads with both orientations, and since
we have fixed a fiducial triad $\e^a_i$, the orientation of the
physical triad $e^a_i$ is captured in the sign of  ${p}$. (As in
the full theory, we also allow degenerate co-triads which now
correspond to ${p}=0$, for which the triad vanishes.)

Finally, note that although we have introduced an elementary cell
$\V$ and restricted all integrals to this cell, the spatial
topology is still $\R^3$ and  not  $\mathbb{T}^3$. Had the
topology been toroidal, connections with non-trivial holonomy
around the three circles would have enlarged the configuration
space and the phase space would then have inherited additional
components.

\subsubsection{Quantum kinematics}
\label{s2.1.2}

To construct quantum kinematics, one has to select a set of
elementary observables which are to have unambiguous operator
analogs. In non-relativistic quantum mechanics they are taken to
be $x,p$. One might first imagine using $c,p$ in their place. This
would be analogous to the procedure adopted in the \WDW theory.
However, unlike in geometrodynamics, LQG now provides a
well-defined kinematical framework for full general relativity,
without any symmetry reduction. Therefore, in the passage to the
quantum theory, we do not wish to treat the reduced theory as a
system in its own right but follow the procedure used in the full
theory. There, the elementary variables are: i) holonomies $h_e$
defined by the connection $A_a^i$ along edges $e$, and, ii) the
fluxes of triads $E^a_i$ across 2-surfaces $S$. In the present
case, this naturally suggests that we use: i) the holonomies
$h^{(\mu)}_k$  along straight edges  $(\mu\, \e^a_k)$  defined
by the connection $A = (c/\sqrt[3]{V_o})\, \w^i\tau_i$ , and ii)
the momentum $p$ itself \cite{abl}. Now, the holonomy along the
$k$th edge is given by:
\be h^{(\mu)}_k\, =\, \cos \f{\mu c}{2} {\mathbb{I}} + 2 \sin \f{\mu
c}{2}\,\, \tau_k  \label{hol}\ee
where $\mathbb{I}$ is the identity 2X2 matrix. Therefore, the
elementary configuration variables can be taken to be $\exp (i\mu
c/2)$ of $c$. These are called \emph{almost periodic} functions of
$c$ because $\mu$ is an arbitrary real number (positive if the
edge is oriented along the fiducial triad vector $\e^a_i$, and
negative if it is oriented in the opposite direction). The theory
of these functions was first analyzed by the mathematician Harold
Bohr (who was Niels' brother).

Thus, in LQC one takes $e^{i\mu c/2}$ and $p$ as the elementary
classical variables which are to have unambiguous operator
analogs. They are closed under the Poisson bracket on
$\Gamma^S_{\rm grav}$. Therefore, as in quantum mechanics, one can
construct an abstract $\star$-algebra $\a$ generated by $e^{i\mu
c/2}$ and $p$, subject to the canonical commutation relations. The
main task in quantum kinematics is to find the appropriate
representation of $\a$.

In this task, one again follows the full theory. There,
surprisingly powerful theorems are now available: By appealing to
the background independence of the theory, one can select an
irreducible representation of the holonomy-flux algebra
\emph{uniquely} \cite{lost,cf}. The unique representation was
constructed already in the mid-nineties and is therefore well
understood \cite{al2,jb1,mm,al3,al4,almmt}. In this
representation, there are well defined operators corresponding to
the triad fluxes and holonomies, but the connection itself does
not lead to a well-defined operator. Since one follows the full
theory in LQC, the resulting representation of $\a$ also has
well-defined operators corresponding to $p$ and almost periodic
functions of $c$ (and hence, holonomies), but there is no operator
corresponding to $c$ itself.

The Hilbert space underlying this representation is $\Hkg =
L^2(\R_{\rm Bohr}, d\mu_{\rm Bohr})$, where $\R_{\rm Bohr}$ is the
Bohr compactification of the real line and $\mu_{\rm Bohr}$ is the
Haar measure on it. It can be characterized as the Cauchy
completion of the space of continuous functions $f(c)$ on the real
line with finite norm, defined by:
\be ||f||^2 \, =\, \lim_{D \rightarrow \infty}\,\, \f{1}{2D}
\int_{-D}^D \bar{f} f\, dc\, . \ee
An orthonormal basis in $\Hkg$ is given by the almost periodic
functions of the connection, $N_{(\mu)}(c) := e^{i\mu c/2}$. (The
$N_{(\mu)} (c)$ are the LQC analogs of the spin network functions
in full LQG \cite{rs2,jb2}). They satisfy the relation
\be \langle N_{(\mu)}|N_{(\mu^\prime)}\rangle \, \equiv \, \langle
e^{\f{i\mu c}{2}}|e^{\f{i\mu^\prime c}{2}} \rangle \,\, = \,\,
\delta_{\mu, \mu^\prime}\, . \ee
Note that, although the basis is of the plane wave type, the right
side has a Kronecker delta, rather than the Dirac distribution.
Therefore a generic element $\Psi$  of $\Hkg$ can be expanded as a
\emph{countable sum} $\Psi (c) = \sum_k\, \alpha_k N_{(\mu_k)}$
where the complex coefficients $\alpha_k$ are subject to $\sum_k
|\alpha_k|^2 < \infty$. Consequently, the intersection between
$\Hkg$ and the more familiar Hilbert space $L^2(\R, dc)$ of
quantum mechanics (or of the \WDW theory) consists only of the
zero function. Thus, already at the kinematic level, the LQC
Hilbert space is \emph{very} different from that used in the \WDW
theory.

The action of the fundamental operators, however, is the familiar
one. The configuration operator acts by multiplication and the
momentum by differentiation:
%
\be \hat{N}_{(\mu)}\Psi (c)= \exp \f{i\mu c}{2} \Psi(c), \quad
{\rm and} \quad \hat{p} \Psi (c) \, =\,  -i \f{8\pi\g \lp^2}{3}\,
\f{\dd\Psi}{\dd c} \ee
where, as usual, $\lp^2 =G\hbar$. The first of these provides a
1-parameter family of unitary operators on $\Hkg$ while the second
is self-adjoint.

It is often convenient to use the Dirac bra-ket notation and set
$e^{i\mu c/2} = \langle c|\mu\rangle$. In this notation, the
eigenstates of $\hat{p}$ are simply the basis vectors
$|\mu\rangle$:
\be \hat{p}|\mu\rangle\, =\, \f{8\pi\g \lp^2}{6}\, \mu\,
|\mu\rangle\, . \ee
Finally, since the operator $\hat{V}$ representing the volume of
the elementary cell $\V$ is given by $\hat{V} =
|\hat{p}|^{\f{3}{2}}$, the basis vectors are also eigenstates of
$\hat{V}$:
\be \hat{V} |\mu\rangle \, =\,  \left(\f{8\pi \g}{6}\,
|\mu|\right)^{\f{3}{2}}\, \lp^3\,\, |\mu\rangle\, . \ee

The algebra $\a$, of course, also admits the familiar
representation on $L^2(\R, dc)$. Indeed, as we will see in section
\ref{s3}, this is precisely the `Schr\"odinger representation'
underlying the \WDW theory. The LQC representation outlined above
is unitarily inequivalent. This may seem surprising at first in
the light of the von-Neumann uniqueness theorem of quantum
mechanics. The LQG representation evades that theorem because
there is no operator $\hat{c}$ corresponding to the connection
component $c$ itself. Put differently, the theorem requires that
the unitary operators $\hat{N}_{(\mu)}$ be \emph{weakly
continuous} in $\mu$, while our operators on $\Hkg$ are not. (For
further discussion, see \cite{afw}.)

\subsection{The Hamiltonian constraint}
\label{s2.2}

\subsubsection{Strategy}
\label{s2.2.1}

Because of spatial flatness, the gravitational part of the
Hamiltonian constraint of full general relativity simplifies and
assumes the form:
\be C_{\rm grav} = -\gamma^{-2}\int_{\cal V}d^3x\,  N\,
\epsilon_{ijk}F_{ab}^i\,\, e^{-1} {E^{aj}E^{bk}} \ee
where $e:=\sqrt{|\det E|}$, and where we have restricted the
integral to our elementary cell $\V$. Because of spatial
homogeneity the lapse $N$ is constant and we will set it to one.

To obtain the corresponding constraint operator, we need to first
express the integrand in terms of our elementary phase space
functions $h_e, p$ and their Poisson brackets. The term involving
triad can be treated using the Thiemann strategy \cite{ttbook,tt}
:
\be \label{cotriad} \epsilon_{ijk}\, e^{-1}\,E^{aj}E^{bk}\, =\,
\sum_k \f{({\rm sgn}\,p)}{2\pi\gamma G\mu_o\, V_o^{\f{1}{3}}}\, \,
{}^o\!\epsilon^{abc}\,\, \w^k_c\,\, \Tr\left(h_k^{(\mu_o)}\,
\{h_k^{(\mu_o)}{}^{-1}, V\}\, \tau_i\right) \ee
where $h_k^{(\mu_0)}$ is the holonomy along the edge parallel to
the $k$th basis vector of length $\mu_o V_o^{1/3}$ with respect to
$\q_{ab}$, and $V= |p|^{3/2}$ is the volume function on the phase
space. While the right side of (\ref{cotriad}) involves $\mu_o$,
it provides an exact expression for the left side which is
independent of the value of $\mu_o$.

For the field strength $F_{ab}^i$, we use the standard strategy
used in gauge theories. Consider a square $\Box_{ij}$ in the
$i$-$j$ plane spanned by two of the triad vectors $\e^a_i$, each
of whose sides has length $\mu_o V_o^{1/3}$ with respect to the
fiducial metric $\q_{ab}$. Then, `the $ab$ component' of the
curvature is given by:
\be \label{F} F_{ab}^k\, = \, -2\,\lim_{\mu_o \rightarrow 0} \,\,
\Tr\, \left(\f{h^{(\mu_o)}_{\Box_{ij}}-1 }{\mu_o^2V_o^{2/3}}
\right) \,\, \tau^k\, \w^i_a\,\, \w^j_b  \ee
where the holonomy $h^{(\mu_o)}_{\Box_{ij}}$ around the square
$\Box_{ij}$ is just the product of holonomies (\ref{hol}) along
the four edges of $\Box_{ij}$:
\be
 h^{(\mu_o)}_{\alpha_{ij}}=h_i^{(\mu_o)} h_j^{(\mu_o)} (h_i^{(\mu_o)})^{-1}
 (h_j^{(\mu_o)})^{-1}\, .
\ee

By adding the two terms and simplifying, $C_{\rm grav}$ can be
expressed as:
\ba \label{reg} C_{\rm grav} &=& \lim_{\mu_o \rightarrow 0}\,\,
C^{(\mu_o)}_{\rm grav},\quad {\rm where}\nonumber\\
C^{(\mu_o)}_{\rm grav} &=& -\frac{4\, ({\rm sgn}\,
p)}{8\pi\gamma^3\mu_o^3 G} \sum_{ijk} \epsilon^{ijk}\, {\rm Tr}\,
(h_i^{(\mu_o)}h_j^{(\mu_o)} h_i^{(\mu_o)-1}h_j^{(\mu_o)-1}
h_k^{(\mu_o)}\{h_k^{(\mu_o)-1},V\}) \ea
Since $C^{(\mu_o)}_{\rm grav}$ is now expressed entirely in terms
of holonomies and $p$ and their Poisson bracket, it is
straightforward to write the corresponding quantum operator on
$\Hkg$:
\ba \label{qh1} \hat{C}_{\rm grav}^{(\mu_o)} &=& \frac{4i\, ({\rm
sgn}\, p)}{8 \pi \gamma^3\mu_o^3\lp^2}\,
\sum_{ijk}\epsilon^{ijk}\,{\Tr}\left(\hat{h}_i^{(\mu_o)}
\hat{h}_j^{(\mu_o)} \hat{h}_i^{(\mu_o)-1}\hat{h}_j^{(\mu_o)-1}
\hat{h}_k^{(\mu_o)}[\hat{h}_k^{(\mu_o)-1},\hat{V}]\right)\nonumber\\
&=& \f{24i\,({\rm sgn}\,p)}{8 \pi \gamma^3\mu_o^3\lp^2}\,
\sin^2\mu_o c \left(\sin\frac{\mu_oc}{2}\hat{V}
\cos\frac{\mu_oc}{2}- \cos\frac{\mu_oc}{2}\hat{V} \sin
\frac{\mu_oc}{2}\right)\, . \ea

However, the limit $\mu_o \rightarrow 0$ of this operator does not
exist. This is not accidental; had the limit existed, there would
be a well-defined operator directly corresponding to the
curvature $F_{ab}^i$ and we know that even in full LQG, while
holonomy operators are well defined, there are no operators
corresponding to connections and curvatures. This feature is
intimately intertwined with the quantum nature of Riemannian
geometry of LQG. The viewpoint in LQC is that the failure of the
limit to exist is a reminder that there is an underlying
\emph{quantum} geometry where eigenvalues of the area operator are
\emph{discrete}, whence there is a smallest non-zero eigenvalue,
$\Delta$, i.e., an \emph{area gap} \cite{rs1,al5}. Thus,
quantum geometry is telling us that it is physically incorrect to
let $\mu_o$ go to zero because that limit corresponds to shrinking
the area enclosed by loops $\Box_{ij}$ to zero. Rather, \emph{the
`correct' field strength operator in the quantum theory
  should be  in fact
non-local, given by setting $\mu_o$ in (\ref{F}) to a non-zero
value, appropriately related to $\Delta$.} In quantum theory, we
simply can not force locality by shrinking the loops $\Box_{ij}$
to zero area. In the classical limit, however, we are led to
ignore quantum geometry, and recover the usual, local field
$F_{ab}$ \cite{abl,jw}.

There are two ways of implementing this strategy. In this paper,
we will discuss the one that has been used in the literature
\cite{mbrev,abl,ab1}. The second strategy will be discussed in
\cite{aps3}; it has a more direct motivation and is more
satisfactory in semi-classical considerations especially in
presence of a non-zero cosmological constant.

In the above discussion, $\mu_o$ enters through holonomies
$h_k^{(\mu_o)}$. Now, it is straightforward to verify that (every
matrix element of) $h_k^{(\mu_o)}$ is an eigenstate of the area
operator $\widehat{Ar} = \widehat{|p|}$ (associated with the face of
the elementary cell $\V$ orthogonal to the $k$th direction):
\be \widehat{Ar}\,\, h_k^{(\mu_o)}(c)\, = \, \left(\f{8\pi\g
\mu_o}{6}\, \lp^2\right)\,\, h_k^{(\mu_o)}(c).\ee
In the first strategy, one fixes $\mu_o$ by demanding that this
eigenvalue be $\Delta = 2\sqrt{3}\pi \g \lp^2$, the area gap
\cite{alrev,al5}, so that $\mu_o = 3\sqrt{3}/{2}$.

To summarize, by making use of some key physical features of
quantum geometry in LQG, we have arrived at a `quantization' of
the classical constraint $C_{\rm grav}$:
\be \label{qh2} \hat{{C}}_{\rm grav} = \hat{C}_{\rm
grav}^{(\mu_o)}\mid_{\mu_o = \f{3\sqrt{3}}{2}}\ee

There are still factor ordering ambiguities which will be fixed in
the section \ref{s2.2.2}.

We will conclude our summary of the quantization strategy by
comparing this construction with that used in full LQG
\cite{tt,alrev,ttbook}. As one would expect, the curvature
operator there is completely analogous to (\ref{F}). However in
the subsequent discussion of the Hamiltonian constraint certain
differences arise: While the full theory is diffeomorphism
invariant, the symmetry reduced theory is not because of gauge
fixing carried out in the beginning of section \ref{s2.1.1}. 
(These differences are discussed in detail in sections 4 and 5 of
\cite{abl}.) As a result, in the dynamics of the reduced theory,
we have to `parachute' the area gap $\Delta$ from the full theory.
From physical considerations, this new input seems natural and the
strategy is clearly viable since that the constraint operator has
the correct classical limit. However, so far there is no systematic
procedure which leads us to the dynamics of the symmetry reduced
theory from that of the full theory. This is not surprising
because the Hamiltonian constraint in the full theory has many
ambiguities and there isn't a canonical candidate that stands out
as being the most satisfactory. The viewpoint in LQC is rather
that, at this stage, it would be more fruitful to study properties
of LQC constraints such as the one introduced in this section and
in \cite{aps3}, and use the most successful of them as a guide to
narrow down the choices in the full theory.

\subsubsection{Constraint operators and their properties}
\label{s2.2.2}

It is easy to verify that, although the classical constraint
function $C_{\rm grav}$ we began with is real on the phase space
$\Gamma^S_{\rm grav}$, the operator $\hat{C}_{\rm grav}^{(\mu_o)}$
is not self-adjoint on $\Hkg$. This came about just because of the
standard factor ordering ambiguities of quantum mechanics and
there are two natural re-orderings that can rectify this
situation.

First, we can simply take the self-adjoint part of (\ref{qh2}) on
$\Hkg$ and use it as the gravitational part of the constraint:
\be \label{qh3} \hat{C}_{\rm grav}^{\prime} = \f{1}{2}\,\,\left[
\hat{{C}}_{\rm grav}+ (\hat{{C}}_{\rm
grav})^\dag \right]\ee
It is convenient to express its action on states $\Psi(\mu) :=
\langle\Psi|\mu\rangle$ in the $\mu$ ---or the triad/geometry---
representation. The action is given by
\be \label{qh4} \hat{C}_{\rm grav}^{\prime}\, \Psi(\mu) \, =
f^\prime_+\, \Psi(\mu + 4\mu_o) + f^\prime_o\, \Psi(\mu) +
f^\prime_-\, \Psi(\mu - 4\mu_o)\ee
where the coefficients $f_{\pm}^\prime, f_o^\prime$ are functions
of $\mu$:
\ba
f^\prime_o (\mu)\, &=& -\sqrt{\frac{4\pi}{3}}
  \frac{\lp}{(\gamma\mu_o^2)^{\frac{3}{2}}}
  \left| |\mu+\mu_o|^{\frac{3}{2}}-|\mu-\mu_o|^{\frac{3}{2}} \right|
  \nonumber\\
f^\prime_+ (\mu)\, &=& -\frac{1}{4}
  (f^\prime_o(\mu)+f^\prime_o(\mu+4\mu_o))   \nonumber\\
f^\prime_- (\mu)\, &=& -\frac{1}{4}
  (f^\prime_o(\mu)+f^\prime_o(\mu-4\mu_o))  \, \ea
By construction, this operator is self-adjoint and one can show
that it is also bounded above (in particular, $\langle\Psi,\,
\hat{C}_{\rm grav}^{\prime}\, \Psi\rangle < \sqrt{\pi/3}\,\lp\,
(\gamma \mu_o)^{-\frac{3}{2}}(5^{\frac{3}{2}}-3^{\frac{3}{2}})
\langle\Psi|\Psi\rangle$). Next, consider the \emph{`parity'}
operator $\Pi$ defined by $\Pi \Psi(\mu) := \Psi(-\mu)$. It
corresponds to the flip of the orientation of the triad and thus
represents a large gauge transformation. It will play an important
role in section \ref{s4}. Here we note that the functions
$f^\prime_\pm, f^\prime_o$ are such that the constraint operator
$\hat{C}_{\rm grav}^{\prime}$ commutes with $\Pi$:
 \be [\,\hat{C}_{\rm grav}^{\prime} , \, \Pi\, ]\, =\, 0 \ee
Finally, for $\mu \gg \mu_o \equiv 3\sqrt{3}/2$, this difference
operator can be approximated, in a well-defined sense, by a second
order differential operator (a connection-dynamics analog of the
\WDW operator of geometrodynamics). Since $\hat{C}_{\rm
grav}^{\prime}$ is just the self-adjoint part of the operator used
in \cite{abl}, we can directly use results of section 4.2 of that
paper. Set $p = 8\pi \g\mu \lp^2/6$ and consider functions
$\Psi(p)$ which, together with their first four derivatives are
bounded.
On these functions, we have
\be \label{wdw1} \hat{C}_{\rm grav}^{\prime}\, \Psi(p) \approx
\f{64\pi^2}{3}\, \lp^4\, [\sqrt{p} \frac{\dd^2}{\dd p^2} +
\f{\dd^2}{\dd p^2} \sqrt{p}]\, \Psi(p) =: \hat{C}_{\rm
grav}^{\prime\,\,{\rm WDW}}\, \Psi(p)\ee
where $\approx$ stands for equality modulo terms of the order
$O(\mu_o)$. That is, had we left $\mu_o$ as a free parameter,
the equality would hold in the limit $\mu_o \rightarrow 0$. This
is the limit in which the area gap goes to zero, i.e., quantum
geometry effects can be neglected.

Let us now turn to the second natural factor ordering. The form of
the expression (\ref{qh1}) of $\hat{{C}}_{\rm grav}$ suggests
\cite{wk} that we simply `re-distribute' the $\sin^2 \mu_o c$ term
in a symmetric fashion. Then this factor ordering leads to the
following self-adjoint gravitational constraint:
\ba \label{qh5}\hat{C}_{\rm grav} &=& \f{24i\, ({\rm sgn}\, p)}{8
\pi \gamma^3\mu_o^3\lp^2}\, \left[\sin\mu_oc\,
\left(\sin\frac{\mu_oc}{2}\hat{V} \cos\frac{\mu_oc}{2}-
\cos\frac{\mu_oc}{2}\hat{V} \sin \frac{\mu_oc}{2}\right)\, \sin
\mu_oc\right]_{\mu_o = \f{3\sqrt{3}}{2}} \nonumber\\
& =: & \left[ \sin\mu_oc\,\, \hat{A}\,\, \sin\mu_oc \right]_{\mu_o
= \f{3\sqrt{3}}{2}} \, \ea
\emph{For concreteness in most of this paper we will work with
this form of the constraint and, for notational simplicity, unless
otherwise stated set $\mu_o = 3\sqrt{3}/2$.} However, numerical
simulations have been performed also using the constraint
$\hat{C}_{\rm grav}^{\prime}$ of Eq. (\ref{qh3}) and the results
are robust.

Properties of $\hat{C}_{\rm grav}$ which we will need in this
paper can be summarized as follows. First, the eigenbasis
$|\mu\rangle$ of $\hat{p}$ diagonalizes the operator $\hat{A}$.
Therefore in the $p$-representation, $\hat{A}$ acts simply by
multiplication. It is easy to verify that
\be \label{A} \hat{A}\, \Psi(\mu ) \, = \, - \,2 \,
\sqrt{\f{8\pi}{6}}\, \f{\lp}{(\g\mu_o^2)^{\f{3}{2}}}\,\, \left|
\, |\mu+\mu_o|^{\f{3}{2}} - |\mu-\mu_o|^{\f{3}{2}} \, \right|
\,\, \Psi(\mu)\ee
By inspection, $\hat{A}$ is self-adjoint and negative definite on
$\Hkg$. The form of $\hat{C}_{\rm grav}$ now implies that it is
also self-adjoint and negative definite. Its action on states
$\Psi(\mu)$ is given by
\be \label{qh6} \hat{C}_{\rm grav}\, \Psi(\mu) \, = f_+\, \Psi(\mu
+ 4\mu_o) + f_o\, \Psi(\mu) + f_-\, \Psi(\mu - 4\mu_o)\ee
where the coefficients $f_{\pm}, f_o$ are again functions of
$\mu$:
\ba f_+ (\mu)\, &=& \f{1}{2}\, \sqrt{\f{8\pi}{6}}\,
\f{\lp}{(\g\mu_o^2)^{\f{3}{2}}}\,\, \left|\,
|\mu+3\mu_o|^{\f{3}{2}} - |\mu+\mu_o|^{\f{3}{2}}\, \right|
\nonumber\\
f_- (\mu)\, &=& f_+(\mu -4\mu_o)\nonumber\\
f_o (\mu)\, &=& -f_+(\mu) - f_-(\mu)\, . \ea

It is clear from Eqs (\ref{qh5}) and (\ref{A}) that $\hat{C}_{\rm
grav}$ commutes with the parity operator $\Pi$ which flips the
orientations of triads:
\be [ \,\hat{C}_{\rm grav},\, \Pi\, ] =0\, .\ee
Finally, the `continuum limit' of the difference operator
$\hat{C}_{\rm grav}$ yields a second order differential
operator. Let us first set
\be \chi(\mu) = f_+(\mu- 2\mu_o) [\Psi(\mu+2\mu_o) -
\Psi(\mu-2\mu_o)] \ee
Then,
\be \hat{C}_{\rm grav}\Psi(\mu) = \chi(\mu+2\mu_o) -
\chi(\mu-2\mu_o) \ee
Therefore, if we again set $p = 8\pi \g\mu \lp^2/6$ and consider
functions $\Psi(p)$ which, together with their first four
derivatives are bounded 
we have
\be \label{wdw2} \hat{C}_{\rm grav}\, \Psi(p) \approx
\f{128\pi^2}{3}\, \lp^4\, [\f{\dd}{\dd p}\sqrt{p} \f{\dd\Psi }{\dd
p}] =: \hat{C}_{\rm grav}^{{\rm WDW}}\, \Psi(p)\ee
where $\approx$ again stands for equality modulo terms of the
order $O(\mu_o)$. That is, in the limit in which the area gap
goes to zero, i.e., quantum geometry effects can be neglected, the
difference operator reduces to a \WDW type differential operator.
(As one might expect, the limiting \WDW operator is independent of
the Barbero-Immirzi parameter $\g$.) We will use this operator
extensively in section \ref{s3}.

In much of computational physics, especially in numerical general
relativity, the fundamental objects are differential equations and
discrete equations are introduced to approximate them. In LQC the
situation is just the opposite. The physical fundamental object is
now the discrete equation (\ref{qh6}) with $\mu_o= 3\sqrt{3}/2$.
The differential equation is the approximation. The leading
contribution to the difference between the two ---i.e., to the
error --- is of the form $O(\mu_o)^2 \Psi^{''''}$ where the second
term depends on the wave function $\Psi$ under consideration.
Therefore, the approximation is not uniform. For semi-classical
states, $\Psi^{''''}$ can be large, of the order of
$10^{16}/\mu^4$ in the examples considered in section \ref{s5}. In
this case, the continuum approximation can break down already at
$\mu \sim 10^4$.

As one might expect the two differential operators, $\hat{C}_{\rm
grav}^{\prime{\rm WDW}}$ and $\hat{C}_{\rm grav}^{{\rm WDW}}$
differ only by  a factor ordering:
\be \label{wdwdiff} \left(\hat{C}_{\rm grav}^{\prime{\rm WDW}} -
\hat{C}_{\rm grav}^{{\rm WDW}}\right) \Psi(p) = -\,
\f{16\pi^2}{3}\,\lp^4 |p|^{-\f{3}{2}}\, \Psi(p)\, .\ee

\textbf{Remark:} As noted in section \ref{s2.2.1}, the `continuum
limit' $\mu_o \rightarrow 0$ of any of the quantum constraint
operators of LQC does not exist on $\Hkg$ because of the quantum
nature of underlying geometry. To take this limit, one has to work
in the setting in which the quantum geometry effects are
neglected, i.e., on the \WDW type Hilbert space $L^2(\R, dc)$. On
this space, operators $\hat{C}_{\rm grav}^{\prime}$, $\hat{C}_{\rm
grav}$,  $\hat{C}_{\rm grav}^{\rm WDW}$  and $\hat{C}_{\rm
grav}^{\prime{\rm WDW}}$ are all densely defined and the limit can
be taken on a suitable dense domain.

\subsection{Open issues and the model}
\label{s2.3}

In section \ref{s2.1} we recalled the kinematical framework used
in LQC and in \ref{s2.2} we extended the existing results by
analyzing two self-adjoint Hamiltonian constraint operators in
some detail. Physical states $\Psi(\mu, \phi)$ can now be
constructed as solutions of the Hamiltonian constraint:
\be \label{qc1} (\hat{C}_{\rm grav} + \hat{C}_{\rm matt})\,
\Psi(\mu, \phi) =0\ee
where $\phi$ stands for matter fields. Given any matter model, one
could solve this equation numerically. However, generically, the
solutions would not be normalizable in the total kinematic Hilbert
space $\Hk$ of gravity plus matter. Therefore, although section
\ref{s2.2} goes beyond the existing literature in LQC, one still
can not calculate expectation values, fluctuations and
probabilities ---i.e., extract physics--- knowing only these
solutions.

To extract physics, then, we still have to complete the following
tasks:

\medskip\noindent
 \b In the classical theory, seek a dynamical variable which
is monotonically increasing on all solutions (or at least `large'
portions of solutions). Attempt to interpret the Hamiltonian
constraint (\ref{qc1}) as an `evolution equation' with respect to
this `internal time'. If successful, this strategy would provide
an `emergent time' in the background independent quantum theory.
Although it is possible to extract physics under more general
conditions, physical interpretations are easier and more direct if
one can locate such an emergent time.\\
\b Introduce an inner product on the space of solutions to
(\ref{qc1}) to obtain the physical Hilbert space $\Hp$. The fact
that the orientation reversal induced by $\Pi$ is a large
gauge transformation will have to be handled appropriately.\\
\b Isolate suitable Dirac observables in the classical theory and
represent them by self-adjoint operators on $\Hp$.\\
\b Use these observables to construct physical states which are
semi-classical at `late times', sharply peaked at a point on the
classical trajectory representing a large classical universe.\\
\b `Evolve' these states using (\ref{qc1}). Monitor the mean
values and fluctuations of Dirac observables. Do the mean values
follow a classical trajectory? Investigate if there is a drastic
departure from the classical behavior. If there is, analyze what
replaces the big-bang.

\medskip
In the rest of the paper, we will carry out these tasks in the
case when matter consists of a zero rest mass scalar field. In the
classical theory, the phase space $\Gamma^S_{\rm grav}$ is now
4-dimensional, coordinatized by $(c,p;\, \phi, p_\phi)$. The basic
(non-vanishing) Poisson brackets are given by:
\be\label{pbs} \{{c},\, {p}\} = \f{8\pi\g G}{3}\, ,\quad {\rm and}
\quad \{\phi,\, p_\phi\} = 1\, .\ee
The symmetry reduction of the classical Hamiltonian constraint is of
the form
\be \label{cc} C_{\rm grav} + C_{\rm matt} \equiv -
\f{6}{\g^2}\, c^2 \sqrt{|p|} + 8\pi G\,
\f{p_\phi^2}{|p|^{\f{3}{2}}}\, = \, 0 ~.\ee
\begin{figure}
\begin{center}
\includegraphics[width=3.5in,angle=0]{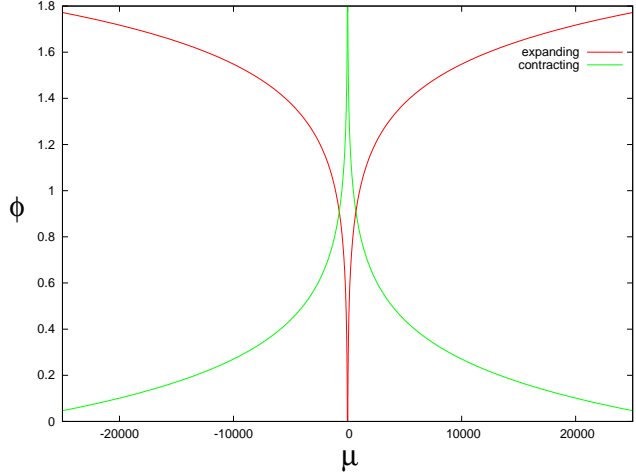}
\caption{Classical phase space trajectories are plotted in the
$\phi, p\sim\mu$ -plane. For $\mu \ge 0$, there is a branch which
starts with a big-bang (at $\mu =0$) and expands out and a branch
which contracts into a big crunch (at $\mu =0$). Their mirror
images appear in the $\mu \le 0$ half plane.} \label{f1}
\end{center}
\end{figure}

Using this constraint, one can solve for $c$ in terms of $p$ and
$p_\phi$. Furthermore, since $\phi$ does not enter the expression
of the constraint, $p_\phi$ is a constant of motion. Therefore,
each dynamical trajectory can be specified on the 2-dimensional
$(p,\, \phi)$ plane. Typical trajectories are shown in Fig. 1.
Because the phase space allows triads with both orientations, the
variable $p$ can take both positive and negative values. At $p=0$
the physical volume of the universe goes to zero and, if the point
lies on any dynamical trajectory, it is an end-point of that
trajectory, depicting a curvature singularity. As the figure
shows, for each fixed value of $p_\phi$, there are four types of
trajectories, two in the $p \ge 0$ half plane and two in the $p\le
0$ half plane. Analytically they are given by:
\be \phi = \pm\, \sqrt{\f{3}{16\pi G}}\, \ln \f{|p|}{|p_\star|} +
\phi_\star \ee
where $p_\star, \phi_\star$ are integration constants. These
trajectories are related by a `parity transformation' on the phase
space which simply reverses the orientation of the physical triad.
As noted before, since the metric and the scalar field are
unaffected, it represents a large gauge transformation.
Therefore, it suffices to focus just on the portion $p \ge 0$ of
figure 1. Then for each fixed value of $p_\phi$, there are two
solutions passing through any given point $(\phi_\star, p_\star)$.
In one, the universe begins with the big-bang and then expands and
in the other the universe contracts into a big crunch. Thus
\emph{in this model, every classical solution meets the
singularity.}

Finally, we can introduce a natural set of  \emph{Dirac
observables}. Since $p_\phi$ is a constant of motion, it is
obviously one. To introduce others, we note that $\phi$ is a
monotonic function on each classical trajectory. Furthermore, in
each solution, the space-time metric takes the form $ds^2 = -dt^2
+ V_o^{-2/3}|p|(t) dS_o^2$ and the time dependence of the scalar
field is given by
\be \f{\dd\phi}{dt} = \f{16\pi G
p_\phi^\star}{|p_\star|^{\f{3}{2}}} \,\, \exp\, [\pm{\sqrt{12\pi
G}\, (\phi - \phi_\star)}]\, , \ee
where $p_\star,\phi_\star$ and $p_\phi^\star$ are constants. Thus,
in every solution $\phi$ is a monotonic function of time and can
therefore serve as a good `internal clock'. This interpretation
suggests the existence of a natural family of Dirac observables: $
p|_{\phi_o}$, the value of $p$ at the `instant' $\phi_o$. The set
$(p_\phi,\, p|_{\phi_o})$ constitutes a complete set of Dirac
observables since their specification uniquely determines a
classical trajectory on the symmetry reduced phase space
$\Gamma^{S}$, i.e., a point in the reduced phase space
$\tilde\Gamma^{S}$. While the interpretation of $\phi$ as
`internal time' motivates this construction and, more generally,
makes physics more transparent, it is not essential. One can do
all of physics on $\tilde\Gamma^{S}$: Since physical states are
represented by points in $\tilde\Gamma^{S}$ and a complete set of
observables is given by $(p_\phi, p|_{\phi_o})$, one can work
just with this structure.

\textbf{Remark:} In the open i.e., $k\! = \! 0$ model now under
consideration, since $p$ is also monotonic along any classical
trajectory, $\phi_{p_o}$ is also a Dirac observable and $p$ could
also be used as an `internal time'. However, in LQC the expression of
the gravitational part of the constraint operator makes it
difficult to regard $p$ as the `emergent time' in quantum theory.
Moreover even in the classical theory, since the universe expands
and then recollapses in the $k\! =\! 1$ case,  $p$ fails to be
monotonic along solutions and cannot serve as `internal time'
globally.

\section{ \WDW theory}
\label{s3}

The \WDW theory of the model has been analyzed in some detail
within geometrodynamics (see especially \cite{ck}). However, that
analysis was primarily in the context of a WKB approximation. More
recently, the group averaging technique was used to construct the
physical Hilbert space in a general cosmological context
\cite{dm,hm2}, an elementary example of which is provided by the
present model. However, to our knowledge a systematic completion
of the program outlined in section \ref{s2.3} has not appeared in
the literature.

In this section we will construct the \WDW type quantum theory in
the \emph{connection dynamics}. This construction will serve two
purposes. First it will enable us to introduce the key notions
required for the completion of the program in a familiar and
simpler context. Second, we will be able to compare and contrast
the results of the \WDW theory and LQC in detail, thereby bringing
out the role played by quantum geometry in quantum dynamics.

\subsection{Emergent time and the general solution to the WDW equation}
\label{s3.1}

Recall that the phase space of the model is 4-dimensional,
coordinatized by $(c,p;\, \phi,p_\phi)$ and the fundamental
non-vanishing Poisson brackets are given by (\ref{pbs}). The
Hamiltonian constraint has the form:
\be C_{\rm grav}+ C_{\phi} \equiv -\frac{6}{\g^2}\, c^2 \sqrt{|p|}
+ 8\pi G\, \f{p_\phi^2}{|p|^{\f{3}{2}}} \, = \, 0\, .  \ee
To make comparison with the standard geometrodynamical \WDW
theory, it is most convenient to work in the $p,\phi$
representation. Then, the kinematic Hilbert space is given by
$\Hkwdw =L^2(\R^2, \dd p\, \dd\phi)$. Operators $\hat{p},
\hat{\phi}$ operate by multiplication while $\hat{c}$ and
$\hat{p}_\phi$ are represented as:
\be \hat{c}\, \Psi = i\hbar\, \f{8\pi\g G}{3}\, \frac{\partial
\Psi}{\partial p} \quad {\rm and} \quad \hat{p}_\phi \Psi =
-i\hbar \f{\partial\Psi}{\partial \phi}\ee
Note that, while in geometrodynamics the scale factor is
restricted to be non-negative, here $p$ ranges over the entire
real line, making the specification of the Hilbert space and
operators easier.

To write down the quantum constraint operator, we have to make a
choice of factor ordering. Since our primary motivation behind the
introduction of the \WDW theory is to compare it with LQC, it is
most convenient to use the factor ordering that comes from the
continuum limit (\ref{wdw2}) of the constraint operator of LQC.
Then, the \WDW equation becomes:
\be \label{wdw3} \f{2}{3} (8\pi G\hbar)^2 \f{\partial}{\partial
p}\sqrt{p}\f{\partial \Psi}{\partial p} = {8\pi G\hbar^2}\,
\widehat{|p|^{-\f{3}{2}}}\,\, \f{\partial^2\Psi}{\partial\phi^2}
=: 8\pi G \hbar^2 \ub{B}(p) \f{\partial^2\Psi}{\partial\phi^2}\,
,\ee
where we have denoted the eigenvalue $|p|^{-\f{3}{2}}$  of
$\widehat{|p|^{-\f{3}{2}}}$ by $\ub{B}(p)$ to facilitate later
comparison with LQC.%
\footnote{Here, and in what follows, quantities with an underbar
will refer to the \WDW theory.}
The operator on the left side of this equation is self-adjoint on
$L^2(\R, dp)$ and the equation commutes with the orientation
reversal operator $\Pi \Psi(p,\phi) = \Psi(-p, \phi)$ representing
a large gauge transformation. Thus, if $\Psi$ is a solution to Eq.
(\ref{wdw3}), so is $\Psi(-p, \phi)$.

For a direct comparison with LQC, it is convenient to replace $p$
with $\mu$ defined by $p = ({8\pi \g G\hbar}/{6})\, \mu$.  Then,
(\ref{wdw3}) becomes:
\ba \label{wdw4}\f{\partial^2 \Psi}{\partial\phi^2} \, &=& \,
\f{16\pi G}{3}\, [\ub{B}(\mu)]^{-1}\,\, \f{\partial}{\partial
\mu}\sqrt{\mu}\f{\partial
\Psi}{\partial \mu} \nonumber\\
 &=:& -\, \ul{\Theta} \Psi\ea
where we have recast the equation in such a way that operators
involving only $\phi$ appear on the left side and operators
involving only $\mu$ appear on the right. The \WDW equation now
has the same form as the Klein-Gordon equation in a static
space-time, $\phi$ playing the role of time and $\ul{\Theta}$ of
the (elliptic operator constructed from the norm of the Killing
field and the) spatial Laplacian. Thus, the form of the quantum
Hamiltonian constraint is such that $\phi$ can be interpreted as
\emph{emergent time} in the quantum theory. In this factor
ordering of the constraint operator, which emerged in the
continuum limit (\ref{wdw2}) of LQC, it is not as convenient to
regard $p \sim \mu$ as emergent time.

Physical states will be suitably regular solutions to
(\ref{wdw4}). Since $\Pi$ is a large gauge transformation, we can
divide physical states into eigenspaces of $\hat{P}$. Physical
observables will preserve each eigenspace. Since $\Pi^2 =1$, there
are only two eigenspaces, one representing the symmetric sector
and the other, anti-symmetric. Since the standard \WDW theory
deals with metrics, it is completely insensitive to the
orientation of the triad. Therefore, it is natural to work with
the symmetric sector. Thus, \emph{the physical Hilbert space will
consist of suitably regular solutions $\Psi(\mu, \phi)$ to
(\ref{wdw4}) which are symmetric under $\mu \rightarrow -\mu$}.

The mathematical similarity with the Klein Gordon equation in
static space-times immediately suggests a strategy to obtain the
general solution of (\ref{wdw4}). We first note that 
$(\dd/\dd\mu) \, \sqrt{\mu}\, (\dd/\dd \mu)$  \,\,is a negative
definite, self-adjoint operator on $L^2_S(\R, d\mu)$, the
symmetric sector of $L^2(\R, d\mu)$. Therefore, $\ul{\Theta}$ is a
positive definite, self-adjoint operator on $L^2_S(\R, \ub{B}(\mu)
d\mu)$. Its eigenfunctions provide us with an orthonormal basis.
It is easy to verify that the eigenvectors $\ub{e}_k(\mu)$ can be
labelled  by $k\in \R$ and are given by
\be \label{eq:ek} \ub{e}_k(\mu) \, := \,
\f{{|\mu|}^{\f{1}{4}}}{4\pi}\, e^{ik\ln |\mu|}\,  \ee
Their eigenvalues are given by:
\be \ul{\Theta}\, \ub{e}_k(\mu) = \omega^2 \ub{e}_k(\mu), \quad
{\rm with}\,\,\,\, \omega^2 = \f{16\pi G}{3}(k^2 + \f{1}{16})\, ;
\ee
(where the factor of $1/16$ is an artifact of the factor ordering
choice which we were led to from LQC). The eigenfunctions satisfy
the orthonormality relations:
\be \int_{-\infty}^\infty d\mu \ub{B}(\mu)\, \bar{\ub{e}}_k(\mu)\,
\ub{e}_{k'}(\mu) = \delta(k,k')\, , \ee
(where the right side is the standard Dirac distribution, not the
Kronecker symbol as on $L^2(\R_{\rm Bohr}, d\mu_{\rm Bohr})$); and
the completeness relation:
\be   \int_{-\infty}^\infty d\mu \ub{B}(\mu)\,
\bar{\ub{e}}_k(\mu)\,\Psi(\mu) =0 \quad \forall k \quad{\rm
iff}\,\, \Psi(\mu) =0 \ee
for any $\Psi(\mu) \in L^2_S(\R, \ub{B}(\mu) d\mu)$.

With these eigenfunctions at hand, we can now write down a
`general' symmetric solution to (\ref{wdw3}). Any solution, whose
initial data at $\phi=\phi_o$ is such that $\mu^{-1/4}
\Psi(\mu,\phi_o)$ and $\mu^{-1/4} \dot\Psi(\mu,\phi_o)$ are
symmetric and lie in the Schwartz space of rapidly decreasing
functions, has the form:
\be \label{sol1} \Psi(\mu,\phi) \, = \, \int_{-\infty}^\infty dk\,
\t\Psi_+(k) \ub{e}_k(\mu) e^{i\omega\phi} + \t\Psi_-(k)
\bar{\ub{e}}_k(\mu) e^{-i\omega\phi} \, ,\ee
for some $\t\Psi_\pm (k)$ in the Schwartz space. Following the
terminology used in the Klein-Gordon theory, if $\t\Psi_\pm(k)$
have support on the negative $k$-axis, we will say the solution is
`outgoing' (or `expanding') while if it has support on the
positive $k$ axis, it is `incoming' (or, `contracting').  If
$\t\Psi_-(k)$ vanishes, the solution will be said to be
\emph{positive frequency} and if $\t\Psi_+(k)$ vanishes, it will
be said to be of \emph{negative frequency}. Thus, every solution
(\ref{sol1}) admits a natural decomposition into positive and
negative frequency parts.  Finally we note that positive
(respectively negative) frequency solutions satisfy a first order
(in $\phi$) equation which can be regarded as the square-root of
(\ref{wdw4}):
\be \label{sch} \mp i\,\f{\partial\Psi_\pm}{\partial \phi} =
\sqrt{\ul{\Theta}}\, \Psi_\pm \ee
where $\sqrt{\ul{\Theta}}$ is the positive, self-adjoint operator
defined via spectral decomposition of $\ul{\Theta}$ on $L^2(\R,
\ub{B}(\mu)d\mu)$. Regarding $\phi$ as time, this is just a first order
Schr\"odinger equation with a \emph{non-local} Hamiltonian
$\sqrt{\ul{\Theta}}$. Therefore, a general `initial datum' $f_\pm
(\mu)$ at $\phi=\phi_o$ can be `evolved' to obtain a solution to
(\ref{sch}) via:
\be\label{sol2} \Psi_{\pm}(\mu,\phi) =
e^{\pm\,i\sqrt{\ul{\Theta}}(\phi-\phi_o)}\, f_\pm(\mu, \phi_o)\ee

\subsection{The physical sector}
\label{s3.2}

Solutions (\ref{sol1}) are not normalizable in $\Hkwdw$ (because
zero is in the continuous part of the spectrum of the \WDW
operator). Our first task is to endow the space of these physical
states with a Hilbert space structure. There are several possible
avenues. We will begin with one that is somewhat heuristic but has
direct physical motivation. The idea \cite{aabook,at} is to
introduce operators corresponding to a complete set of Dirac
observables and select the required inner product by demanding
that they be self-adjoint. In the classical theory, such a set is
given by $p_\phi$ and $\mu|_{\phi_o}$. Since $\hat{p}_\phi$
commutes with the \WDW operator in (\ref{wdw4}), given a
(symmetric) solution $\Psi(\mu,\phi)$ to (\ref{wdw4}),
\be \label{dirac1}\hat{p}_\phi \Psi(\mu, \phi) := -i\hbar
\f{\partial \Psi}{\partial \phi} \ee
is again a (symmetric) solution. So, we can just retain this
definition of $\hat{p}_\phi$ from $\Hkwdw$. The Schr\"odinger type
evolutions (\ref{sol2}) enable us to define the other Dirac
observable $\widehat{|\mu|_{\phi_o}}$, where the absolute
value suffices because the states are symmetric under $\Pi$.
 Given a (symmetric)
solution $\Psi(\mu,\phi)$ to (\ref{wdw4}), we can first decompose
it into positive and negative frequency parts
$\Psi_\pm(\mu,\phi)$, freeze them at $\phi=\phi_o$, multiply this
`initial datum' by $|\mu|$ and evolve via (\ref{sol2}):
\be \label{dirac2} \widehat{|\mu|_{\phi_o}}\,\Psi (\mu,\,\phi)
= e^{i\sqrt{\ul{\Theta}}(\phi-\phi_o)}\, |\mu| \,
\Psi_+(\mu,\,\phi_o) + e^{-i\sqrt{\ul{\Theta}}(\phi-\phi_o)}\,
|\mu| \, \Psi_-(\mu,\,\phi_o) \ee
The result is again a (symmetric) solution to (\ref{wdw4}). Now,
we see that both these operators have the further property that
they preserve the positive and negative frequency subspaces. Since
they constitute a complete family of Dirac observables, we have
\emph{superselection}. In quantum theory we can restrict ourselves
to one superselected sector. In what follows, for definiteness
\emph{we will focus on the positive frequency sector and, from now on,
drop the suffix $+$}.

We now seek an inner product on positive frequency solutions
$\Psi(\mu,\phi)$ to (\ref{wdw4}) (invariant under the $\mu$
reflection) which makes $\hat{|p|}_\phi$ and
$\hat{|\mu|}_{\phi_o}$ self-adjoint. Each of these solutions is
completely determined by its initial datum $\Psi(\mu,\phi_o)$ and
the Dirac observables have the following action on the datum:
\be \widehat{|\mu|_{\phi_o}}\, \Psi(\mu,\phi_o) = |\mu|
\Psi(\mu,\phi_o),\quad {\rm and} \quad \hat{p}_\phi
\Psi(\mu,\phi_o) = \hbar \sqrt{\ul{\Theta}} \Psi(\mu,\phi_o)\, .
\ee
Therefore, it follows that (modulo an overall rescaling,) the
unique inner product which will make these operators self-adjoint
is just:
\be \label{ip1} \langle\Psi_1|\Psi_2\rangle_{\rm phy} =
\int_{\phi=\phi_o} d\mu \ub{B}(\mu)\, \bar{\Psi}_1\, \Psi_2 \ee
(see e.g. \cite{aabook,at}). Note that the inner product is
conserved, i.e., is independent of the choice of the `instant'
$\phi=\phi_o$. Thus, \emph{the physical Hilbert space $\Hpwdw$ is
the space of positive frequency wave functions $\Psi(\mu,\phi)$
which are symmetric under $\mu$ reflection and have a finite norm
(\ref{ip1}).} The procedure has already provided us with a
representation of our complete set of Dirac observables on this
$\Hpwdw$:
\be \label{dirac3}\widehat{|\mu|_{\phi_o}}\, \Psi(\mu,\phi) =
e^{i\sqrt{\ul{\Theta}}(\phi-\phi_o)}\,|\mu|\, \Psi(\mu,\phi_o),
\quad {\rm and} \quad
 \hat{p}_\phi \Psi(\mu,\phi) = \hbar \sqrt{\ul{\Theta}} \Psi(\mu,\phi)
 \, . \ee

We will now show that the same representation of the algebra of
Dirac observables can be obtained by the more systematic group
averaging method \cite{dm,hm2} which also brings out the
mathematical inputs that go in this choice. (The two methods have
been applied and compared for a non-trivially constrained system
in \cite{lr}). Here, one first notes that the total constraint
operator is self-adjoint on an auxiliary Hilbert space $\Hawdw :=
L^2_S(\R^2, \ub{B}(\mu)\dd\mu\,\dd\phi)$, (where, as before the
subscript $S$ denotes restriction to functions which are symmetric
under $\mu$-reflection). One must then select an appropriate dense
subspace $\Phi$ of $\Hawdw$. A natural candidate is the Schwartz
space of rapidly decreasing functions. One then `averages'
elements of $\Phi$ under the 1-parameter family group generated by
the constraint operator $\hat{\ul{C}}= \p_\phi^2 +
{\underline\Theta}$ to produce a solution to (\ref{wdw4}):
\ba \Psi_f(\mu,\phi) :&=& \int_{-\infty}^{\infty}\, \dd\alpha\,\,
e^{i\alpha\hat{\ub{C}}}\, f(\mu,\phi)\nonumber\\
&=& \int_{-\infty}^\infty \f{\dd k}{2|\omega|}\,\,
\left(\t{f}(k,\omega)\,\ub{e}_k(\mu)\,e^{i\omega\phi}
+\t{f}(k,-\omega)\, \ub{e}_k(\mu)\,e^{-i\omega\phi}\right) \ea
where to arrive at the second step we expanded $f$ in the
eigenbasis of $\ul{\Theta}$ and $\hat{p}_\phi$ with $\tilde{f}$ as
the coefficients. Thus, the group averaging procedure reproduces
the solution (\ref{sol1}) with $\t\Psi_+(k)=
\t{f}(k,\omega)/2|\omega|$ and $\t\Psi_- (k)=
\t{f}(k,-\omega)/2|\omega|$. Solutions $\Psi_f$ are regarded as
`distributions' or elements of the dual $\Phi^\star$ of $\Phi$ and
the physical norm is given by the action of this distribution,
$\Psi_f$, on the `test function' $f$. However, there is some
freedom in the specification of this action which generally
results in seemingly different but unitarily equivalent
representations of the algebra of Dirac observables. For us the
most convenient choice is:
\be ||\Psi||^2 := \Psi_f\,(f) := \int_{-\infty}^\infty \dd\phi
\int_{-\infty}^\infty \dd\mu \ub{B}(\mu) \bar{\Psi}_f (\mu,\phi)
\sqrt{\ul{\Theta}} f(\mu,\phi)\ee
where $\Psi_f(f)$ is the action of the distribution $\Psi_f$ on
the test field $f$. Then, the inner product coincides with
(\ref{ip1}) and the representation of the Dirac observables is the
same as in (\ref{dirac1}) and (\ref{dirac2}). Had we chosen to
drop the factor of $\sqrt{\ul{\Theta}}$ in defining the action of
$\Psi_f$ on $f$, we would have obtained a unitarily equivalent
representation in which the action of $\hat{\mu}|_{\phi_o}$ is
more complicated.

Finally, with the physical Hilbert space and a complete set of
Dirac observables at hand, we can now introduce semi-classical
states and study their evolution. Let us fix a `time'
$\phi=\phi_o$ and construct a semi-classical state which is peaked
at $p_\phi = p_\phi^\star$ and $|\mu|_{\phi_o} = \mu^\star$. We
would like the peak to be at a point that lies on a large
classical universe. This implies that we should choose $\mu^\star
\gg 1$ and (in the natural classical units $c$=$G$=1),\,\,
$p_\phi^\star \gg \hbar$. In the closed ($k\!=\! 1$) models for
example, the second condition is necessary to ensure that the
universe expands out to a size much larger than the Planck scale.
At `time' $\phi=\phi_o$, consider the state
\be \label{sc} \Psi(\mu,\phi_o) = \int_{-\infty}^\infty dk
\t\Psi(k)\, \ub{e}_k(\mu)\, e^{i\omega(\phi_o-\phi^\star)}, \quad
{\rm where}\,\, \t\Psi(k) = e^{-\f{(k-k^\star)^2}{2\sigma^2}}\,
. \ee
where $k^\star = -\sqrt{3/16\pi G\hbar^2}\,\, p_\phi^\star$ and
$\phi^\star = -\sqrt{3/16\pi G} (\ln |\mu^\star|) + \phi_o$. It is
easy to evaluate the integral in the approximation $\omega =
-\sqrt{(16\pi G/3)}\, k$ (which is justified because $k^\star \ll
-1$) and calculate mean values of the Dirac observables and their
fluctuations. One finds that, as required, the state is sharply
peaked at values $\mu^\star, p_\phi^\star$. The above construction
is closely related to that of coherent states in non-relativistic
quantum mechanics. The main difference is that the observables of
interest are not $\mu$ and its conjugate momentum but rather $\mu$
and $p_\phi$ ---the momentum conjugate to `time', i.e., the
analog of the Hamiltonian in non-relativistic quantum mechanics.

We can now ask for the evolution of this state. Does it remain
peaked at the classical trajectory defined by $p_\phi =
p_\phi^\star$ passing through $\mu^\star$ at $\phi =\phi_o$? This
question is easy to answer because (\ref{sol2}) implies that the
(positive frequency) solution  $\Psi(\mu,\phi)$ to (\ref{wdw4})
defined by `initial data' (\ref{sc}) is obtained simply by
replacing $\phi_o$ by $\phi$ in (\ref{sc})! Since the measure of
dispersion $\sigma$ in (\ref{sc}) does not depend on $\phi$, it
follows that the initial state $\Psi(\mu,\phi_o)$ which is the
semi-classical, representing a large universe at `time' $\phi_o$
continues to be peaked at a trajectory defined by:
\be \phi =  \sqrt{\f{3}{16\pi G}}\,\,  \ln{\f{|\mu|}{|\mu^\star|}}
+  \phi_o\, .\ee
This is precisely the classical trajectory with $p_\phi =
p_\phi^\star$, passing through $\mu^\star$ at $\phi =\phi_o$. This
is just as one would hope during the epoch in which the universe
is large. However, this holds also in the Planck regime and, in
the backward evolution, the semi-classical state simply follows
the classical trajectory into the big-bang singularity. (Had we
worked the positive $k^\star$, we would have obtained a
contracting solution and then the forward evolution would have
followed the classical trajectory into the big-crunch
singularity.) In this sense, the \WDW evolution does not resolve
the classical singularity.

\textbf{Remark:} In the above discussion for simplicity we
restricted ourselves to eigenfunctions $\ub{e}_k(\mu)$ which are
symmetric under $\mu\,\rightarrow\, -\mu$ from the beginning. Had
we dropped this requirement, we would have found that there is a
4-fold (rather than 2-fold) degeneracy in the eigenfunctions of
$\ul{\Theta}$. Indeed, if $\theta(\mu)$ is the step function
($\theta(\mu) =0$ if $\mu < 0$ and $\, =1$ if $\mu >0$), then
$\theta(\mu)\ub{e}_{|k|},\, \theta(\mu) \ub{e}_{-|k|},\,
\theta(-\mu)\ub{e}_{|k|},\, \theta(-\mu) \ub{e}_{-|k|}$ are all
continuous functions of $\mu$ which satisfy the eigenvalue
equation (in the distributional sense) with eigenvalue $\omega^2 =
(16\pi G/3)(k^2 + 1/16)$. This fact will be relevant in the next
section.

\section{Analytical issues in Loop quantum cosmology}
\label{s4}

We will now analyze the model within LQG. We will first observe
that the form of the Hamiltonian constraint is such that the
scalar field $\phi$ can again be used as emergent time. Since the
form of the resulting `evolution equation' is very similar in the
\WDW theory, we will be able to construct the physical Hilbert
space and Dirac observables following the ideas introduced in
section \ref{s3.2}.

\subsection{Emergent time and the general solution to the LQC
Hamiltonian constraint} \label{s4.1}

The quantum constraint has the form
\be \hat{C}_{\rm grav} + \hat{C}_{\rm \phi} =0 \ee
where $\hat{C}_{\rm grav}$ is given by (\ref{qh6}). Since
$\hat{C}_{\rm \phi} = (8\pi G)\, (\widehat{1/p^{3/2}})\,
(\hat{p}_\phi^2)$, the constraint becomes:
\be \label{eq:main} 8\pi G \hat{p}_\phi^2\, \Psi(\mu,\phi) =
[\tilde B(p)]^{-1}\, \hat{C}_{\rm grav} \Psi(\mu,\phi) \ee
where  $\tilde B(p)$ is the eigenvalue of the
operator $\widehat{1/|p|^{3/2}}$:
\be \label{eq:bp} \tilde B(p) =: \left(\f{6}{8 \pi \gamma \lp^2}\right)^{3/2} \,
B(\mu), \, \, {\rm where} \, \, B(\mu) =  \left(\f{2}{3\mu_o}\right)^6 \,\left[|\mu +
\mu_o|^{\f{3}{4}} - |\mu-\mu_o|^{\f{3}{4}}\right]^6 ~.\ee
Thus, we now have a separation of variables. Both the classical
and the \WDW theory suggests that $\phi$ could serve as emergent
time. To implement this idea, let us introduce an appropriate
kinematical Hilbert space for both geometry and the scalar field:
$\Hk := L^2(\R_{\rm Bohr}, B(\mu)\dd\mu_{\rm Bohr}) \otimes
L^2(\R, \dd \phi)$. Since $\phi$ is to be thought of as `time' and
$\mu$ as the genuine, physical degree of freedom which evolves
with respect to this `time', we chose the standard Schr\"odinger
representation for $\phi$ but the `polymer representation' for
$\mu$ which correctly captures the quantum geometry effects. This
is a conservative approach in that the results will directly
reveal the manifestations of quantum geometry; had we chosen a
non-standard representation for the scalar field, these effects
would have been mixed with those arising from an unusual
representation of `time evolution'. Comparison with the \WDW
theory would also become more complicated. (However, the use of a
`polymer representation' for $\phi$ may become necessary to treat
inhomogeneities in an adequate fashion.)

On $\Hk$, the constraint takes the form:
\ba \label{qh7} \f{\p^2\Psi}{\p \phi^2} &=&
[B(\mu)]^{-1} \, \left( C^+(\mu)\Psi(\mu+4\mu_o, \phi) +
C^o(\mu)\Psi(\mu, \phi) + C^-(\mu) \Psi(\mu-4\mu_o, \phi)
\right)\nonumber\\
&=& - \Theta \Psi(\mu, \phi)\, \ea
were the functions $C^{\pm}, C^o$ are given by:%
\footnote{Note that this fundamental evolution equation makes no
reference to the Barbero-Immirzi parameter $\gamma$. If we set
$\t\mu = \mu/\mu_o$ and $\t\Psi(\t\mu, \phi) = \Psi(\mu,\phi)$, the
equation satisfied by $\t\Psi(\t\mu,\phi)$ makes no reference to
$\mu_o$ either. \emph{This is the equation used in numerical
simulations.} To interpret the results in terms of scale factor,
however, values of $\gamma$ and $\mu_o$ become relevant.}
\ba C^+(\mu) &=& \f{\pi G}{9|\mu_o|^{3}}\,\, \mid|\mu
+3\mu_o|^{\f{3}{2}} - |\mu+\mu_o|^{\f{3}{2}}\mid \nonumber\\
C^-(\mu) &=& C^+ (\mu -4\mu_o)\nonumber\\
C^o(\mu) &=& - C^+(\mu) - C^-(\mu)\, . \ea
The form of (\ref{qh7}) is the same as that of the \WDW constraint
(\ref{wdw4}), the only difference is that the $\phi$-independent
operator $\Theta$ is now a difference operator rather than a
differential operator. Thus, the the LQC quantum Hamiltonian
constraint can also be regarded as an `evolution equation' which
evolves the quantum state in the emergent time $\phi$.

However, since $\Theta$ is a difference operator, an important
difference arises from the \WDW analysis. For, now the space of 
physical states, i.e. of appropriate solutions to the constraint
equation, is naturally divided into sectors  each of which is
preserved by the `evolution' and by the action of our Dirac
observables. Thus, there is super-selection. Let
$\La_{|\epsilon|}$ denote the `lattice' of points
$\{|\epsilon|+4n\mu_o,\, n\in \Z\}$ on the $\mu$-axis,
$\La_{-|\epsilon|}$ the `lattice' of points
$\{-|\epsilon|+4n\mu_o,\, n\in \Z\}$ and let $\La_{\epsilon} =
\La_{|\epsilon|} \cup \La_{-|\epsilon|}$. Let
$\H_{|\epsilon|}^{\rm grav},\H_{-|\epsilon|}^{\rm grav}$ and
$\H_{\epsilon}^{\rm grav}$ denote the subspaces of $L^2(\R_{\rm
Bohr}, B(\mu)d\mu_{\rm Bohr})$ with states whose support is
restricted to lattices $\La_{|\epsilon|}, \La_{-|\epsilon|}$ and
$\La_\epsilon$. Each of these three subspaces is mapped to itself
by $\Theta$. Since $\hat{C}_{\rm grav}$ is self-adjoint and
positive definite on $\Hkg \equiv L^2(\R_{\rm Bohr}, \dd\mu_{\rm
Bohr})$, it follows that $\Theta$ is self-adjoint and positive
definite on all three Hilbert spaces.

Note however that since $\H_{|\epsilon|}^{\rm grav}$ and
$\H_{-|\epsilon|}^{\rm grav}$ are mapped to each other by the
operator $\Pi$, only $\H_\epsilon^{\rm grav}$ is left invariant by
$\Pi$. Now, because $\Pi$ reverses the triad orientation, it
represents a large gauge transformation. In gauge theories, we
have to restrict ourselves to sectors, each consisting of an
eigenspace of the group of large gauge transformations. (In QCD in
particular this leads to the $\theta$ sectors.) The group
generated by $\Pi$ is just $\Z_2$, whence there are only two
eigenspaces, with eigenvalues $\pm 1$. Since there are no fermions
in our theory, there are no parity violating processes whence we
are led to choose the symmetric sector with eigenvalue $+1$.
(Also, in the anti-symmetric sector all states are forced to
vanish at the `singularity' $\mu=0$  while there is no such a
priori restriction in the symmetric sector.) Thus, we are
primarily interested in the symmetric subspace of
$\H_\epsilon^{\rm grav}$; the other two Hilbert spaces will be
useful only in the intermediate stages of our discussion.

Our first task is to explore properties of the operator $\Theta$.
Since it is self-adjoint and positive definite, its spectrum is
non-negative. Therefore, as in the \WDW theory, we will denote its
eigenvalues by $\omega^2$. Let us first consider a generic
$\epsilon$, i.e., not equal to $0$ or $2\mu_o$. Then, on each of the
two Hilbert spaces $\H_{\pm|\epsilon|}^{\rm grav}$, we can solve
for the eigenvalue equation $\Theta\, e_\omega(\mu) = \omega^2\,
e_\omega (\mu)$, i.e.,
\be  \label{eq:eigen} C^+(\mu)e_\omega(\mu+4\mu_o) +
C^o(\mu)e_\omega(\mu) + C^-(\mu) e_\omega(\mu-4\mu_o)  = \omega^2
B(\mu) e_\omega(\mu)\, \ee
Since this equation has the form of a recursion relation and since
the coefficients $C^\pm(\mu)$ never vanish on the `lattices' under
consideration, it follows that we will obtain an eigenfunction by
freely specifying, say,  $\Psi(\mu^\star)$ and
$\Psi(\mu^\star+4\mu_o)$ for any $\mu^\star$ on the `lattice'
${\cal L}_{|\epsilon|}$ or ${\cal L}_{-|\epsilon|}$. Hence the
eigenfunctions are 2-fold degenerate on each of
$\H_{|\epsilon|}^{\rm grav}$ and $\H_{-|\epsilon|}^{\rm grav}$. On
$\H_\epsilon^{\rm grav}$ therefore, the eigenfunctions are 4-fold
degenerate as in the \WDW theory. Thus, $\H_\epsilon^{\rm
grav}$ admits an orthonormal basis $e_\omega^I$ where the
degeneracy index $I$ ranges from $1$ to $4$, such that
\be \langle e_\omega^I|e_{\omega'}^{I'}\rangle\,\, =\,\,
\delta_{I,I'}\, \delta(\omega,\, \omega'). \ee
(The Hilbert space $\H_\epsilon^{\rm grav}$ is separable and the
spectrum is equipped with the standard topology of the real line.
Therefore we have the Dirac distribution
$\delta(\omega,\,\omega')$ rather than the Kronecker delta
$\delta_{\omega,\,\omega'}$.) As usual, every element $\Psi(\mu)$
of $\H_\epsilon^{\rm grav}$ can be expanded as:
\be \Psi (\mu) = \int_{{\rm sp}\Theta} d\omega\, \t\Psi_I(\omega)
e^I_\omega(\mu) \quad {\rm where}\,\,\,\, \t\Psi_I(\omega)\, =\,
\langle e^I_\omega|\Psi\rangle\, , \ee
where the integral is over the spectrum of $\Theta$. The numerical
analysis of section \ref{s5} and comparison with the \WDW theory
are facilitated by making a convenient choice of this basis in
$\H_\epsilon^{\rm grav}$, i.e., by picking specific vectors from
each 4 dimensional eigenspace spanned by $e^I_\omega$. To do so,
note first that, as one might expect, every eigenvector
$e_\omega^I(\mu)$ has the property that it approaches unique
eigenvectors $\ub{e}^\pm(\omega)$ of the \WDW differential
operator $\ul\Theta$ as $\mu \rightarrow \pm \infty$. The precise
rate of approach is discussed in section \ref{s5.1}. In general,
the two \WDW eigenfunctions $\ub{e}^\pm(\omega)$ are distinct.
Indeed, because of the nature of the \WDW operator $\ul\Theta$,
its eigenvectors can be chosen to vanish on the entire negative
(or positive) $\mu$-axis; their behavior on the two half lines is
uncorrelated. (See the remark at the end of section \ref{s3.2}.)
Eigenvectors of the LQC $\Theta$ on the other hand are rigid;
their values at any two lattice points determine their values on
the entire lattice $\La_{\pm|\epsilon|}$. Second, recall that the
spectrum of the \WDW operator $\ul\Theta$ is bounded below by
$\omega^2 \ge \pi G/3$, whence $\ub{e}_\omega$ with $\omega^2 <
\pi G/3$ does not appear in the spectral decomposition of
$\ul\Theta$. Note however that solutions to the eigenvalue
equation $\ul\Theta\, \ub{e}_\omega = \omega^2 \ub{e}_\omega$
continue to exist even for $\omega^2 < \pi G/3$. But such
eigenfunctions diverge so fast as $\mu \rightarrow \infty$ or as
$\mu \rightarrow -\infty$ that $\langle\ub{e}_\omega|\Psi\rangle$
fails to converge for all $\Psi \in L^2(\R, \ub{B}(\mu) d\mu)$,
whence they do not belong to the basis. What is the situation with
eigenvectors of the LQC $\Theta$? Since eigenvectors $e_\omega$ of
$\Theta$ approach those of $\ul\Theta$, $\langle
e_\omega|\Psi\rangle$ again fails to converge for all $\Psi \in
\H_\epsilon^{\rm grav}$ if $\omega^2 < \pi G/3$. Thus the spectrum
of $\Theta$ is again bounded below by $\pi
G/3$.%
\footnote{A rigorous version of this argument can be constructed
e.g.  by using the Gel'fand triplet \cite{gs} associated with the
operator $\Theta$. However, this step has not been carried out.}
Therefore, to facilitate comparison with the \WDW theory we will
introduce a variables $k$ via $\omega^2 - \pi G/3 = (16\pi G/3)
k^2$ and use $k$ in place of $\omega$ to label the orthonormal
basis. To be specific, let us
\begin{quote}
i) Denote by $e_{-|k|}^\pm(\mu)$ the basis vector in
$\H_{\pm|\epsilon|}^{\rm grav}$ with eigenvalue $\omega^2$, which
is proportional to the \WDW $\ub{e}_{-|k|}$ as $\mu \rightarrow
\infty$; (i.e., it has only `outgoing' or `expanding' component
in this limit);\\
ii) Denote by  $e_{|k|}^\pm(\mu)$ the basis vector in
$\H_{\pm|\epsilon|}^{\rm grav}$ with eigenvalue $\omega^2$ which
is orthogonal to $e_{-|k|}^\pm(\mu)$ (since eigenvectors are
2-fold degenerate in each of $\H_{\pm|\epsilon|}^{\rm grav}$, the
vector $e_{|k|}^\pm (\mu)$ is uniquely determined up to a
multiplicative phase factor.)
\end{quote}
As we will see in section \ref{s5.1}, this basis is well-suited
for numerical analysis.

We thus have an orthonormal basis $e_k^\pm$ in $\H_{\epsilon}^{\rm
grav}$ with $k \in \R$: $\langle e_k^\pm|e_{k'}^\pm\rangle =
\delta(k,k')$, and $\langle e_k^+|e_{k'}^-\rangle = 0$. The four
eigenvectors with eigenvalue $\omega^2$ are now $e^+_{|k|},
e^+_{-|k|}$ which have support on the `lattice' $\La_{|\epsilon|}$,
and $e^-_{|k|}, e^-_{-|k|}$ which have support on the `lattice'
$\La_{-|\epsilon|}$. We will be interested only in the symmetric
combinations:
\be \label{eq:e-symm} e^{(s)}_k (\mu) = \f{1}{{2}}\,
\left(e^+_k(\mu) + e^+_k(-\mu) + e^-_k(\mu) + e^-_k(-\mu) \right)
\ee
which are invariant under $\Pi$. Finally we note that any
symmetric element $\Psi(\mu)$ of $\H_{\epsilon}^{\rm grav}$ can be
expanded as
\be \label{sym} \Psi(\mu) = \int_{-\infty}^{\infty} dk\,
\t\Psi(k)\, e^{(s)}_k(\mu) \ee

We can now write down the general symmetric solution to the
quantum constraint (\ref{qh7}) with initial data in
$\H_{\epsilon}^{\rm grav}$ :
\be \label{sol3}\Psi(\mu,\phi) = \int_{-\infty}^{\infty} dk\,
[\t\Psi_+(k) e^{(s)}_k(\mu) e^{i\omega\phi} + \t\Psi_- (k)
\bar{e}^{(s)}_k(\mu) e^{-i\omega\phi}]\ee
where $\t{\Psi}_\pm(k)$ are in $L^2(\R, dk)$. As $\mu \rightarrow
\pm \infty$, these approach solutions (\ref{sol1}) to the \WDW
equation. However, the approach is not uniform in the Hilbert
space but varies from solution to solution. As indicated in
section \ref{s2.2}, the LQC solutions to (\ref{qh7}) which are
semi-classical at late times can start departing from the \WDW
solutions for relatively large values of $\mu$, say $\mu \sim 10^4
\mu_o$.

As in the \WDW theory, if $\Psi_-(k)$ vanishes, we will say that
the solution is of positive frequency and if $\Psi_+(k)$ vanishes
we will say it is of negative frequency. Thus, every solution to
(\ref{qh7}) admits a natural positive and negative frequency
decomposition. The positive (respectively negative) frequency
solutions satisfy a Schr\"odinger type first order differential
equation in $\phi$:
\be \mp i\f{\p\Psi_\pm}{\p\phi} = \sqrt{\Theta} \Psi_\pm \ee
but with a Hamiltonian $\sqrt{\Theta}$ (which is non-local in
$\mu$). Therefore the solutions with initial datum $\Psi(\mu,
\phi_o) = f_\pm(\mu)$ are given by:
\be \Psi_\pm(\mu,\phi) \,=\, e^{\pm i\sqrt{\Theta}(\phi-\phi_o)}\,
f_\pm(\mu,\phi)\ee

\textbf{Remark:} In the above discussion, we considered a generic
$\epsilon$. We now summarize the situation in the special cases,
$\epsilon= 0$ and $\epsilon =2\mu_o$. In these cases, differences
arise because the individual lattices are invariant under the
reflection $\mu \rightarrow -\mu$, i.e., the lattices
$\La_{|\epsilon|}$ and $\La_{-|\epsilon|}$ coincide. As before,
there is a 2-fold degeneracy in the eigenvectors of $\Theta$ on
any one lattice. For concreteness, let us label the Hilbert spaces
$\H_{|\epsilon|}^{\rm grav}$ and choose the basis vectors
$e_k^+(\mu)$, with $k\in \R$ as above. Now, symmetrization can be
performed on each of these Hilbert spaces by itself. So, we have:
\be \label{sb} e_k^{(s)}(\mu) = \f{1}{\sqrt{2}}\, (e^+_k(\mu) +
e^+_k (-\mu))\ee
However, the vector $e_{|k|}^{(s)}(\mu)$ coincides with the vector
$e_{-|k|}^{(s)}(\mu)$ so there is only one symmetric eigenvector
per eigenvalue. This is not surprising: the original degeneracy
was 2-fold (rather than 4-fold) and so there is one symmetric and
one anti-symmetric eigenvector per eigenvalue. Nonetheless, it is
worth noting that there is a precise sense in which the Hilbert
space of symmetric states is only `half as big' in these
exceptional cases as they are for a generic $\epsilon$.

For $\epsilon=2\mu_o$, there is a further subtlety because $C^+$
vanishes at $\mu= -2\mu_o$ and $C^-$ vanishes at $\mu= 2\mu_o$.
Thus, in this case, as in the \WDW theory, there is a decoupling
and the knowledge of the eigenfunction $e^+_k(\mu)$ on the
positive $\mu$-axis does not suffice to determine it on the
negative $\mu$ axis and vice-versa. However, the degeneracy of the
eigenvectors does not increase but remains 2-fold because
(\ref{qh7}) now introduces two new constraints: $C^\pm(2\mu_o)
e^+_k(\pm 6\mu_o) = [\omega^2 B(\pm 2\mu_o) - C^o(\pm 2\mu_o)]
e^+_k(\pm 2\mu_o) =0$.   Conceptually, this difference is not
significant; there is again a single symmetric eigenfunction for
each eigenvalue.

\subsection{The Physical sector}
\label{s4.2}

Results of section \ref{s4.1} show that while the LQC operator
$\Theta$ differs from the \WDW operator $\ul\Theta$ in interesting
ways, the structural form of the two Hamiltonian constraint
equations is the same. Therefore, apart from the issue of
superselection sectors which arises from the fact that $\Theta$ is
discrete, introduction of the Dirac observables and determination
of the inner product either by demanding that the Dirac
observables be self-adjoint or by carrying out group averaging is
completely analogous in the two cases. Therefore, we will not
repeat the discussion of section \ref{s3.2} but only summarize the
final structure.

The sector of physical Hilbert space $\Hp^\epsilon$ labelled by
$\epsilon \in [0,\, 2 \mu_o]$ consists of positive frequency solutions
$\Psi(\mu,\phi)$ to (\ref{qh7}) with initial data $\Psi (\mu,
\phi_o)$ in the symmetric sector of $\H^\epsilon_{\rm grav}$. Eq.
(\ref{sol3}) implies that they have the explicit expression in
terms of our eigenvectors $e^{(s)}_k(\mu)$
\be \label{eq:psi-int} \Psi(\mu,\phi) = \int_{-\infty}^\infty
dk\, \t\Psi(k) \,e^{(s)}_k(\mu)\, e^{i\omega\phi}\, , \ee
where, as before, $\omega^2 = (16\pi G/3)(k^2 + 1/16)$ and
$e^{(s)}_k(\mu)$ is given by (\ref{eq:e-symm}) and (\ref{sb}). By
choosing appropriate functions $\t\Psi(k)$, this expression will
be evaluated in section \ref{s5.1} using Fast Fourier Transforms.
The resulting $\Psi(\mu,\phi)$ will provide, numerically, quantum
states which are semi-classical for large $\mu$. The physical
inner product is given by:
\be \label{ip2} \langle\Psi_1|\Psi_2\rangle_\epsilon\,\, =
\sum_{\mu \in \{\pm|\epsilon| + 4n\mu_o;\, n\in \Z\}}\,\, B(\mu)
\bar{\Psi}_1(\mu,\phi_o)\, \Psi_2(\mu,\phi_o) \ee
for any $\phi_o$.
The action of the Dirac observables is independent of $\epsilon$,
and has the same form  as in the \WDW theory:
\be \label{dirac4}\widehat{|\mu|_{\phi_o}}\, \Psi(\mu,\phi) =
e^{i\sqrt{{\Theta}}(\phi-\phi_o)}\, |\mu|
\,\Psi(\mu,\phi_o),\quad {\rm and} \quad \hat{p}_\phi
\Psi(\mu,\phi) = - \, i\hbar\, \f{\p \Psi(\mu,\phi)}{\p\phi} \, .
\ee

The kinematical Hilbert space $\Hk$ is non-separable but, because
of super-selection, each physical sector $\Hp^\epsilon$ is
separable. Eigenvalues of the Dirac observable
$\widehat{|\mu|_{\phi_o}}$ constitute a discrete subset of the
real line in each sector. In the kinematic Hilbert space $\Hkg$,
the spectrum of $\hat{p}$ is discrete in a subtler sense: while
every real value is allowed, the spectrum has discrete topology,
reflecting the fact that each eigenvector has a finite norm in
$\Hkg$. Thus, the more delicate discreteness of the spectrum of
$\hat{p}$ on $\Hkg$ descends to the standard type of discreteness
of Dirac observables. Question is often raised whether the
kinematic discreteness in LQG will have strong imprints in the
physical sector or if they will be washed away in the passage to
the physical Hilbert space. A broad answer ---illustrated by the
area eigenvalues of isolated horizons \cite{abck,abk}--- is that
the discreteness will generically descend to the physical sector
at least in cases where one can construct Dirac observables
directly from the kinematical geometrical operators \cite{alrev}.
The present discussion provides another illustration of this
situation.

Note that the eigenvalues of $\widehat{|\mu|_{\phi_o}}$ in
distinct sectors are distinct. Therefore which sector  actually
occurs is a question that can be in principle answered
experimentally, provided one has access to microscopic
measurements which can distinguish between values of the scale
factor which differ by $\sim 10 \lp$. This will not be feasible in
the foreseeable future. Of greater practical interest are the
coarse-grained measurements, where the coarse graining occurs at
significantly greater scales. For these measurements, different
sectors would be indistinguishable and one could work with any
one.

The group averaging procedure used in this section is quite
general in the sense that it is applicable for a large class of
systems, including full LQG if, e.g., its dynamics is formulated
using Thiemann's master constraint program \cite{ttmc}. In this
sense, the physical Hilbert spaces $\Hp^\epsilon$ constructed here
are natural. However, using the special structures available in
this model, one can also construct an inequivalent representation
which is closer to that used in the \WDW theory. The main results
on the bounce also hold in that representation. Although that
construction appears to have an ad-hoc element, it may well admit
extensions and be useful in more general models. Therefore, it is
presented in Appendix \ref{a3}.

\section{Numerics in loop quantum cosmology}
\label{s5}

In this section, we will find physical states of LQG and analyze
their properties numerically. This section is divided into three
parts. In the first we study eigenfunctions $e^\pm_k (\mu)$ of
$\Theta$ and then use them to directly evaluate the right side of
(\ref{eq:psi-int}), thereby obtaining a `general' physical state.
In the second part we solve the initial value problem starting
from initial data at $\phi=\phi_o$, thereby obtaining a `general'
solution to the difference equation (\ref{qh7}). In the third we
summarize the main results and compare the outcome of the two
methods. Readers who are not interested in the details of
simulations can go directly to the third subsection.

A large number of simulations were performed within each of the
approaches by varying the parameters in the initial data and
working with different lattices ${\cal L}_\epsilon$. They show
that the final results are robust. To avoid the making the paper
excessively long, we will only show illustrative plots.

\subsection{Direct evaluation of the integral representation
(\ref{eq:psi-int}) of solutions} \label{s5.1}

The goal of this sub-section is to evaluate the right side of
(\ref{eq:psi-int}) using suitable momentum profiles $\t\Psi(k)$.
This calculation requires the knowledge of eigenfunctions
$e_k^{(s)}(\mu)$ of $\Theta$. Therefore, we will first have to
make a somewhat long detour to numerically calculate the basis
functions $e^\pm_k(\mu)$ and $e_k^{(s)}(\mu)$ introduced in
section \ref{s4.1}. The integral in (\ref{eq:psi-int}) will be
then evaluated using a fast Fourier transform.

\subsubsection{General eigenfunctions of the $\Theta$ operator:
asymptotics} \label{sec:eig-as}

We will first establish properties of the general eigenfunctions
$e_\omega(\mu)$ of $\Theta$ that were used in section \ref{s4.1}.

Let us fix a `lattice', say  $\La_{|\epsilon|}$. Since the left
side of (\ref{eq:eigen}) approaches
$\hat{C}_{\text{grav}}^{\text{WDW}} e_{\omega}(\mu)$ as
$|\mu|\to\infty$, in this limit one would expect each $e_{\omega}(\mu)$
to converge to an eigenfunction $\ul{e}_{\omega}(\mu)$ of $\ul{\Theta}$
with the same eigenvalue. Numerical simulations have shown that
this expectation is correct and have also provided the rate of
approach.

\begin{figure}[tbh!]
\begin{center}
\includegraphics[width=5in,angle=0]{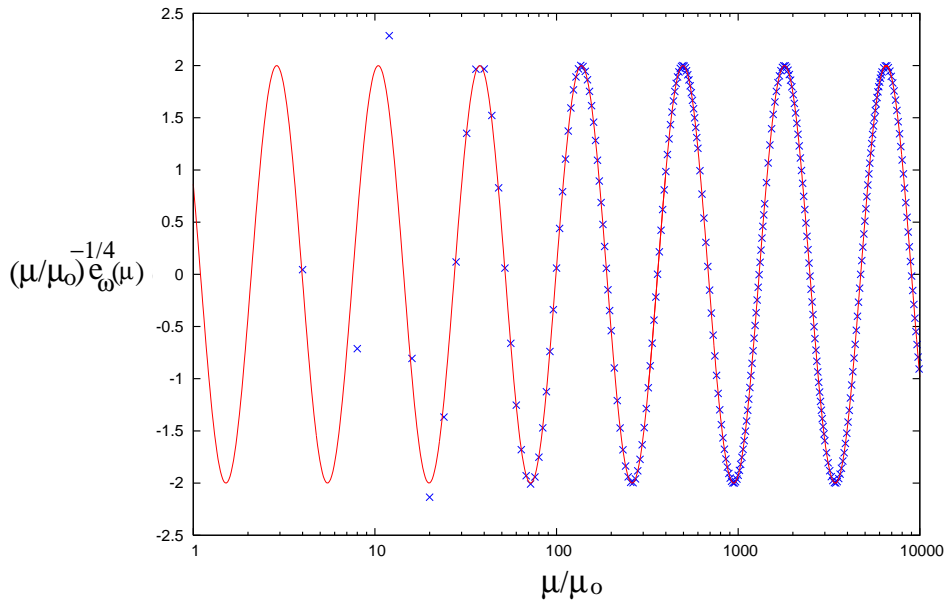}
\caption{Crosses denote the values of an eigenfunction
$e_{\omega}(\mu)$ of $\Theta$ for $\epsilon=0$ and $\omega=20$.
The solid curve is the eigenfunction $\ub{e}_\omega(\mu)$ of the
\WDW $\ul\Theta$ to which $e_\omega(\mu)$ approaches at large
positive $\mu$. As $\mu$ increases, the set of points on ${\cal
L}_\epsilon$ becomes denser and fill the solid curve. For visual
clarity only some of these points are shown for $\mu >100$.}
\label{fig:e-LQC-WdW}
\end{center}
\end{figure}

Recall that each \WDW eigenfunction $\ul{e}_{\omega}(\mu)$ is a linear
combination of basis functions $\ub{e}_{|k|}(\mu)$, $\ub{e}_{-|k|}(\mu)$
defined in section \ref{s3.1}. Therefore, given an $e_{\omega}(\mu)$ it
suffices to calculate the coefficients of the decomposition of
$\ub{e}_{\omega}(\mu)$ with respect to this basis. The method of
finding these coefficients is presented in detail in Appendix
\ref{a2}.%
\footnote{If $\omega^2 < \pi G/3$, then $\omega^2$ is not part of
the spectrum of the self-adjoint operator $\ul\Theta$.
Nonetheless, by directly solving the eigenvalue equation
$\ul\Theta \ub{e}_\omega = \omega^2 \ub{e}_\omega$ one can
introduce an analogous decomposition onto fixed eigenfunctions,
$\ul{e}_{\pm|k'|} := |\mu|^{\frac{1}{4}\pm k'} \ ,
\quad {k'}^2 = 1/16 - 3/(16\pi G)\omega^2$\, and also write the
limit of $e_{\omega}$ in terms of coefficients in this `basis'.}
Once the limiting  $\ub{e}_{\omega}(\mu)$ were found, they were
compared with the original eigenfunction $e_{\omega}(\mu)$ for a
variety of values of $\omega$. An illustrative plot comparing
${e}_{\omega}(\mu)$ with its limit is shown in Fig.
\ref{fig:e-LQC-WdW}. In general, each $e_\omega(\mu)$ approaches
distinct eigenfunctions $\ub{e}_{\omega,\,\pm}(\mu)$ of
$\ul\Theta$ in the limits $\mu \rightarrow \pm \infty$. The rate
of approach is given by:
\begin{equation}\label{eq:e-asympt}
  \mu^{-\frac{1}{4}} e_{\omega}(\mu)\ = \begin{cases}
    \mu^{-\frac{1}{4}} \underline{e}_{\omega,\, +} (\mu)
      + O\left( \frac{1}{\mu^2} \right) \ ,
      & \text{for }\mu>0 \ , \\
    \mu^{-\frac{1}{4}} \underline{e}_{\omega,\, -} (\mu)
      + O\left( \frac{1}{\mu^2} \right) \ ,
      & \text{for }\mu<0 \ .
  \end{cases}
\end{equation}
Numerical tests were performed up to $|\mu| = 10^6\mu_o$. The
quantity $\mu^{7/4}\mid e_\omega -\ub{e}_{\omega,\,\pm} \mid
(\mu)$ was found to be bounded. The bound decreases with $\mu$.
For the case $\epsilon=0,\, \omega =20$ depicted in Fig.
\ref{fig:e-LQC-WdW} the absolute bound in the $\mu$-interval
$(10^2\mu_o,\, 10^6\mu_o)$ was less than 90.
%
%

Because the eigenfunctions $e_\omega(\mu)$ of $\Theta$ are
determined on the entire lattice $\La_{|\epsilon|}$ by their values
on (at most) two points, the WDW limits for positive and negative
$\mu$ are not independent. Thus, if the limits are expressed as
\begin{subequations}\label{eq:asympt-coeffs}\begin{align}
  e_{\omega}(\mu) &\xrightarrow{\mu\gg 1}
    A\,\ub{e}_{|k|}(\mu) + B\,\ub{e}_{-|k|}(\mu), &
  e_{\omega}(\mu) &\xrightarrow{\mu\ll -1}
    C\,\ub{e}_{|k|}(\mu) + D\, \ub{e}_{-|k|}(\mu)
  \tag{\ref{eq:asympt-coeffs}}
\end{align}\end{subequations}
the coefficients $C,D$ are uniquely determined by values of $A,B$
(and vice versa). One relation, suggested by analytical
considerations involving the physical inner product, was verified
in detail numerically:
\begin{equation}\label{eq:e-norm-pres}
  |A|^2 - |B|^2\ =\ |C|^2 - |D|^2 \ .
\end{equation}
It will be useful in the analysis of basis functions in the next
two sub-sections.

\subsubsection{Construction of the basis $e^{\pm}_{-|k|}$ }
  \label{sec:basis-pm}

In section \ref{s4.1} we introduced a specific basis of
$\H_{\epsilon}$ which is well adapted for comparison with the \WDW
theory. We will now use numerical methods to construct this basis
and analyze its properties. Our investigation will be restricted
to the vectors $e^{\pm}_{-|k|}(\mu)$ because, the physical states
$\Psi(\mu,\phi)$ of Eq. (\ref{eq:psi-int}) we are interested will
have negligible projections on the vectors $e^{\pm}_{|k|}(\mu)$.
(Recall that in general $\Psi \in \H_\epsilon$ has support on
$\La_{|\epsilon|}\cup \La_{-|\epsilon|}$. $e^+_{-|k|}(\mu)$ has
support on $\La_{+|\epsilon|}$ and $e^-_{-|k|}(\mu)$ on
$\La_{-|\epsilon|}$.)

Each of the eigenfunctions $e^{\pm}_{-|k|}$ is calculated as
follows. To solve (\ref{eq:eigen}), we need to specify initial
conditions at two points on each of the two lattices. We fix large
positive $\mu_{\pm}^\star\in\La_{\pm|\epsilon|}$ and demand that
the values of $e^{\pm}_{-|k|}(\mu)$ agree with those of
$\ub{e}_{-|k|}(\mu)$ at the points $\mu_{\pm}^\star$ and
$\mu_{\pm}^\star +4\mu_o$. Then $e^\pm_{-|k|}$ are evaluated
separately on finite domains $\La_{\pm|\epsilon|} \cap
[-\mu_{\pm}^\star,\mu_{\pm}^\star]$ of each of the two lattices.
For large negative $\mu$, these eigenfunctions are linear
combinations of the \WDW basis functions $C^\pm\, \ub{e}_{|k|}+
D^\pm\, \ub{e}_{-|k|}$. The coefficients $C^\pm, D^\pm$ are
evaluated using the method specified in Appendix \ref{a2}.

\begin{figure}[tbh!]
  \begin{center}
    \includegraphics[width=5in,angle=0]{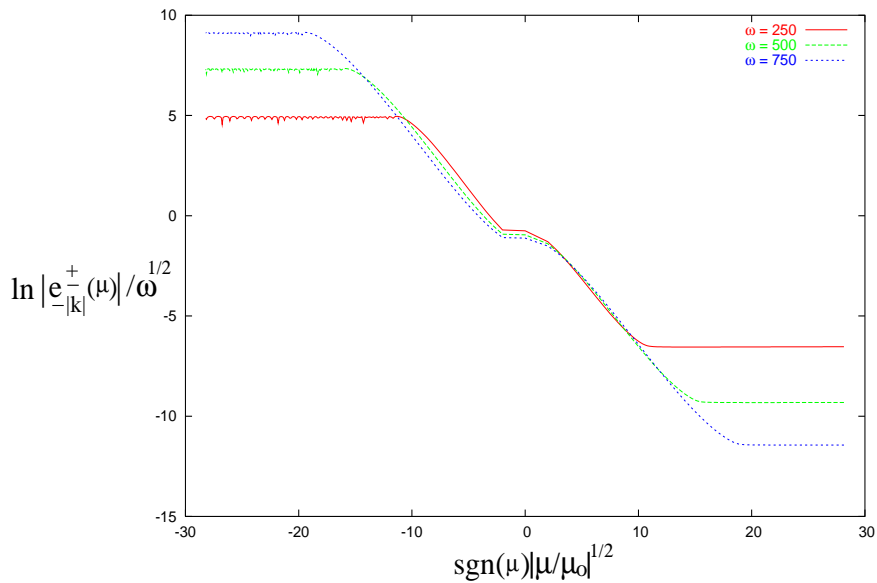}
    \caption{The exponential growth of $|e^{\pm}_{-|k|}|(\mu)$ in the
`genuinely quantum region' is shown for three different values of
$\omega$, where $\omega$ is given by $\Theta\, e^{\pm}_{-|k|} =
\omega^2\, e^{\pm}_{-|k|}$.}
    \label{fig:ek-log}
  \end{center}
\end{figure}

\begin{figure}[tbh!]
  \begin{center}
    \includegraphics[width=5in,angle=0]{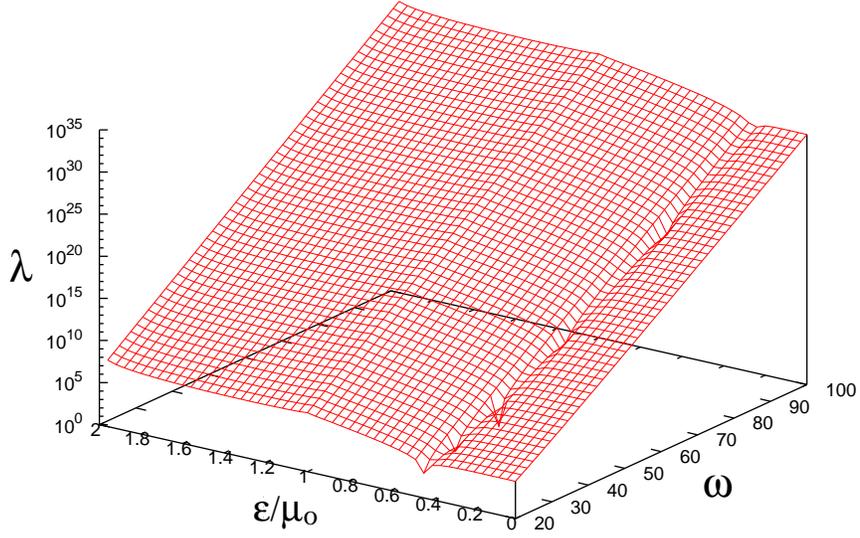}
    \caption{The amplification factor $\lambda^{\pm}$
    in the `genuinely quantum region is shown as a function of the
    parameter $\epsilon$ labeling the lattice and $\omega$.} \label{fig:ampl-3d}
  \end{center}
\end{figure}

\begin{figure}[tbh!]
  \begin{center}
    \includegraphics[width=5in,angle=0]{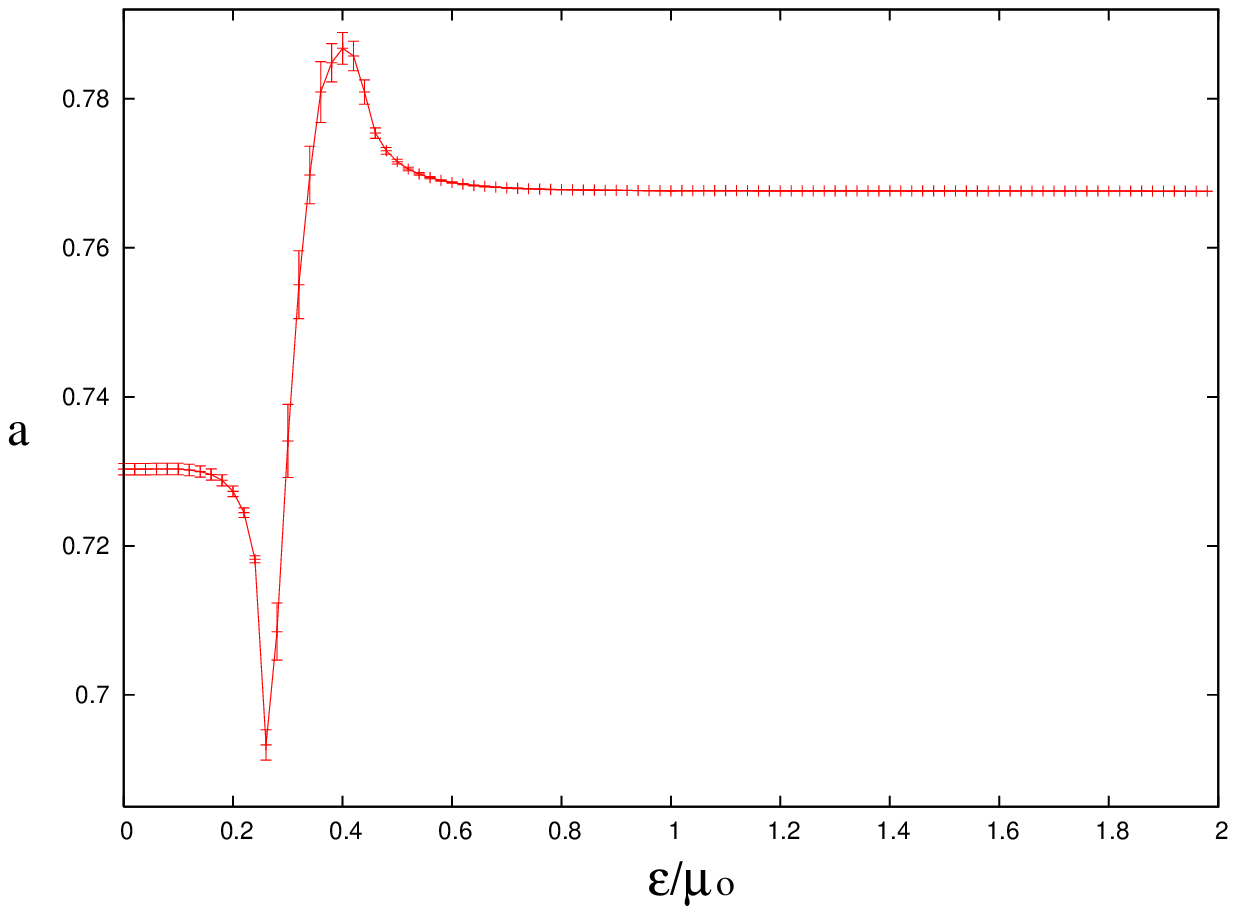}
    \caption{The function $a(\epsilon)$ of Eq
    \ref{eq:lambda-fit} is plotted by connecting
    numerically calculated data points.}
    \label{fig:ampl-a}
  \end{center}
\end{figure}

\begin{figure}[tbh!]
  \begin{center}
    \includegraphics[width=5in,angle=0]{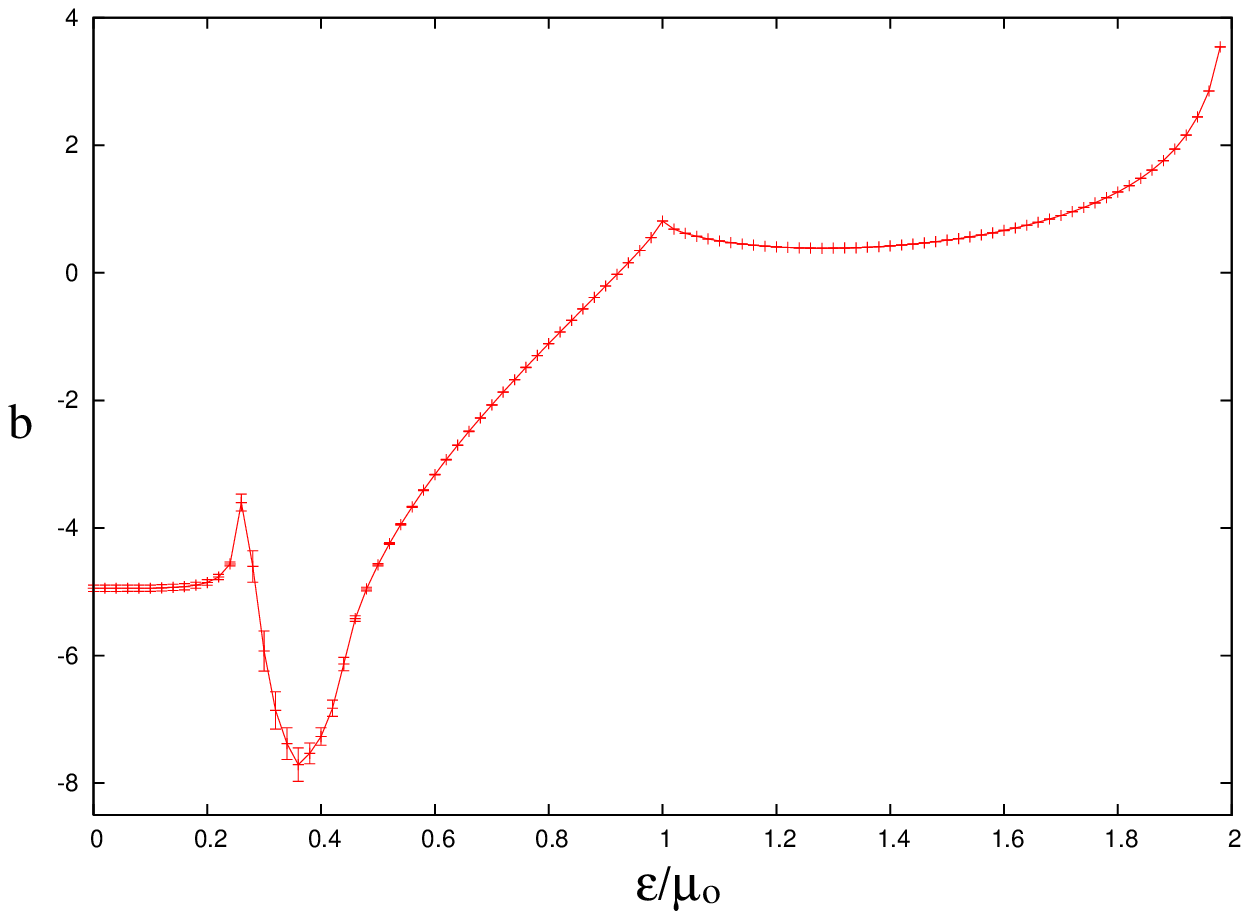}
    \caption{The function $b(\epsilon)$ of Eq
    \ref{eq:lambda-fit} is plotted by connecting
    numerically calculated data points.}
     \label{fig:ampl-b}
  \end{center}
\end{figure}

Eigenfunctions $e^{\pm}_{-|k|}(\mu)$ were calculated for
approximately $2\times 10^4$ different $|k|$'s in the range $5\le
\omega \le 10^3$. They revealed the following properties:
\begin{enumerate}[ (i)]
  \item Each $e^{\pm}_{-|k|}(\mu)$ is well approximated by a \WDW
eigenfunction until one reaches the `genuinely quantum region'. In
this region the absolute value $|e^{\pm}_{-|k|}|$ grows very
quickly as $\mu$ decreases: $|e^{\pm}_{-|k|}|\,\propto\, e^{-{\rm
sgn}(\mu) \,\alpha\sqrt{\omega\,|\mu|}}$, where $\alpha \equiv
\alpha (\epsilon)$ is a constant on any given lattice. This
property is illustrated by Fig. \ref{fig:ek-log}. This region of
rapid growth is symmetric about $\mu=0$ and its size depends
linearly on $\omega$ (the square-root of the eigenvalue of
${\Theta}$); its boundary lies at  $\mu\approx 0.5\omega\,\mu_o$).
(However, this region excludes the interval $[-4\mu_o,\,4\mu_o]$
where $B(\mu)$ decreases and goes to zero, departing significantly
from its \WDW analog $\ub{B}(\mu)$.)
    \item After leaving this region of growth, the basis function
$e^{\pm}_{-|k|}(\mu)$ again approaches some WDW eigenfunction
    \begin{equation}
      e^{\pm}_{-|k|} \xrightarrow{\mu << -1 }
      C^{\pm}\, \ub{e}_{|k|} + D^{\pm}\, \ub{e}_{-|k|} \ .
    \end{equation}
where the coefficients $C^{\pm},\, D^{\pm}$ are large. Their
absolute values grow exponentially with $\omega$. To investigate
this property qualitatively we defined an `amplification factor'
    \begin{equation}
      \lambda^{\pm}\ := |C^{\pm}| + |D^{\pm}|
    \end{equation}
The numerical calculations show that $\lambda^+ = \lambda^-$, so
parts of $e^\pm_{-|k|}$  supported on $\La_{+|\epsilon|}$ and
$\La_{-|\epsilon|}$ are amplified equally. The dependence of
$\lambda^{\pm}$ on $\omega$ and $\epsilon$ is shown in Fig.
\ref{fig:ampl-3d}. Almost everywhere it can be well approximated
by the function
    \begin{equation} \label{eq:lambda-fit}
      \lambda^{\pm} (\omega, \epsilon) \approx e^{a\omega+b} \ .
    \end{equation}
where $a,b$ are rather complicated functions of $\epsilon$. The
fits of $a(\epsilon),b(\epsilon)$ are presented in Fig.
\ref{fig:ampl-a} and Fig. \ref{fig:ampl-b}. The actual simulations
were carried out for various values of $p_\phi = \hbar \omega$, up
to $p_\phi=10^3$.
\item The general relation (\ref{eq:e-norm-pres}) holds in our
case with $|B| =0$. Existence of the tremendous amplification now
implies that the absolute values of $C$ and $D$ are almost equal
(with differences of the order of $1/\lambda^{\pm}$)
    \begin{equation}
      |C^{\pm}| \approx |D^{\pm}| \ .
    \end{equation}
Thus for negative $\mu$, eigenfunctions asymptotically approach
\WDW eigenfunctions and are almost equally composed of incoming
and outgoing waves.
\end{enumerate}

\subsubsection{Basis for the symmetric sector}
  \label{sec:basis-sym}

Once the basis functions $e^{\pm}_{-|k|}$ are known one can
readily use \eqref{eq:e-symm} to construct the basis
$e^{(s)}_{-|k|}$ for solutions to (\ref{eq:eigen}) which are
symmetric under $\mu \rightarrow -\mu$. Due to strong
amplification in the region around $\mu=0$ the behavior of
$e^{(s)}_{-|k|}$ is dominated by properties of $e^{\pm}_{-|k|}$
for $\mu<0$. The numerical calculations show the following
properties:
\begin{enumerate}[ (i)]
  \item Each symmetric basis eigenfunction is strongly suppressed in
the `genuinely quantum region'  around to $\mu=0$. 
The behavior of $e^{\pm}_{-|k|}$ in this region implies that
$|e^{(s)}_{-|k|}|\propto\cosh(\alpha\sqrt{\omega\mu})$, where $\alpha$
is a function of $\epsilon$ only.
  \item Outside the `genuinely quantum region' $e^{(s)}_{-|k|}$ quickly
approaches the \WDW eigenfunction almost equally composed of
incoming and outgoing `plane waves' ($\ub{e}_{|k|}$ and
$\ub{e}_{-|k|}$). In the exceptional cases $\epsilon = 0$ {\rm or}
$2\mu_o$ the two contributions are exactly equal. To establish
this result, note first that the symmetry requirement and the fact
that $C^{+}(-2\mu_o) = C^{-}(2\mu_o) = 0$ imply that in both cases
the value of $e^{(s)}_{-|k|}$ at one point already determines the
complete `initial data' (i.e., values at some $\mu_*$ and
$\mu_*+4\mu_o$) and hence the eigenfunction on the entire
$\La_{\epsilon}$:
\begin{subequations}\label{eq:02-symm-cond}\begin{align}
      e^{(s)}_{-|k|}(4\mu_o)\ &=\ e^{(s)}_{-|k|}(-4\mu_o)\
      =\ e^{(s)}_{-|k|}(0) \ , &
      \text{for }\epsilon &= 0 \ , \\
      e^{(s)}_{-|k|}(6\mu_o)\
      &=\ \frac{\omega^2B(2\mu_o)-C^o(2\mu_o)}{C^{+}(2\mu_o)}
        e^{(s)}_{-|k|}(2\mu_o) \ , &
      \text{for }\epsilon &= 2\mu_o \ .
    \end{align}\end{subequations}
Because of the reality of coefficients of Eq. (\ref{eq:eigen}),
this implies that the phase of $e^{(s)}_{-|k|}$ is exactly
constant, whence contributions of $\ub{e}_{|k|}$ and
$\ub{e}_{-|k|}$ are also exactly equal.
  \item On each lattice $\La_{\pm|\epsilon|}$ the incoming and outgoing
components are rotated with respect to each other by an angle
$\alpha^{\pm}$
    \begin{equation}\label{eq:phases}
      e^{(s)}_{-|k|}\mid_{\La_{\pm|\epsilon|}} \xrightarrow{\mu\to\infty}
      z^{\pm} ( e^{i\alpha^{\pm}}\ub{e}_{|k|} +
      e^{-i\alpha^{\pm}}\ub{e}_{-|k|} ) \ ,
    \end{equation}
where $z^{\pm}$ are some complex \emph{constants} satisfying
$|z^+|\approx|z^-|$, while the phases $\alpha^{\pm}$ are functions
of $\epsilon$ and $\omega$. In general, for $\epsilon \not= 0$ or
$\epsilon \not= 2\mu_o$, $\alpha_{+}$ need not equal $\alpha_{-}$.
\end{enumerate}

\begin{figure}[tbh!]
  \begin{center}
    \includegraphics[width=5in,angle=0]{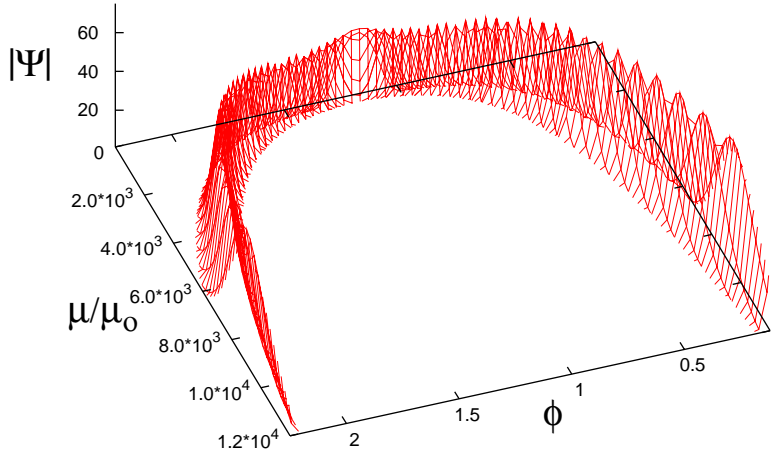}
    \caption{Plot of the wave function $\Psi(\mu,\phi)$
    obtained by directly evaluating the right side of
    (\ref{eq:psi-int}). Parameters are $p_{\phi}=500$,
    $\Delta p_{\phi}/p_{\phi} = 0.05$ and $\epsilon = \mu_o$.}
    \label{fig:direct-3d}
  \end{center}
\end{figure}

\begin{figure}[tbh!]
  \begin{center}
    \includegraphics[width=5in,angle=0]{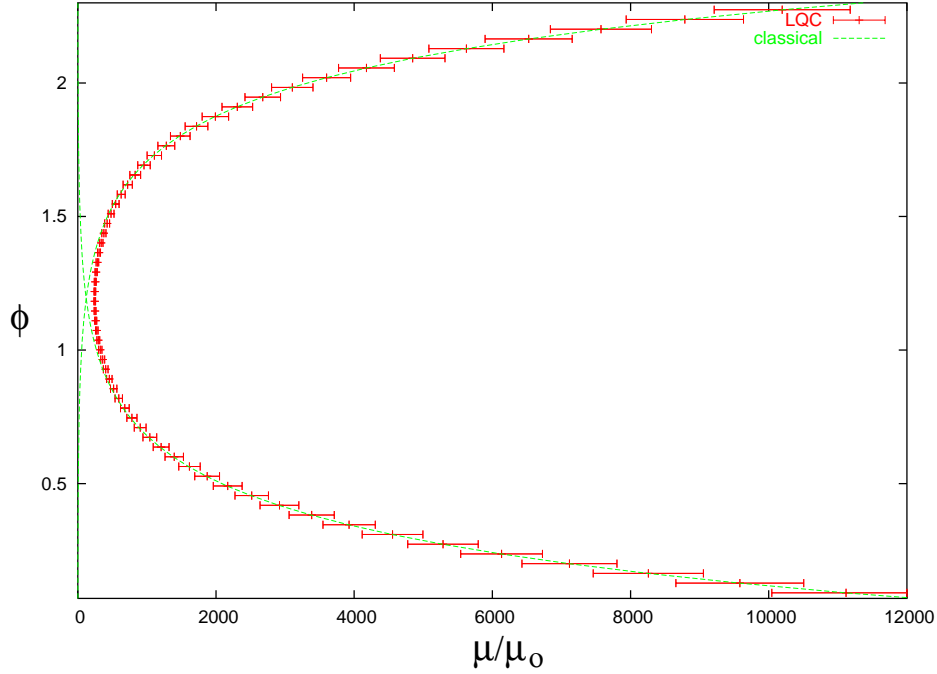}
    \caption{Expectation values and dispersion of
    $\widehat{|\mu|_{\phi}}$ for wave function presented in
    Fig. \ref{fig:direct-3d} are compared with classical
    trajectories.}
    \label{fig:direct-mu}
  \end{center}
\end{figure}

\subsubsection{Evaluation of the integral in (\ref{eq:psi-int})}
  \label{sec:state-direct}

Now that we have the symmetric basis functions $e^{(s)}_{-|k|}$ at
our disposal, we can obtain the desired physical states by
directly evaluating the integral in (\ref{eq:psi-int}).

We wish to construct physical states which are sharply peaked at a
phase space point on a classical trajectory of the expanding
universe at a late time (e.g., `now'). The form of the integrand
in (\ref{eq:psi-int}), the expression (\ref{ip2}) of the physical
inner product, the functional form of $\omega(k)$ and standard
facts about coherent and squeezed states in quantum mechanics
provide a natural strategy to select an appropriate $\t\Psi(k)$.
If we set
\be \t\Psi(k) =  e^{-\f{(k - k^\star)^2}{2 \sigma^2}}\,
e^{-i\omega \phi^\star} \ee
with suitably small $\sigma$, the final state will be sharply
peaked at $p_\phi = p_\phi^\star$
\be p_\phi^\star= -\left(\f{16\pi G\hbar^2}{3}
\right)^{\f{1}{2}}\,\, k^\star
\ee
and the parameter $\phi^\star$ will determine the value
$\mu^\star$ of the Dirac observable $\widehat{|\mu|_{\phi_o}}$
at which the state will be peaked at `time' $\phi_o$. As mentioned
in section \ref{s3.2}, to obtain a state which is semi-classical
at a late time, we need a large value of $p_\phi$: $p_\phi^\star
\gg \hbar$ in the classical units, $c$=$G$=1. Therefore, we need
$k^\star \ll -1$, whence the functions $\t\Psi(k)$ of interest
will be negligibly small for $k>0$. Therefore, without loss of
physical content, we can set them to zero on the positive $k$
axis. This is why the explicit form of eigenfunctions
$e^{(s)}_{|k|}$ is not required in our analysis.

Thus, the integral we wish to evaluate is:
\begin{equation}\label{eq:state-direct}
\Psi(\mu, \phi) \ =\ \int_0^{\infty} \, d k \, e^{-\f{(k -
k^\star)^2}{2 \sigma^2}} \, e^{(s)}_{k}(\mu)\, e^{i \omega(k)(\phi
- \phi^\star)} \ ,
\end{equation}
where $\sigma$ is the spread of the Gaussian. The details of the
numerical evaluation can be summarized as follows.
\begin{itemize}
  \item For a generic $\epsilon$, the $e^{(s)}_{-|k|}$ were found
numerically following the procedure specified in sections
\ref{sec:basis-pm} and \ref{sec:basis-sym}. For the exceptional
cases, $\epsilon= 0,2\mu_o$, in order to avoid loss of precision
in the region where $e^{(s)}_{-|k|}$ is very small, we provided
`initial values' of $e^{(s)}_{k_j}$ at $\mu=\pm\epsilon$ and $\mu
= \pm\epsilon+4\mu_o$ using \eqref{eq:02-symm-cond}. On the $k$
axis we chose a set $\{k_j\}$ of points which are uniformly
distributed across the interval $[k^\star-10\sigma,
k^\star+10\sigma]$. In numerical simulations, the number $l$ of
points in the set $\{k_j\}$ ranged between $2^{11}$ and $2^{13}$.
  \item Next, for each $k_j$, we calculated $e_{-|k_j|}^{(s)}(\mu_i)$ for
$\mu\in \{\pm\epsilon+4n\mu_o:\, n\in\{-N,\ldots,N\}\}$ where $N$
is a large constant $(\sim\, 50 p_\phi^\star)$.
  \item Finally, we evaluated \eqref{eq:state-direct} using fast
Fourier transform. The result was a set of profiles
$\Psi(\mu_i,\phi_j)$ where $\phi_j=\sqrt{3\pi/(4G)}(j-l/2)/(k_l-k_1)$
This is a positive frequency solution to the LQC equation
(\ref{qh7}).
\end{itemize}

Our next task is to analyze properties of these solution. Given
any one $\Psi(\phi_j,\mu_i)$, we chose `instants of time' $\phi$
and calculated the norm and the expectation values of our Dirac
observable $|\hat{\mu}|_{\phi}$ and $\hat{p}_{\phi}$ using:
\begin{subequations}\label{eq:expect}\begin{align}
  ||\Psi||^2\ &=\ \sum_{\mu_i\in\La_{\epsilon}} B(\mu_i)
                  |\Psi(\phi,\mu_i)|^2 \ , \\
  \left<\widehat{|\mu|_{\phi}}\right>\
    &=\ \frac{1}{||\Psi||^2}
    \sum_{\mu_i\in\La_{\epsilon}}
    B(\mu_i)\,|\mu_i|\,|\Psi(\phi,\mu_i)|^2 \ , \\
  \left<\hat{p}_{\phi}\right>\ &=\ \frac{1}{||\Psi||^2}
    \sum_{\mu_i\in\La_{\epsilon}}
    B(\mu_i)\,\bar{\Psi}(\phi,\mu_i)\,\,\f{\hbar}{i} \partial_{\phi}
    \Psi(\phi,\mu_i)
    \, .
\end{align}\end{subequations}
Finally, the dispersions were evaluated using their definitions:
\begin{subequations}\label{eq:disp}\begin{align}
  \left<\Delta \widehat{|\mu|_{\phi}}\right>^2\ &=\
    |\left<\widehat{|\mu^2|_{\phi}}\right>
      - \left<\widehat{|\mu|_{\phi}}\right>^2| \ , &
  \left<\Delta \hat{p}_{\phi}\right>^2\ &=\
    |\left<\hat{p}_{\phi}^2\right>
      - \left<\hat{p}_{\phi}\right>^2| \ .
      \tag{\ref{eq:disp}}
\end{align}\end{subequations}

These calculations were performed for 16 different choices of
$\epsilon$ and for ten values of $p_\phi$ up to a maximum of
$p_\phi= 10^3$, and for 5 different choices of the dispersion
parameter $\sigma$. Fig. \ref{fig:direct-3d} provides an example
of a state constructed via this method. The expectation values of
$\left<{\widehat{|\mu|_{\phi}}}\right>$ are shown in Fig.
\ref{fig:direct-mu}.

Results are discussed in section \ref{s5.3} below.

\subsection{Evolution in $\phi$}
  \label{s5.2}

We can also regard the quantum constraint (\ref{qh7}) as an
initial value problem in `time' $\phi$ and solve it by carrying
out the $\phi$-evolution. Conceptually, this approach is simpler
since it does not depend on the properties of the eigenfunctions
of $\Theta$. However, compared to the direct evaluation of the
integral (\ref{eq:psi-int}), this method is technically  more
difficult because it entails solving a large number of coupled
differential equations. Nonetheless, to demonstrate the
robustness of results, we carried out the $\phi$-evolution as
well. This sub-section summarizes the procedure.

\subsubsection{Method of integration}
  \label{sec:int-method}

At large $|\mu|$ the difference equation  is well approximated by
the \WDW equation which is a hyperbolic partial differential
equation (PDE). However since $\Theta$ couples $\mu$ in discrete
steps, for a given $\epsilon$-sector \eqref{qh7} is just a system
of a countable number of coupled ordinary differential equations
(ODEs) of the 2nd order.%
\footnote{Unfortunately for the $\epsilon=0$ sector the equation
is singular at $\mu=0$, so the analysis of this sub-section will
not go through. This sector was handled by the direct evaluation
of the integral representation of the solution, presented in the
last sub-section.}

Technical limitations require restriction of the domain of
integration to a set $|\mu-\epsilon|\leq 4N\mu_o$, where $1 \ll N
\in \mathbb{Z}$. This restriction makes the number of equations
finite. However one now needs to introduce appropriate boundary
conditions. The fundamental equation on the boundary is $i\p_\phi
\Psi = s\sqrt{\Theta}\Psi$, where $s=+1$ ($-1$) for the forward
(backward) evolution in $\phi$. It is difficult to calculate
$\sqrt{\Theta}\Psi$  at each time step. Therefore, \emph{just on
the boundary itself}, this equation was simplified to:
\begin{equation}\label{eq:boundary}
  \partial_{\phi} \Psi(\mu,\phi)\ =\ s\sqrt{\pi  G/3}(|\mu|-2\mu_o)
  ( \Psi(\mu,\phi) - \Psi(\mu-4\sgn(\mu)\mu_o,\phi) ) \ .
\end{equation}
This is the discrete approximation of the continuum operator
$\p_\phi\Psi  = s(\sqrt{16\pi G/3})\mu\,\p_\mu \Psi$ which itself
is an excellent approximation to the fundamental equation when the
boundary is far. The boundary condition requires the solution to
leave the domain of integration. (For, to make the evolution
deterministic in the domain of interest, it is important to avoid
waves entering the integration domain from the boundary.) The
boundary was chosen to lie sufficiently far from the location (in
$\mu$) of the peak of the initial wave packet. Its position was
determined by requiring that the value of the wave function at the
boundary be less than $10^{-n}$ times that at its value of the
peak, and $n$ ranged between $9$ and $24$ in different numerical
simulations.

\begin{figure}[tbh!]
  \begin{center}
    \includegraphics[width=5in,angle=0]{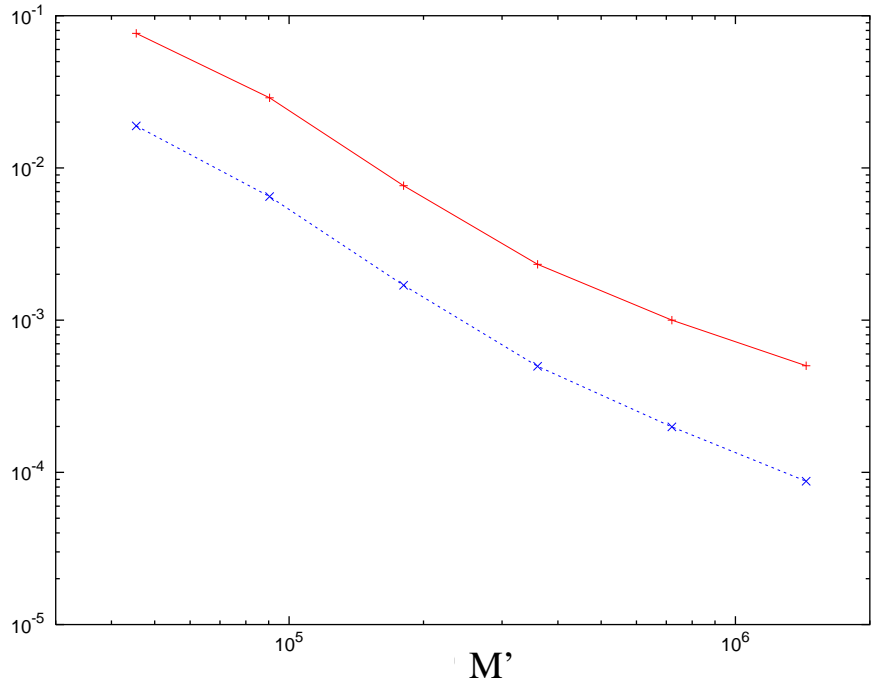}
    \caption{Error functions $|\Psi_{(M')} - \Psi_{(M)}|_{I}$
(upper curve) and $|\Psi_{(M')} - \Psi_{(M)}|_{II}$ (lower curve)
are plotted as a function of time steps. Here $\Psi_{(M')}$ refers
to final profile of wave function for simulation with $M'$ time
steps. $\Psi_{(M)}$ refers to the final profile for the finest
evolution (approximately $2.88 \times 10^6$ time steps). In both
cases, the evolution started at $\phi=0$ and the final profile
refers to $\phi = -1.8$.} \label{fig:conv-test}
\end{center}
\end{figure}

Three different methods were used to specify the initial data
$\Psi$ and $\partial_{\phi}\Psi$ at $\phi=\phi_o$. These are
described in section \ref{sec:phi-init-data}. The data were then
evolved using the fourth order adaptive Runge-Kutta method
(RK4). To estimate the numerical error due to discretization of
time evolution, two sup-norms were used:
\begin{equation}\label{eq:psi-psi2_1}
|\Psi_1 -\Psi_2|_{I}\, (\phi) = \frac{\sup_{|\mu_i-\epsilon|\leq
N\mu_o}\,|\Psi_1-\Psi_2|(\phi) } {\sup_{|\mu_i-\epsilon|\leq
N\mu_o} |\Psi_2|(\phi) } \ .
\end{equation}
and
\begin{equation}\label{eq:psi-psi2_2}
|\Psi_1 -\Psi_2|_{II}\,(\phi) = \frac{\sup_{|\mu_i-\epsilon|\leq
N\mu_o} \left| |\Psi_1| -|\Psi_2|\right|(\phi)}
{\sup_{|\mu_i-\epsilon|\leq N\mu_o} |\Psi_2|(\phi)} \ .
\end{equation}
Fig. \ref{fig:conv-test} shows an example of the results of
convergence tests for the solution corresponding to $p_\phi =
10^3$ with initial spread in $p_\phi$ of $3\%$ (used in figures
\ref{fig:rel-dmu} --- \ref{fig:zoom-symm}). One can see that the
phase of $\Psi$ is more sensitive to numerical errors than its
absolute value. Therefore, although the accuracy is high for the
solution $\Psi(\mu,\phi)$ itself, it is even higher for the mean
values and dispersions of $\widehat{|\mu|_{\phi_o}}$ and
$\hat{p}_\phi$.

\subsubsection{Initial data}
  \label{sec:phi-init-data}

Although we are interested in positive frequency solutions, to
avoid having to take square-roots of $\Theta$ at each `time step',
the second order evolution equation (\ref{qh7}) was used. Thus,
the initial data, consist of the pair $\Psi$ and its time
derivative $\p_{\phi} \Psi$, specified at some `time' $\phi_o$.
The positive frequency condition was incorporated by specifying,
in the initial data, the time derivative of $\Psi$ in terms of
$\Psi$.  Since $\Psi(\mu) = \Psi(-\mu)$, we will restrict
ourselves to the positive $\mu$ axis.  
The idea is to choose semi-classical initial data peaked
at a point $(p_\phi^\star,\, \mu^\star)$ on a classical trajectory
at `time' $\phi=\phi_o$, with $(p_\phi^\star \gg
\sqrt{G\hbar^2},\, \mu^\star\gg 1)$, and evolve them. To avoid
philosophical prejudices on what the state should do at or near
the big bang, we specify the data on the expanding classical
trajectory at `late time' (i.e., `now') and ask it to be
semi-classical like the observed universe.

The idea that the data be semi-classical was incorporated in three
related but distinct ways.
\begin{enumerate}
\item \textit{Method I}: This procedure mimics standard quantum
mechanics. Since $(c,\mu)$ are canonically conjugate on the phase
space, we chose $\Psi(\mu)$ to be a  Gaussian (with respect to
the measure defined by the \WDW inner product) and peaked at
large $\mu^\star$ and the value $c^\star$ of $c$ at the point on
the classical trajectory determined by
$(\mu^\star,p_{\phi}^\star,\phi_o)$):
    \begin{equation}
      \Psi\mid_{\phi_o} := N \, |\mu|^{\f{3}{4}} \, e^{-\frac{(\mu -
      \mu^*)^2}{2\tilde{\sigma}^2}} \,  e^{-i \f{c^\star
      (\mu - \mu^*)}{2}} \ .
    \end{equation}
where $N$ is a normalization constant. The initial value of
$\partial_\phi\Psi|_{\phi_o}$ was calculated using the classical
Hamilton's equations of motion
   \begin{equation}
      \partial_\phi\Psi|_{\phi_o}\
      =\  \left( -\sgn(k^\star)\sqrt{\frac{16\pi G}{3}}\,
        \frac{(\mu-\mu^{\star})\mu^{\star}}{\tilde{\sigma}^2}
          + i \, \omega^\star\,\f{\mu}{\mu^\star} \right)
          \Psi\mid_{\phi_o} \ ,
    \end{equation}
where $\omega^\star = p^\star_\phi/\hbar$.

\item \textit{Method II}: This procedure takes advantage of the
fact that \eqref{sc} provides a \WDW physical state which is
semi-classical at late times. The idea is to calculate $\Psi$ and
$\p_\phi\Psi$ at $\phi=\phi_o$ and use their restrictions to
lattices ${\cal L}_\epsilon$ as the initial data for LQC, setting
$\omega = \sqrt{16\pi G/3} \, |k|$ as in the discussion following
\eqref{sc}. Thus the initial data used in the simulations were of
the form:
\begin{eqnarray}
\Psi\mid_{\phi_o}\
       &=&\ \left(\frac{\mu}{\mu^\star}\right)^{\frac{1}{4}}
         e^{ -\frac{\sigma^2}{2} \ln^2 \frac{|\mu|}{|\mu^{\star}|}
         }\,\,
         e^{i k^\star \ln \frac{|\mu|}{|\mu^\star|} } \ , \\
       \partial_{\phi}\Psi|_{\phi_o}\
       &=&\ \left(-\sgn(k^\star)\sqrt{\frac{16\pi}{3}}
         \,\sigma^2\ln\frac{|\mu|}{|\mu^\star|}
       + i \, \omega(k^\star) \right)
     \Psi|_{\phi_o} \
\end{eqnarray}
This choice is best suited to comparing the LQC results with those
of the \WDW theory.

The spreads $\t\sigma$ of \emph{Method I} and ${\sigma}$ of
\emph{Method II} are related to the initial spread
$\Delta\mu|_{\phi_o}$ as follows
\begin{equation}
  \tilde{\sigma}\ =\ \frac{\mu^\star}{\sigma}\
  =\ \sqrt{2}\Delta\mu|_{\phi_o} \ .
\end{equation}

\item \textit{Method III}: To facilitate comparison with the
direct evaluation of the integral solution described in section
\ref{s5.1}, a variation was made on \textit{Method II}.
Specifically, in the expression \eqref{sc} of $\Psi$, the \WDW
basis eigenfunctions $\ub{e}_k(\mu)$ were rotated by multiplying
them with a $k$ and $\epsilon$ dependent phase factor defined in
Eq. \eqref{eq:phases}
\begin{equation}\label{eq:base-rot}
  \ub{e}_{-|k|}\mapsto e^{-i\alpha^{+}} \ub{e}_{-|k|} \ .
\end{equation}
These phases were first found numerically using the method
specified in appendix \ref{a2} and then functions of the form
\begin{equation}
  \alpha^{+}\, =\, A \, \ln(Bk+C)k+D \ ,
\end{equation}
(where $A,B,C,D$ are real constants) were fitted to the results.
After subtraction of the $0$th and the $1$st order terms in the
expansion in $k$ around $k^\star$ (which respectively correspond
to a constant phase and a shift of origin of $\phi$) the resulting
function was used to rotate the basis $\ub{e}_k(\mu)$ appearing in
\eqref{sc}) via \eqref{eq:base-rot}). The expression on the right
side of Eq \eqref{sc} and its $\phi$-derivative were then
integrated numerically at $\phi=\phi_o$.
\end{enumerate}

In the simulations, 15 different values of $p_\phi^\star$ were
used ranging between $10^2$ and $10^5$. $\mu^\star$ was always
greater than $2.5p_\phi$. The dispersion $\sigma$ was allowed to
have five different values. These simulations involved four
different, randomly chosen values of $\epsilon$.

\begin{figure}[tbh!]
  \begin{center}
    \includegraphics[width=5in,angle=0]{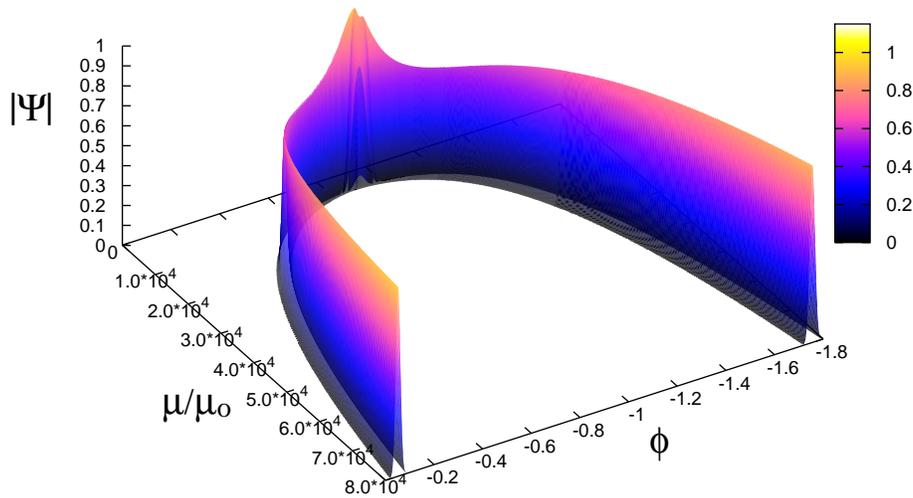}
    \caption{The absolute value of the wave function obtained
    by evolving an initial data of \emph{Method II}. For clarity
    of visualization, only the values of $|\Psi|$ greater than
    $10^{-4}$ are shown. Being a physical state, $\Psi$ is
    symmetric under $\mu \rightarrow -\mu$. In this simulation, the
    parameters were: $\epsilon=2 \mu_o$, $p_{\phi}^{\star}=10^4$, and
    $\Delta p_\phi/p_\phi^{\star} = 7.5 \times 10^{-3}$.}
    \label{fig:l-3d}
  \end{center}
\end{figure}

\begin{figure}[tbh!]
  \begin{center}
    \includegraphics[width=5in,angle=0]{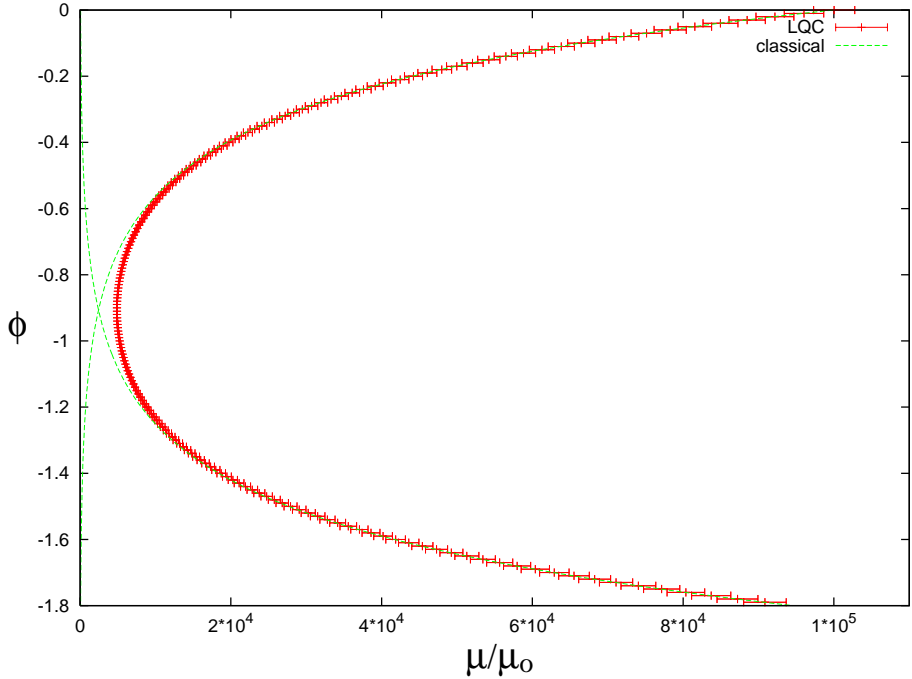}
    \caption{The expectation values (and dispersions) of
     $\widehat{|\mu|_{\phi}}$ are plotted for the wavefunction in
     Fig. \ref{fig:l-3d} and compared with expanding and contracting
     classical trajectories.}
    \label{fig:l-mu}
  \end{center}
\end{figure}

\begin{figure}[tbh!]
  \begin{center}
    \includegraphics[width=5in,angle=0]{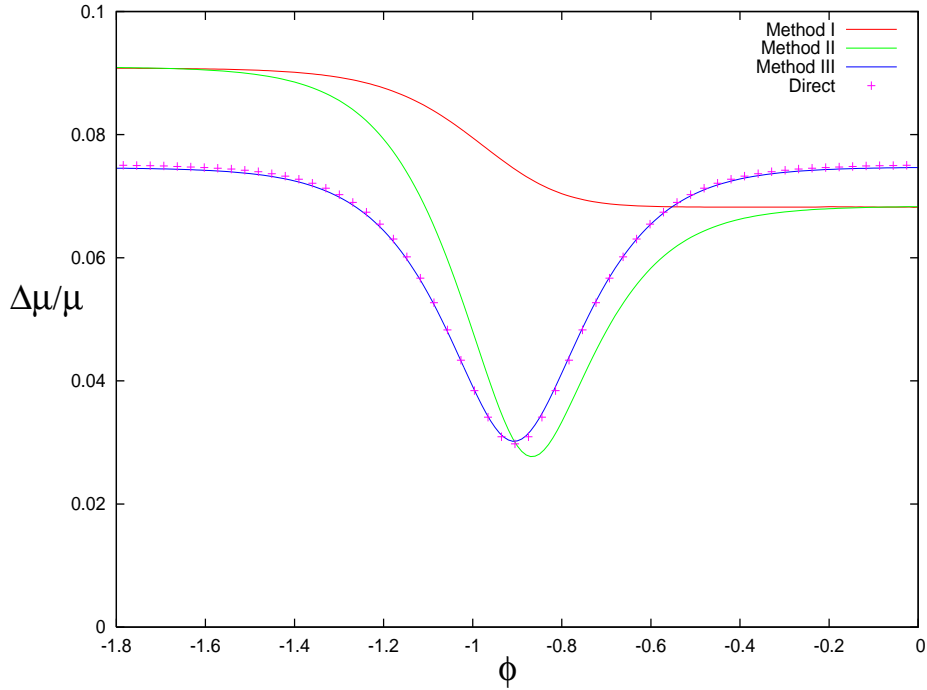}
    \caption{Comparisons between the relative dispersions
    $\Delta\mu/\mu$ as functions of $\phi$ for all three methods
    specifying initial data and the result of direct construction.}
    \label{fig:rel-dmu}
  \end{center}
\end{figure}

\begin{figure}[tbh!]
  \begin{center}
    \includegraphics[width=5in,angle=0]{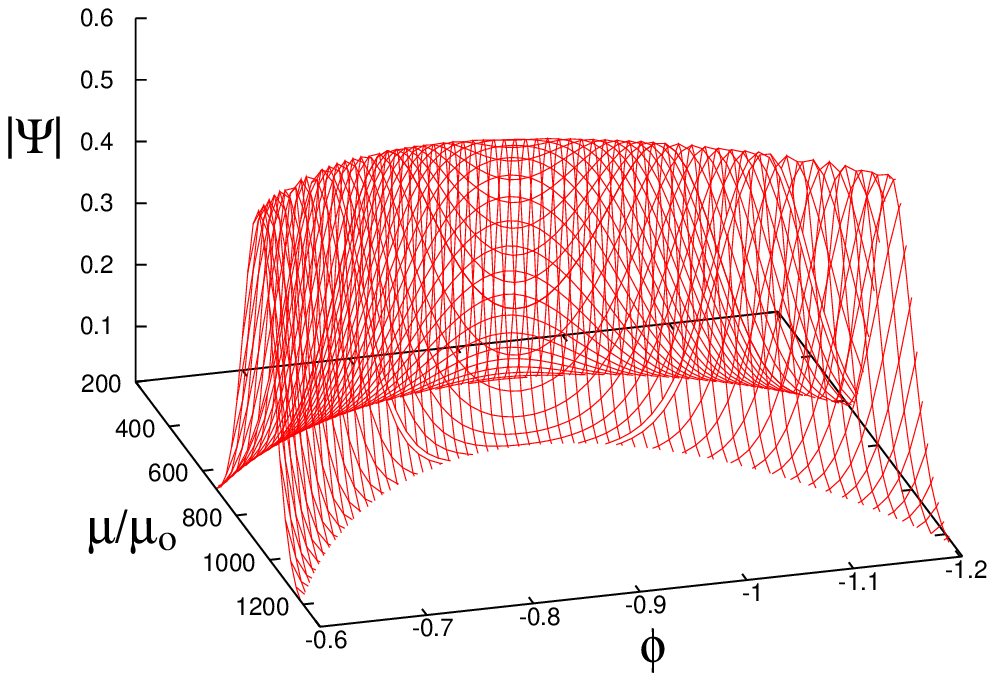}
    \caption{A zoom on the absolute value of the wave function near
    the bounce point. Initial data was specified using
    \emph{Method I}.}
    \label{fig:zoom-gauss}
  \end{center}
\end{figure}

\begin{figure}[tbh!]
  \begin{center}
    \includegraphics[width=5in,angle=0]{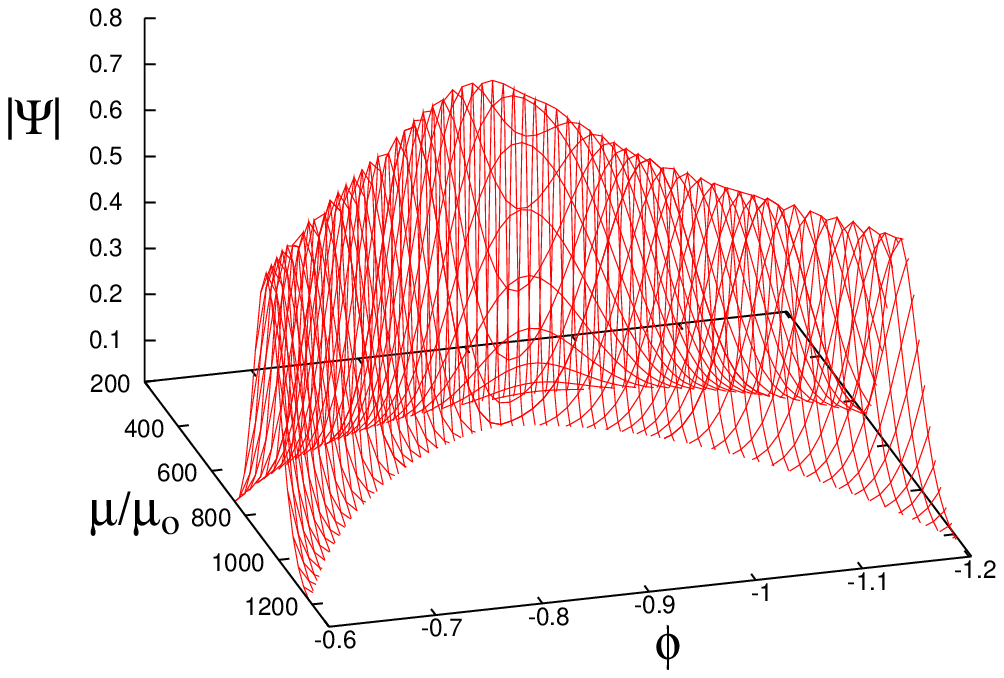}
    \caption{A zoom on the absolute value of the wave function near
    the bounce point. Initial data was specified using
    \emph{Method II}.}
    \label{fig:zoom-wdw}
  \end{center}
\end{figure}

\begin{figure}[tbh!]
  \begin{center}
    \includegraphics[width=5in,angle=0]{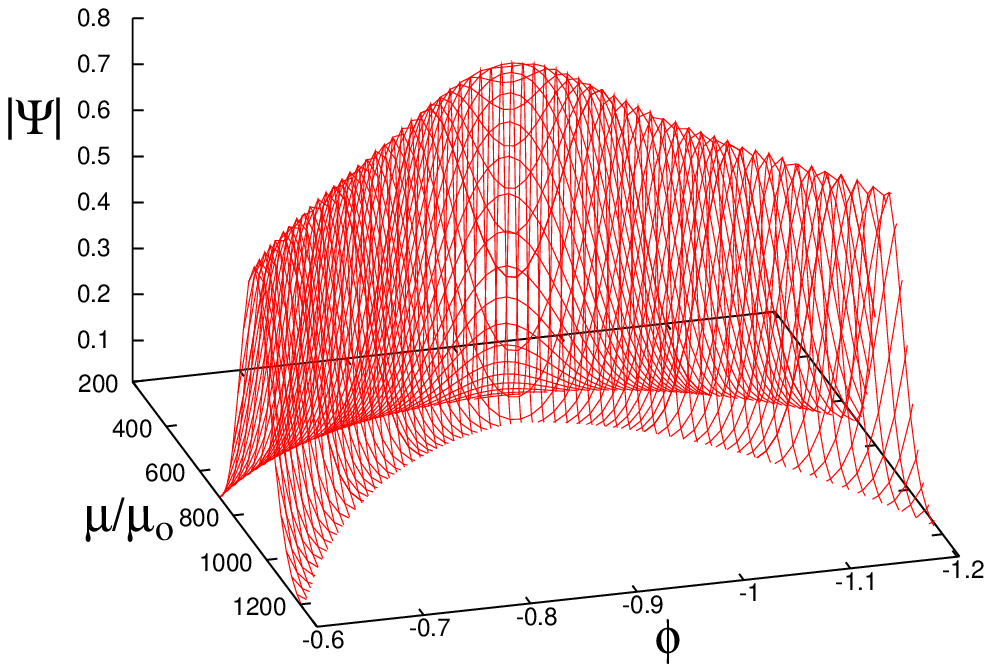}
    \caption{A zoom on the absolute value of the wave function near
    the bounce point. Initial data was specified using
    \emph{Method III}.}
    \label{fig:zoom-symm}
  \end{center}
\end{figure}

\begin{figure}[tbh!]
  \begin{center}
    \includegraphics[width=5in,angle=0]{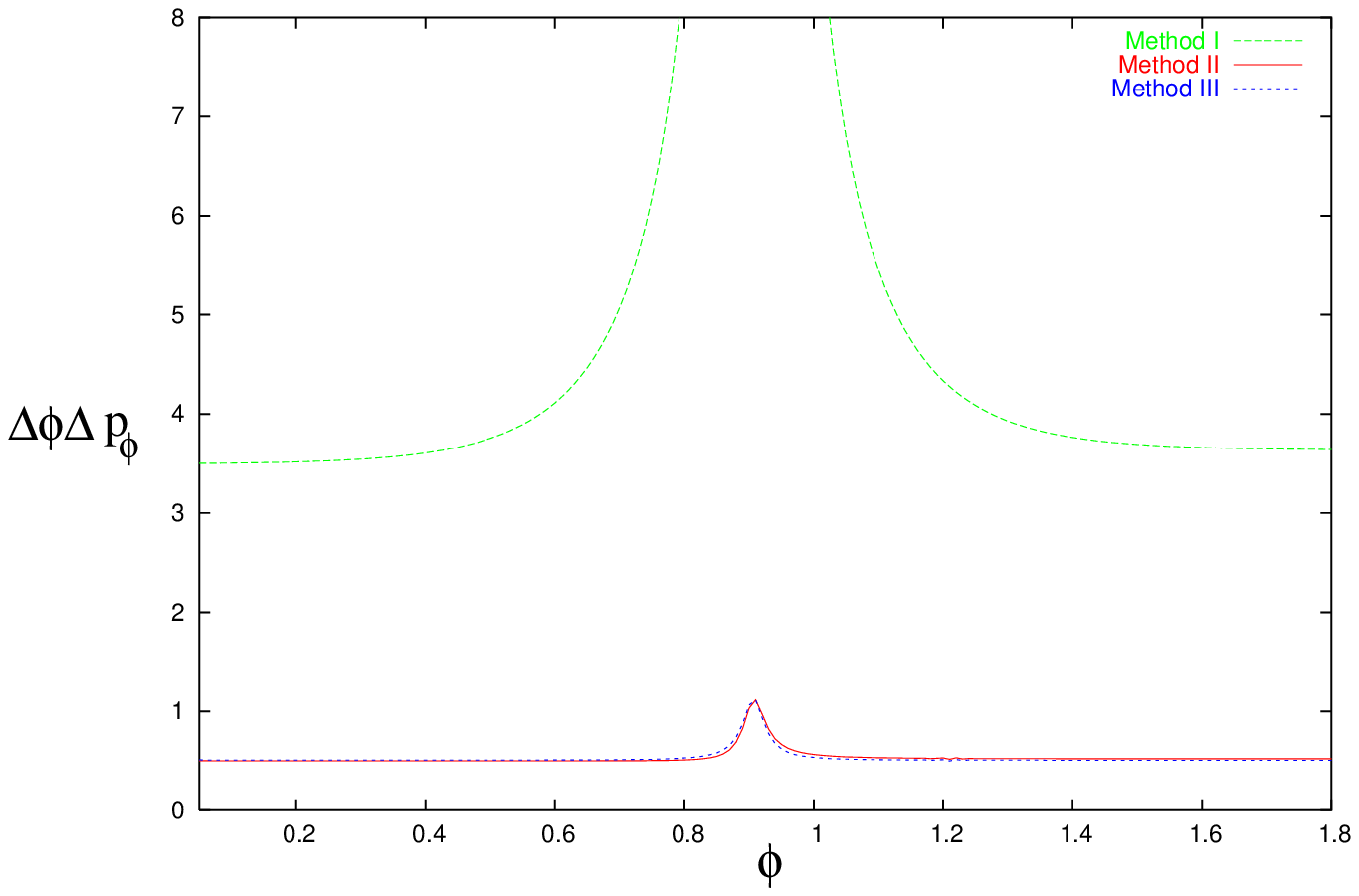}
    \caption{The uncertainty product $\Delta\phi \Delta p_\phi$ for
three methods of specifying the initial data.}
    \label{fig:dphi1}
  \end{center}
\end{figure}

\subsection{Results and Comparisons}
  \label{s5.3}

We now summarize the results obtained by using the two
constructions specified in sections \ref{s5.1} and \ref{s5.2}. The
qualitative results are robust and differences lie in the finer
structure. In particular, evolutions of initial data constructed
from the three methods described in section
\ref{sec:phi-init-data} yield physical states $\Psi(\mu,\phi)$
which are virtually identical except for small differences in the
behavior of relative dispersions of the Dirac observables. The
numerical evaluation of the integral (\ref{eq:psi-int}) in section
\ref{sec:state-direct} yielded results very similar to those
obtained in \ref{sec:phi-init-data} using initial data of
\textit{Method III}. An example of results is presented in Fig.
\ref{fig:l-3d} and Fig. \ref{fig:l-mu}.

Highlights of the results can be summarized as follows:
\begin{enumerate}[ (1)]
\item The state remains sharply peaked throughout the evolution.
However, as shown in Fig. \ref{fig:dphi1}, while the product
$\Delta \phi\,\Delta p_\phi$ is nearly constant for large $\mu$,
there is a substantial increase near $\mu=0$.
\item The expectation values of $\hat \mu_\phi$ and $\hat p_\phi$
are in good agreement with the classical trajectories, until the
increasing matter density approaches a critical value. Then, the
state bounces from the expanding branch to a contracting branch
with the same value of $\langle \hat{p}_\phi\rangle$. (See Fig.
\ref{fig:l-mu}). This phenomena occurs {\it universally}, i.e.,
in every $\ep$-sector, for all three methods of choosing the
initial data and for any choice of $p_\phi\gg \sqrt{G\hbar^2}$.
\emph{In this sense the classical big-bang is replaced by a
quantum bounce.} Note that this is in striking contrast with the
situation with the \WDW theory we encountered in section
\ref{s3.2}, even when the initial data is chosen using
\textit{Method II} which is tailored to the \WDW theory. As
indicated in Appendix \ref{a1}, the existence of the bounce can be
heuristically understood from an `effective theory'. The detailed
numerical work supports that description, thereby providing a
justification for the approximation involved.
\item If the state is peaked on the expanding branch $\mu(\phi) =
\mu^\star \exp (\sqrt{16\pi G/3}\, (\phi-\phi_o))$ in the distant
future, due to the bounce it is peaked on a contracting branch in the
distant past, given by $\mu(\phi) = D(p_\phi^\star)\, \mu^\star
\exp (-\sqrt{16\pi G/3}\, (\phi-\phi_o))$, where $D(p_\phi) =
\mu_o^2p_\phi^2/12\pi G\hbar^2$. Thus, for large $|\mu|$ the
solution $\Psi(\mu,\phi)$ exhibits reflection symmetry (about
$\phi = \phi_o -\f{1}{2}\ln D(p_\phi^\star)$). However, it is not
exactly reflection symmetric (compare \cite{shw}).
\item As a consistency check, we verified that the norm and the
expectation value $\langle \hat{p}_\phi \rangle$ are preserved
during the entire evolution. Furthermore, the dispersion also
remains small throughout the evolution, although the precise
behavior depends on the method of specification of the initial
data. Differences arise primarily near the bounce point and
manifest themselves through the behavior of the relative
dispersion $\Delta \mu/\mu$ as a function of $\phi$. These are
illustrated  (for all three methods as well as for the direct
evaluation of section \ref{s5.1}) in Fig. \ref{fig:rel-dmu}.
Finally, as argued in Appendix \ref{a1}, since the state is
sharply peaked,  the value of $\Delta\mu/\mu$ can be related to
$\Delta\phi$ via Eq \ref{dp_dm_m}. One finds that the product
$\Delta\phi\Delta p_{\phi}$ is essentially constant in the region
away from the bounce but grows significantly near the bounce.

The differences can be summarize as follows:
\begin{enumerate}[ (i)]
\item In \emph{Method I} the initial state is a minimum
uncertainty state in $(\mu, c)$ but doesn't minimize the
uncertainty in $(\phi, p_\phi)$ at any value of $\phi$. The
relative spread in $\mu$ remains approximately constant in the
regions where $\langle\hat{\mu}_{\phi}\rangle$ is large and
increases near the bounce point monotonically. The wave function
interpolates `smoothly' between expanding and contracting branches
(see Fig. \ref{fig:zoom-gauss}). On the other hand the product
$\Delta \phi \Delta p_\phi$ of uncertainties has a value much
higher than 1/2, grows quickly near the bounce and settles down to
constant value after it. See Fig. \ref{fig:dphi1}.
\item The state obtained by evolving the initial data constructed
from \emph{Method II} has minimal uncertainty in $(\phi, p_\phi)$.
$\Delta \mu/\mu$ is approximately constant for large
$\langle\widehat{\mu_{\phi}}\rangle$ and it decreases quickly
near the bounce point, reaching its minimal value shortly before
the bounce point. After the bounce it grows and stabilizes at the
value of the relative spread found for data constructed using
\textit{Method I} (for the same values of $p_{\phi}^\star$ and
initial $\Delta\mu/\mu$). Behavior of the wave function is also
different from that in the previous case. Near the bounce point
its value grows to form a bulge (see Fig. \ref{fig:zoom-wdw}). The
product $\Delta\phi\Delta p_{\phi}$ remains almost constant for as
long as the results of the numerical measurement are reliable (see
Fig. \ref{fig:dphi1}) approaching a constant after the bounce,
with a somewhat higher value. The heuristic estimate on
$\Delta\phi\Delta p_{\phi}$ (see Appendix \ref{a1}) agrees with
these results. Finally, we also found that the increase of the
relative spread depends on $p_{\phi}^\star$ and initial
$\Delta\mu/\mu$.
\item  The state obtained by evolving the initial data constructed
from \emph{Method III} does not have minimum uncertainty in
$(\phi, p_\phi)$. The behavior of $\Delta\mu/\mu$ is similar to
the previous case except that it becomes equal to $\Delta
p_{\phi}/p_{\phi}$ at the bounce point and its asymptotic value in
the contracting branch is same as its starting value. Thus the
spread is symmetric with respect to reflection in $\phi$ around
the bounce point. The difference between the value of
$\Delta\mu/\mu$ for large
$\langle\widehat{\mu_{\phi}}\rangle$ and the one
corresponding to the minimum uncertainty state (for the same
$p_{\phi}^\star$ and $\Delta p_{\phi}$) is a function of
$p_{\phi}$ and $\Delta p_{\phi}$. The wave function forms a
symmetric bulge near the bounce point and the value
$\Delta\phi\Delta p_{\phi}$ remains constant within regime of
validity of its estimation (see Fig.\ref{fig:dphi1}).
\item The direct construction of section \ref{s5.1} yields results
similar to those obtained by using the third method to choose the
initial data for the $\phi$ evolution.
\item The differences between the relative dispersion $\Delta
\mu/\mu$ resulting from different methods of choosing initial data
can also be estimated using the solutions to the effective
dynamical equations (for details of the method see Appendix
\ref{a1}).
\end{enumerate}
\end{enumerate}

We will conclude with two remarks.
\begin{enumerate}[(i)]
\item Let us return to the comparison between the results of LQC
and the \WDW theory in light of our numerical results. As remarked
in section \ref{s2.2}, on functions $\Psi(\mu)$ which we used to
construct the semi-classical initial data, the leading term in the
difference $(\hat{C}_{\rm grav} - \hat{C}_{\rm grav}^{\rm
wdw})\Psi$ between the actions of the two constraint operators
goes as $O(\mu_o^2) \Psi''''$. Now, in LQC $\mu_o$ is fixed,
($\mu_o= 3\sqrt{3}/2$), and on semi-classical states $\Psi''''
\sim k^{\star 2}/\mu^4\, \Psi$. Hence the difference is negligible
only in the regime $k^{\star 4}/\mu^4 \ll 1$. In our simulations,
$k^\star \sim 10^4$ whence the differences are guaranteed to be
negligible only for $\mu \gg 10^4$, i.e., well away from the
bounce. But let us probe the situation in greater detail. Let us
regard $\mu_o$ as a mathematical parameter which can be varied and
shrink it. In the limit $\mu_o \rightarrow 0$, \, $(\hat{C}_{\rm
grav} - C^{\rm wdw}_{\rm grav})\Psi$ should tend to zero. Since
there is no bounce in the \WDW theory, we are led to ask: Would
the LQC bounce continue to exist all the way to $\mu_o=0$ or is
there a critical value at which the bounce stops? The answer is
that for any finite value of $\mu_o$ there is a bounce. However,
if we keep the \emph{physical} initial data the same, we find that
as we decrease $\mu_o$ the solution follows the classical
trajectory into the past more and more and bounce is pushed
further and further in to the past. In the limit as $\mu_o$ goes
to zero, the wave function follows the classical trajectory into
infinite past, i.e., the bounce never occurs. This is the sense in
which the \WDW result is recovered in the limit $\mu_o \rightarrow
0$.

\item Numerical simulations show that the matter density at the
bounce points is inversely proportional to the expectation value
$\langle \hat{p}_\phi\rangle \equiv p_\phi^\star$ of the Dirac
observable $\hat{p}_\phi$: Given two semi-classical states with
$\langle \hat{p}_\phi \rangle = p_\phi^\star$ and
$\bar{p}_\phi^\star$, we have $\rho_{\rm crit}/ \bar{\rho}_{\rm
crit} = \bar{p}^\star_\phi/p^\star_\phi$. Therefore, this density
can be made small by choosing sufficiently large $p_\phi^\star$.
Physically, this is unreasonable because one would not expect
departures from the classical theory until matter density becomes
comparable to the Planck density. This is a serious weakness of
our framework. Essentially every investigations within LQC we are
aware of has this ---or a similar--- drawback but it did not
become manifest before because the physics of the singularity
resolution had not been analyzed systematically. The origin of
this weakness can be traced back to details of the construction of
the Hamiltonian constraint operator, specifically the precise
manner in which the operator corresponding to the classical field
strength $F_{ab}^i$ was introduced. The physical idea that in LQC
the operator corresponding to the field strength $F_{ab}^i$ should
be defined through holonomies, and that quantum geometry does not
allow us to shrink the loop to zero size, seem compelling.
However, the precise manner in which the value of $\mu_o$ was
determined using the area gap $\Delta$ is not as systematic and
represents only a `first stab' at the problem. In \cite{aps3} we
will discuss an alternate and more natural way of implementing
this idea. The resulting Hamiltonian constraint has a similar form
but also important differences. Because of similarities  the
qualitative conclusions of this analysis ---including the
occurrence of the quantum bounce--- are retained but the
differences are sufficiently important to replace the expression
of the critical density by $\rho_{\rm crit}^\prime = (3/8\pi
G\gamma^2 \Delta)$ where, as before, $\Delta = 2\sqrt{3}\pi\gamma
\lp^2$ is the area gap. Since $\rho_{\rm crit}^\prime$ is of
Planck scale and is independent of parameters associated with the
semi-classical state, such as $p_\phi^\star$, the departures from
classical theory now appear only in the Planck regime. This issue
is discussed in detail in \cite{aps3}.

\end{enumerate}

\section{Discussion} \label{s6}

We will first present a brief summary and then compare our
analysis with similar constructions and results which have
appeared in the literature. However, since this literature is
vast, to keep the discussion to a manageable size, these
comparisons will be illustrative rather than exhaustive.

In general relativity, gravity is encoded in space-time geometry.
A basic premise of LQG is that geometry is fundamentally quantum
mechanical and its quantum aspects are central to the
understanding of the physics of the Planck regime. In the last
three sections, we saw that LQC provides a concrete realization of
this paradigm. In our model every classical solution is singular
and the singularity persists in the \WDW theory. The situation is
quite different in LQC. As in full LQG, the kinematical framework
of LQC forces us to define curvature in terms of holonomies around
closed loops. The underlying quantum geometry of the full theory
suggests that it is physically incorrect to shrink the loops to
zero size because of the `area gap' $\Delta$. This leads to a
replacement of the \WDW differential equation by a difference
equation whose step size are dictated by $\Delta$. Careful
numerical simulations demonstrated in a robust fashion that the
classical big-bang is now replaced by a quantum bounce. Thus, with
hindsight, one can say that although the \WDW theory is quite
similar to LQC in its structure, the singularity persists in the
\WDW theory because it ignores the quantum nature of geometry.%
\footnote{Differences arise in two places. The first occurs in the
matter part of the constraint and stems from the fact that the
functions $B(\mu)$ representing the eigenvalues of the operator
$\widehat{1/|p|^{3/2}}$ in LQG is different from the corresponding
$\ub{B}(\mu)$ of the \WDW theory. The second comes from the role
of quantum geometry in the gravitational part of the Hamiltonian
constraint, emphasized above. In our model, qualitatively new
features of the LQG quantum dynamics can be traced back to the
second. In particular, the bounce would have persisted even if we
had used $\ub{B}(\mu)$ in place of $B(\mu)$ in the analysis
presented in this paper.}

\bigskip
\centerline{Emergent time}
\medskip

In this paper we isolated the scalar field $\phi$ as the emergent
time and used it to motivate and simplify various constructions.
However, we would like to emphasize that this ---or indeed any
other--- choice of emergent time is not \emph{essential} to the
final results. In the classical theory (for any given value of the
constant of motion $p_\phi$), we can draw dynamical trajectories
in the $\mu-\phi$ plane without singling out an internal clock. A
complete set of  Dirac observables can be taken to be $p_\phi$,
and either $|\mu|_{\phi_o}$ \emph{or} $\phi|_{\mu_o}$.%
\footnote{In the closed models, care is needed to specify the
latter because they can not be defined globally. But this issue is
well-understood in the literature, especially through Rovelli's
contributions \cite{crtime}. See also \cite{dm}.}
What these observables measure is correlations and their
specification singles out a point in the reduced phase space.
However, we do not have to single out a time variable to define
them. The same is true in quantum theory. To have complete control
of physics, we need to construct the physical Hilbert space $\Hp$
and introduce on it a complete set of Dirac observables. Again,
both these steps can be carried out without singling out $\phi$ as
emergent time. For example, the scalar product can be constructed
using group averaging which requires only the knowledge of the
full quantum constraint and its properties, and not its
decomposition into a `time evolution part' $\p_\phi^2$ and an
operator $\Theta$ on the `true degrees of freedom'. Once the
scalar product is constructed, we can introduce a complete set of
Dirac observables consisting of $\hat{p}_\phi$, and
$\widehat{|\mu|_{\phi_o}}$ \emph{or} $\widehat{\phi|_{\mu_o}}$.
Again, what matters is the correlations. This and related issues
have been discussed exhaustively in the quantum gravity literature
in relativity circles. In particular, a major part of a conference
proceedings \cite{asbook} was devoted to it in the late eighties
and several exhaustive reviews also appeared in the nineties (see
in particular \cite{cji,kk}).

However, thanks to our knowledge of how quantum theory works in
static space-times, singling out $\phi$ as the emergent time
turned out to be extremely useful in practice because it provided
guidance at several intermediate steps. In particular,  it
directly motivated our choice of $L^2(\R_{\rm Bohr}, B(\mu)
\dd\mu_{\rm Bohr}) \otimes L^2(\R, \dd\phi)$ as our auxiliary
Hilbert space; stream-lined the detailed definition of operators
representing the Dirac observables; and facilitated the subsequent
selection of the inner product by demanding that these be
self-adjoint. More importantly, by enabling us to regard the
constraint as an evolution equation, it transformed the `frozen
formalism' to a familiar language of `evolution' and enabled us to
picture and interpret the bounce and associated physics more
easily. Indeed, following the lead of early LQC papers, initially
we tried to use $\mu$ as time and ran in to several difficulties:
specification of physically interesting data became non-intuitive
and cumbersome; one could not immediately recognize the occurrence
of the bounce; and the physics of the singularity resolution
remained obscure.

Our specific identification and use of emergent time differs in
some respects from that introduced earlier in the literature. For
example, in the context of the \WDW theory, there is extensive
work on isolating time in the WKB approximation (see e.g.,
\cite{ck,wkb}). By contrast, a key feature of our emergent time is
that it is not restricted to semi-classical regimes: We first
isolated the scalar field $\phi$ as a variable which serves as a
good internal clock away from the singularity in the classical
theory, but then showed that the form of the \emph{quantum}
Hamiltonian constraint is such that $\phi$ can be regarded as
emergent time in the \emph{full} quantum theory, without a
restriction that the states be semi-classical or stay away from
the singularity. The idea of identifying an emergent time and
exploiting the resulting `deparametrization' to select an inner
product on the space of solutions to the Hamiltonian constraint is
not new \cite{asbook,aabook,cji,kk,dm,hm2}. However, several of
the concrete proposals turn out to have serious deficiencies (for
a further discussion see, e.g., \cite{greensite,kiefertime}). The
idea of using a matter field to define emergent time is rather
old. In the framework of geometrodynamics, it was carried out in
detail for dust in \cite{bktime}. A proposal to use a massless
scalar field as time was also made in the framework of LQG
\cite{lstime} but its implementation remained somewhat formal. In
particular, it is unlikely that the required gauge conditions can
be imposed globally in the phase space and  the modifications in
the construction of the physical scalar product that are necessary
to accommodate a more local constructions were not spelled out.
More recently, a massless scalar field was used as the internal
clock in quantum cosmology in the connection dynamics framework
\cite{kodama}. However, the focus of discussion there is on the
Kodama state in inflationary models. Because of the inflationary
potential, the quantum constraint has explicit time dependence and
the construction of the physical inner product is technically much
more subtle. In particular, a viable inner product cannot depend
on any auxiliary structure such as the choice of an `instant of
time'. These issues do not appear to have been fully addressed in
\cite{kodama}.

\bigskip
\centerline{Resolutions of the Big-Bang Singularity}
\medskip

The issue of obtaining singularity free cosmological models has
drawn much attention over the years. The discovery of singularity
theorems sharpened this discussion and there is a large body of
literature on how one may violate one or more assumptions of these
theorems, thereby escaping the big bang. Proposals include the use
of matter which violate the standard energy conditions, addition
of higher derivative corrections to the Einstein-Hilbert action
and introduction of higher dimensional scenarios inspired by
string theory. To facilitate comparison with the model discussed
in this paper, we will restrict ourselves to spatially non-compact
situations.

Already in the seventies, Bekenstein investigated a model where
the matter source consisted of incoherent radiation and dust,
interacting with a conformal massless scalar field (which can have
a negative energy density). He showed that Einstein's equations
admit solutions which are free of singularities \cite{bek}. In the
eighties, Narlikar and Padmanabhan found a singularity free
solution to Einstein's equation with radiation and a
\emph{negative energy} massless scalar field (called the `creation
field') as source, and argued that the resulting model was
consistent with the then available observations
\cite{narlikar_paddy}. Such investigations were carried out
entirely in the paradigm of classical relativity and the key
difference from the standard Friedmann-Robertson-Walker models
arose from the use of `non-standard' matter sources. Our analysis,
by contrast, uses a standard massless scalar field and
\emph{every} solution is singular in the classical theory. The
singularity is resolved because of quantum effects.

Another class of investigations starts with actions containing
higher derivative terms which are motivated by suitable
fundamental considerations. For example, to guide the search for
an effective theory of gravity which is viable close to the Planck
scale, Mukhanov and Brandenberger proposed an action with higher
order curvature terms for which all isotropic cosmological
solutions are non-singular, even when coupled to matter
\cite{mukh_bran}. The modifications to Einstein's equations are
thought of as representing quantum corrections. However one
continues to work with differential equations formulated in the
continuum. By contrast, our investigation is carried out in the
framework of a genuine quantum theory with a physical Hilbert
space, Dirac observables and detailed calculations of expectation
values and fluctuations. Departures from classical general
relativity arise directly from the quantum nature of geometry. The
final results are also different: While solutions in
\cite{mukh_bran} asymptotically approach de Sitter space, in our
analysis the classical big-bang is replaced by a quantum bounce.

Perhaps the most well-known discussions of bounces come from the
pre-big-bang cosmology and Ekpyrotic/Cyclic models. The
pre-big-bang model uses the string dilaton action and exploits the
scale factor duality to postulate the existence of a
super-inflating pre-big-bang branch of the Universe, joined to the
radiation dominated post-big-bang branch \cite{pbb1}. However, the
work was carried out in the framework of perturbative string
theory and the transition from the pre-big-bang to post-big-branch
was \emph{postulated}. The initial hope was that non-perturbative
stringy effects would enforce such a transition. However, as of
now, such mechanisms have not been found \cite{pbb2}. Although
subsequent investigations have shown that a bounce can occur in
simplified models \cite{pbb3} or by using certain effective
equations (see, e.g., \cite{pbb4}), it is not yet clear that this
is a consequence of the fundamental theory. The Ekpyrotic and the
more recent Cyclic models \cite{ekp1,ekp2} are motivated by
certain compactifications in string theory and feature a five
dimensional bulk space-time with a 4-d branes as boundaries. In
the Ekpyrotic model, the collision between a bulk brane with a
boundary brane is envisioned as a big bang. A key difficulty is
the singularity associated with this collision \cite{kklt} (which
can be avoided but at the cost of violating the null energy
condition \cite{ekp2}). In the Cyclic model, collision occurs
between the boundary branes \cite{ekp2}, however it has been shown
that the singularity problems persists \cite{ffkl}. Thus, a common
limitation of these models is that the branch on `our side' of the
big-bang is not joined deterministically to the branch on the
`other side'. In LQC by contrast, the quantum evolution is fully
deterministic. This is possible because the approach is
non-perturbative and does not require a space-time continuum in
the background.

Finally, the idea of a bounce has been pursued also in the context
of braneworld models. In the original Randall-Sundrum scenario,
the Friedmann equation \emph{is} modified by addition of a term on
the right side which is quadratic in density: $\dot{a}^2/a^2 =
(8\pi G/3)\, \rho(1 + \rho/(2 \sigma))$ where $\sigma$ is brane
tension. However, since $\sigma >0$, the sign of the quadratic
term in $\rho$ is positive whence $\dot{a}$ can not vanish and
there is no bounce. To obtain a bounce, the correction should be
negative, i.e., make a `repulsive' contribution. One way to
reverse the sign is to introduce a second time-like dimension in
the bulk \cite{shtanov_sahni}. However, this strategy does not
appear to descend from fundamental considerations and the physical
meaning of the second time-like direction is also unclear. Another
avenue is to consider a bulk with a charged black hole. A
non-vanishing charge leads to terms in the modified Friedmann
equation which are negative. $\dot{a}$ can now vanish and a bounce
can occur \cite{mp_brane}. However it was shown that transition
from contraction to expansion for the brane trajectory occurs in
the Cauchy horizon of the bulk which is unstable to small
excitations, thus the brane encounters singularity before bouncing
\cite{hov_myers}. In LQC, by contrast, while the Friedmann
equation \emph{is} effectively modified, the corrections come from
quantum geometry and they are automatically negative.

\bigskip
\centerline{Extensions}
\medskip

A major limitation of our analysis---shared by all other current
investigations in quantum cosmology--- is that the theory is not
developed by a systematic truncation of full quantum gravity. This
is inevitable because we do not have a satisfactory quantum
gravity theory which can serve as an unambiguous starting point.
The viewpoint is rather that one should use lessons learned from
mini and midi superspace analysis to work one's way up to more
general situations, especially to reduce the large number of
ambiguities that exist in the dynamics of the full theory.

Even within quantum cosmology, our detailed analysis was
restricted to a specific mini-superspace model. In more
complicated models, differences are bound to arise.  For example,
the full solution is not likely to remain so sharply peaked on the
classical trajectory till the bounce point and even the existence
of a bounce is not a priori guaranteed, especially when
inhomogeneities are added. However, the \emph{methods} developed
in the paper can be applied to more general situations. First, one
could consider anisotropies. Now, the main structural difference
is that the operator $\Theta$ will no longer be positive definite.
However, a detailed analysis shows that what matters is just the
operator $|\Theta|$, obtained by projecting the action of $\Theta$
to the positive eigenspace in its spectral decomposition.
Therefore, our analytical considerations should go through without
a major modification. The numerical simulations will be more
complicated because we have to solve a higher dimensional
difference equation (involving 4 variables in place of 2). Another
extension will involve the inclusion of non-trivial potentials for
the scalar field. Now, generically $\phi$ will no longer be a
monotonic function on the classical trajectories and one would not
be able to use it as `internal time' globally. In the quantum
theory, the operator $\Theta$ becomes `time-dependent' (i.e.
depends on $\phi$) and the mathematical analogy between the
quantum constraint and the Klein-Gordon equation in a static
space-time is no longer valid. Nonetheless, one can still use the
group averaging procedure \cite{dm,hm2} to construct the physical
Hilbert space. For a general potential, a useful notion of time
will naturally emerge only in the semi-classical regimes. For
specific potentials (such as the quadratic one used in chaotic
 inflation) one should be able to use methods that have been
successfully employed in the quantization of model systems
\cite{at,lr} (in particular, a pair of harmonic oscillators
constrained to have a fixed total energy).

Incorporation of spherical inhomogeneities seems to be within
reach since significant amount of technical groundwork has already
been laid \cite{mb-ss}. Incorporation of general inhomogeneities,
on the other hand, will be substantially more difficult.
Background dependent treatments have suggested that results
obtained in the mini-superspace approximation may be qualitatively
altered once field theoretical complications are unleashed (see,
e.g., \cite{hh1}). However, already in the anisotropic case, there
is a qualitative difference between perturbative and
non-perturbative treatments. Specifically, if anisotropies are
treated as perturbations of a background isotropic model, the
big-bang singularity is not resolved while if one treats the whole
problem non-perturbatively, it is \cite{mb-aniso}. Therefore
definitive conclusions can not be reached until detailed
calculations have been performed in inhomogeneous models.
However, if a quantum bounce does generically replace the big
bang singularity, it would be possible to explore the relation
between the effective descriptions of LQG and the Hartle-Hawking
`no boundary' proposal \cite{hh2}. For, in the effective
description, the extrinsic curvature would vanish at the bounce.
Therefore generically it may be possible to attach to the
Lorentzian, post-bounce effective solution representing the
universe at late times, a \emph{Riemannian} pre-bounce solution
without boundary. If so, it would be very interesting to analyze
the sense in which this Riemannian solution captures the physics
of the pre-bounce branch of the full quantum evolution.

Finally, it is instructive to  recall the situation with
singularities in classical general relativity. There,
singularities first appeared in highly symmetric situations. For a
number of years, arguments were advanced that this is an artifact
of symmetry reduction and generic, non-symmetric solutions will be
qualitatively different. However, singularity theorems by Penrose,
Hawking, Geroch and others showed that this is not correct. An
astute use of the \emph{differential geometric} Raychaudhuri
equation revealed that singularities first discovered in the
simple, symmetric solutions are in fact a generic feature of
classical general relativity. A fascinating question is whether
the singularity resolution due to quantum geometry is also generic
in an appropriate sense \cite{nd}. Is there a general equation in
\emph{quantum geometry} which implies that gravity effectively
becomes repulsive near generic space-like singularities, thereby
halting the classical collapse? If so, one could construct robust
arguments, now establishing  general `singularity resolution
theorems' for broad classes of situations in quantum gravity,
without having to analyze models, one at a time.

\bigskip
\textbf{Acknowledgments:} We would like to thank Martin Bojowald,
Jim Hartle and Pablo Laguna  for discussions. This work was
supported in part by the NSF grants PHY-0354932 and PHY-0456913,
the Alexander von Humboldt Foundation, and the Eberly research
funds of Penn State.

\begin{appendix}

\section{Heuristics}
\label{a1}

Quantum corrections to the classical equations can be calculated
using ideas from a geometric formulation of quantum mechanics
where the Hilbert space is regarded as (an infinite dimensional)
phase space, the symplectic structure being given by the imaginary
part of the Hermitian inner product (see, e.g., \cite{as}). This
`quantum phase space' has the structure of a bundle with the
classical phase space as the base space, and all states with the
same expectation values for the canonically conjugate operators
$(\hat{q}^i,\, \Hat{p}_i)$ as an (infinite dimensional) fiber.
Thus, any horizontal section
provides an embedding of the classical phase space into the
`quantum phase space'. In the case of a harmonic oscillator (or
free quantum fields) coherent states constitute horizontal
sections which are furthermore preserved by the full quantum
dynamics. In the semi-classical sector defined by these coherent
states, the effective Hamiltonian coincides with the classical
Hamiltonian and there are no quantum corrections to classical
dynamics. For more general systems, using suitable semi-classical
states one may be able to find horizontal sections which are
preserved by the quantum Hamiltonian flow to a desired accuracy
(e.g. in an $\hbar$ expansion). The effective Hamiltonian
governing this flow ---the expectation value of the quantum
Hamiltonian operator in the chosen states, calculated to the
desired accuracy--- is generally different from the classical
Hamiltonian. In this case, dynamics generated by the effective
Hamiltonian provides systematic quantum corrections to the
classical dynamics \cite{jw} (see also \cite{dh1,dh2,sv}).

This procedure has been explicitly carried out in LQC for various
matter sources \cite{jw,vt}. For a massless scalar field, the
leading order quantum corrections are captured in the following
effective Hamiltonian constraint \cite{vt}:
\be\label{heff0} C_{\mathrm{eff}} = - \f{6}{\gamma^2 \mu_o^2}
 |p|^{\f{1}{2}}\,\sin^2(\mu_o c) + 8\pi G\,B(p) \, {p_\phi^2}
\ee
where $B(p)$ is the eigenvalue of $\widehat{1/|p|^{3/2}}$ operator
given by (\ref{eq:bp}).
\footnote{In the literature, eigenvalues $B(p)$ often contain a
half-integer $j$, a parameter representing a quantization
ambiguity. In view of the general consistency arguments advanced
in \cite{ap}, we have set its value to its minimum, i.e. $j =
1/2$.}
For $|\mu| \gg \mu_o$, $B(p)$ can be approximated as
\be B(p) = \left(\f{6}{8 \pi \gamma \lp^2}\right)^{3/2} \,
|\mu|^{-3/2} \, \left(1 + \f{5}{96} \f{\mu_o^2}{\mu^2} + {
O}\left(\f{\mu_o^4}{\mu^4}\right) \right) ~. \ee
The leading order term is $1/p^{3/2}$, thus $B(p)$ quickly
approaches its classical value for $|\mu| \gg\mu_o$, corrections
being significant only in the `genuinely quantum region' in the
vicinity of $\mu=0$. \emph{From now on, we will ignore the quantum
corrections to $B(p)$.}

\begin{figure}[tbh!]
  \begin{center}
    \includegraphics[width=5in,angle=0]{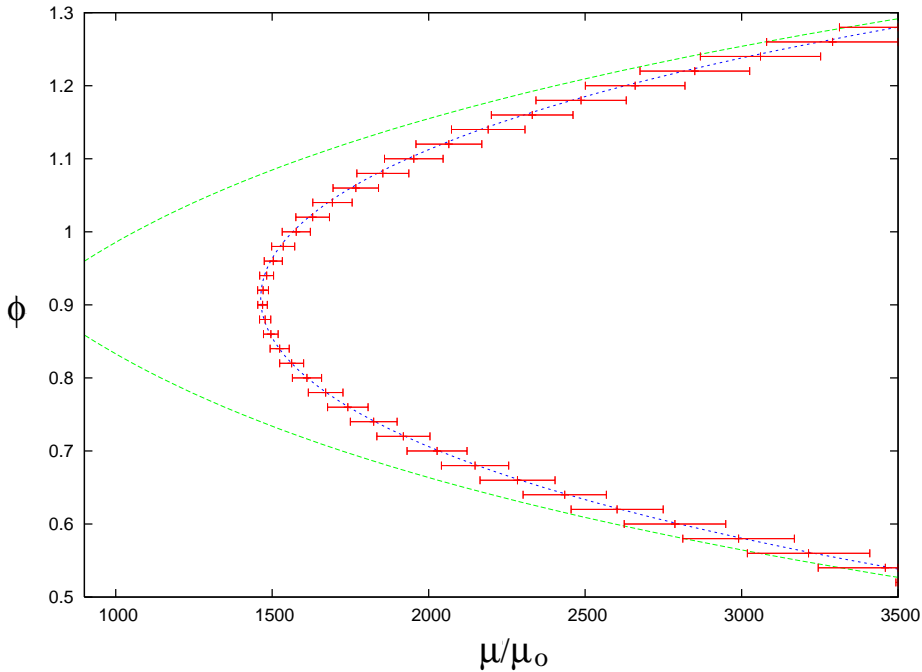}
    \caption{Expectation values (and dispersions) of
    $\widehat{|\mu|_{\phi}}$ are plotted near the bounce point,
    together with with classical and effective trajectories
    (fainter and darker dots, respectively). While the classical
    description fails in this region, the effective description
    provides an excellent approximation to the exact quantum
    evolution. In this plot, $p_\phi = 3000$ and $\epsilon = 2
    \mu_o$.}
    \label{fig:mu-traj-zoom}
  \end{center}
\end{figure}

To obtain the equations of motion, we need the effective
Hamiltonian $\heff$. As usual it is obtained simply by a rescaling
of $C_{\rm eff}$ which gives $\heff$ the dimensions of energy and
ensures that the matter contribution to it is the standard matter
Hamiltonian:
\be\label{heff1} \heff =  \f{C_{\rm eff}}{16\pi G} =  - \f{3}{8\pi
G\gamma^2 \mu_o^2} \, |p|^{1/2} \, \sin^2(\mu_o c) +
 \f{p_\phi^2}{2 p^{3/2}}. \ee
Then, the Hamilton's equation for $\dot p$ become:
\be \label{dotp} \dot p = \{p,\,\heff\} = - \f{8 \pi \gamma
G}{3}\, \f{\partial\heff}{\partial c}  = \frac{2 |p|^{1/2}}{\gamma
\mu_o} \, \snf \, \csf~ .  \ee
Further, since the Hamiltonian constraint implies that $\heff$ of
(\ref{heff1}) vanishes, we have:
\be \snfs = \frac{8 \pi  \gamma^2 \mu_o^2 G}{6 \, |p|^2} \,
p_\phi^2\, , \ee
which, on using Eq.(\ref{dotp}),  provides the \emph{modified
Friedmann equation} for the Hubble parameter $H$:
\be \label{mod_fried} H^2 \equiv \f{\dot p^2}{4 p^2} = \frac{8 \pi
G}{3} \, \rho \left( 1 - \f{\rho}{\rcr} \right), \quad {\rm where}
\quad \rcr = \left(\f{3}{8 \pi G \gamma^2 \mu_o^2}\right)^{3/2}
\f{\sqrt{2}}{ p_\phi} ~. \ee

To obtain the dynamical trajectory, we also need the Hamilton's
equation for $\phi$,
\be\label{dotphi} \dot \phi = \{\phi, \heff\} =
\f{p_\phi}{p^{3/2}}\, . \ee
By combining Eq. (\ref{dotp}) and (\ref{dotphi}) we obtain the
effective equation of motion in the $\mu$-$\phi$ plane:
\be \label{mod_dmdf} \f{d\mu}{d \phi} = \sqrt{\f{16 \pi G}{3}} \,
\left(1 - \f{\rho}{\rcr}\right)^{1/2} \, \mu ~. \ee
The classical Friedmann dynamics results if we set $\rcr =\infty$.
Eqs. (\ref{mod_fried}) and (\ref{mod_dmdf}) suggest that the LQC
effects significantly modify the Friedmann dynamics
once the matter density reaches a critical value, $\rcr$. In the
classical dynamics, the Hubble parameter $H$ can not vanish
(except in the trivial case with $p_\phi =0$). In the modified
dynamics, on the other hand, $H$ vanishes at $\rho =\rcr$. At this
point, the Universe bounces. Thus, the bounce predicted by
Eq.(\ref{mod_fried}) has its origin in quantum geometry. (The
critical value $\mu$ at which this bounce occurs is given by
$\mu_{\mathrm{crit}} = (\sqrt{6/8 \pi G \hbar^2})\, p_\phi \, \mu_o$.) As
pointed out in the main text, a physical limitation of the present
framework is that if $p_\phi$ is chosen to be sufficiently large,
the critical density $\rcr$ can be small.

To obtain the effective equations, several approximations were
made \cite{jw,vt} which are violated in the deep Planck regime.
Nonetheless, the resulting picture of the bounce is consistent
with the detailed numerical analysis. In fact, within numerical
errors the trajectory
\be \label{mu_phi} \mu(\phi) = \f{1}{2} \left( \exp
\left(\sqrt{\f{16 \pi G}{3}} (\phi - \phi_o)\right) + D(p_\phi) \,
\exp\left(-\sqrt{\f{16 \pi G}{3}} (\phi - \phi_o)\right) \right)
\ee
obtained by integrating Eq (\ref{mod_dmdf}) approximates the
expectation values of $\widehat{|\mu|_\phi}$ quite well. (As in
the main text, $D(p_\phi) = \mu_o^2p_\phi^2/12\pi G\hbar^2$). An
illustrative plot of this generic behavior is shown in Fig.
\ref{fig:mu-traj-zoom}. Therefore, in retrospect, this analysis
can be taken as a justification for the validity of the
approximation throughout the evolutionary history of
semi-classical states used in this paper. However, by its very
nature, the effective description can not reproduce the
interesting features exhibited by quantum states captured in
Figs. \ref{fig:zoom-gauss}-\ref{fig:dphi1}.
\begin{figure}[tbh!]
  \begin{center}
    \includegraphics[width=5in,angle=0]{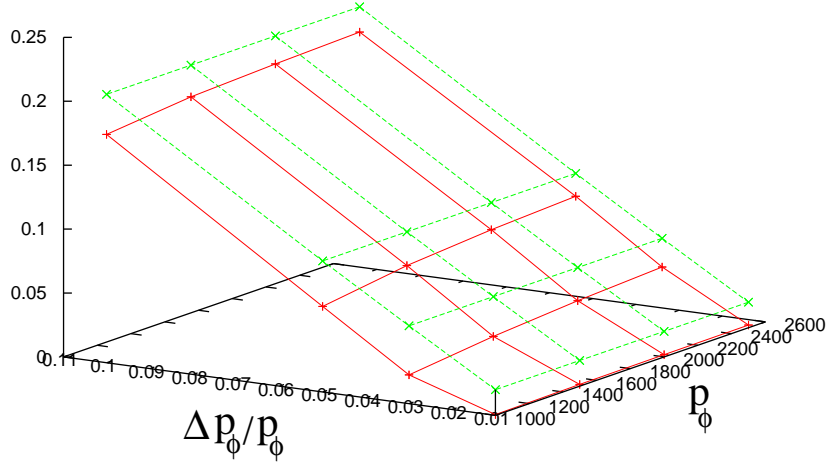}
    \caption{The darker `lattice' show the difference between final
and initial relative spreads  $\Delta\mu/\mu$ for states obtained
by evolving initial data of $\emph{Method II}$. The upper, fainter
`lattice' shows the heuristic bound $\delta\mu/\mu$ given by Eq
(\ref{est_wdw}).}
    \label{fig:bound-wdw}
  \end{center}
\end{figure}
\begin{figure}[tbh!]
  \begin{center}
    \includegraphics[width=5in,angle=0]{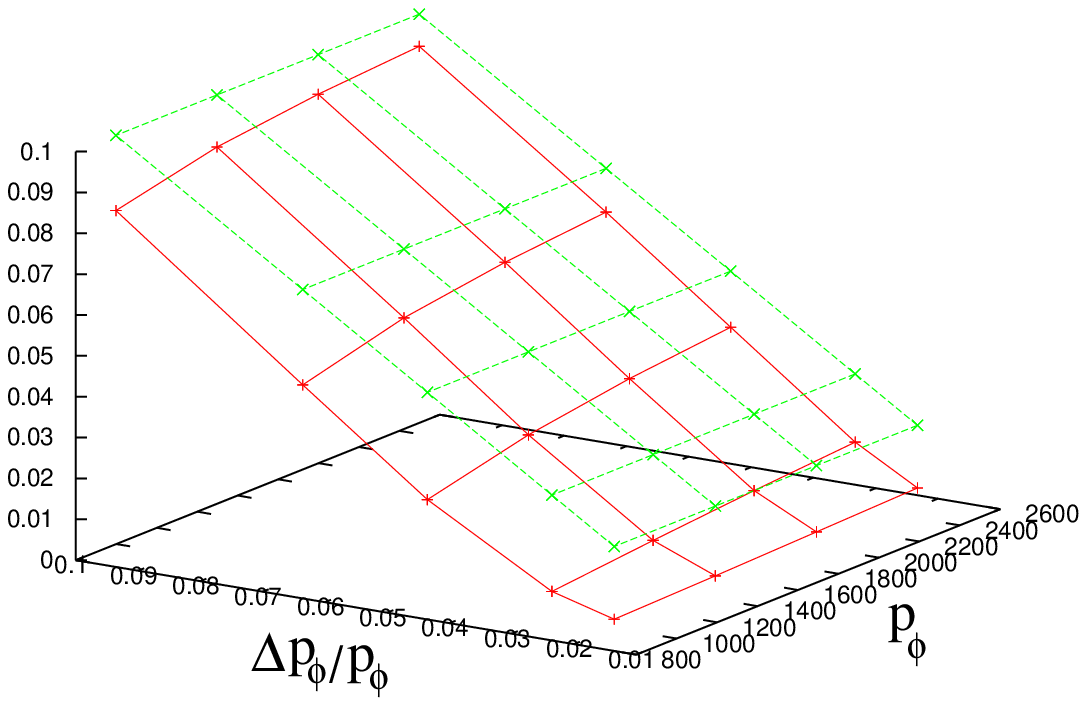}
\caption{The darker `lattice' show the difference between final
and initial relative spreads  $\Delta\mu/\mu$ for states obtained
by evolving initial data of $\emph{Method III}$. The upper,
fainter `lattice' shows the heuristic bound $\delta\mu/\mu$ given
by Eq (\ref{est_sym_wdw}).}
    \label{fig:bound-symm}
  \end{center}
\end{figure}

However, the effective description can be used to provide an
intuitive understanding of the behavior of various uncertainties
discovered through numerical analysis. Let us first note that the
position of the bounce point depends linearly on the value of
$p_\phi$. Next, consider two near-by solutions with slightly
different $p_\phi$ which asymptote to the same \WDW solution for
the expanding branch in the distant future. We wish to know the
way in which $(\delta\mu/\mu)(\phi) :=
((\mu_1-\mu_2)/\mu_2)(\phi)$ changes in the backward evolution as
the two wave functions asymptote to \WDW solutions in the distant
past. This relative difference can be found using Eq
(\ref{mu_phi}) and is given by
\be \label{est_wdw} \f{\delta \mu}{\mu} = 2 \f{\delta
p_\phi}{p_\phi} + \left(\f{\delta p_\phi}{p_\phi} \right)^2 ~, \ee
where $\delta p_\phi$ is the difference between values of $p_\phi$
of the two classical trajectories. A heuristic estimate on the
relative difference in $\mu$ can be compared with the relative
dispersion $\Delta \mu/\mu$ obtained from the \emph{Method II} in
the $\phi$ evolution of section \ref{s5.2}. It turns out that
estimate in Eq (\ref{est_wdw}) provides a reasonably good upper
bound to the relative dispersion found numerically (see Fig.
\ref{fig:bound-wdw}).
A similar comparison can be made for \emph{Method III} of section
\ref{s5.2}. In this case the corresponding solution to Eq
(\ref{mod_dmdf}) is
 \be \mu(\phi) = \f{1}{2}
\left(D(p_\phi)^{-1/2} \, \exp\left(\sqrt{\f{16 \pi G}{3}} (\phi -
\phi_o)\right) + D(p_\phi)^{1/2} \, \exp\left(-\sqrt{\f{16 \pi G}{3}}
(\phi - \phi_o)\right) \right) \ee with relative difference \be
\label{est_sym_wdw} \f{\delta \mu}{\mu} = \f{\delta
p_\phi}{p_\phi} ~. \ee
A comparison with the relative dispersions in numerical analysis
is shown in Fig. \ref{fig:bound-symm}. As in the case of
construction of coherent states via \emph{Method II}, above
estimate serves as an upper bound for relative dispersions
computed by numerical analysis.

Finally, in the numerical analysis an important issue concerns
with the behavior of dispersions of our Dirac observables $\hat
\mu_\phi$ and $\hat p_\phi$ and the product $\Delta \phi \Delta
p_\phi$. Intuitive understanding of our numerical results of Fig.
\ref{fig:dphi1} can be gained by casting Eq. (\ref{mod_dmdf}) in
the form
\be\label{dp_dm_m} \Delta \phi = \sqrt{\f{3}{16\pi G}} \, \left(1
- \f{\rho}{\rcr}\right)^{-1/2} \, \f{\Delta \mu}{\mu} ~. \ee
Since $\Delta \mu/\mu$ can be determined numerically, we can then
estimate $\Delta \phi$ throughout the evolution. The factor
$\left(1 - \f{\rho}{\rcr}\right)^{-1/2}$ is approximately equal to
unity for $\rho \ll \rcr$ or equivalently for $\mu \gg
\mu_{\mathrm{crit}}$. However near the bounce point, $\left(1 -
\f{\rho}{\rcr}\right)^{-1/2} \gg 1$. In \emph{Method II} and
\emph{Method III} of constructing initial states described in Sec.
\ref{sec:phi-init-data}, this change compensates the corresponding
decrease in $\Delta \mu/\mu$ and leads to a nearly constant value
of $\Delta \phi$. However, for \emph{Method I}, since $\Delta
\mu/\mu$ increases monotonically, the fluctuation $\Delta \phi$
increases significantly near the bounce point.

\section{Issues in numerical analysis} \label{a2}

In this Appendix we will spell out the way in which the \WDW limit
of the eigenfunctions of $\Theta$ were found.

Consider a general eigenfunction $\ub{e}_{\omega}$ of
$\ul{\Theta}$ ( for $\omega^2 \ge \pi G/3$). It is always a
linear combination of basis functions $\ub{e}_{|k|},
\ub{e}_{-|k|}$ (where $k^2 = 3/(16\pi G)\omega^2 - 1/16$) defined
in Eq. (\ref{eq:ek}). For later convenience, let us express the
linear combination as:
\begin{equation}\label{eq:e-dec}
  \ub{e}_{\omega} = r^+ e^{i(\beta+\alpha)} \ub{e}_{|k|}
  + r^- e^{i(\beta-\alpha)} \ub{e}_{-|k|} \, ,
\end{equation}
where $r_\pm,\alpha,\beta$ are real numbers. Since each
$\ub{e}_{\pm|k|}$ is a product of an `amplitude'
$|\mu|^{\frac{1}{4}}/4\pi$ and a `phase' $e^{\pm i|k|\ln|\mu|}$, it
is natural to rescale $\ub{e}_{\omega}$:
\begin{equation}
  \tilde{\ub{e}}_{\omega}(\mu)
  := 4\pi|\mu|^{-\frac{1}{4}}\ub{e}_{\omega} \ .
\end{equation}
In terms of coefficients defined in \eqref{eq:e-dec}, we have
\begin{equation}
  \tilde{\ub{e}}_{\omega}(\mu)\ =\ e^{i\beta} \left(
  (r^{+}+r^{-})\cos(\alpha+|k|\ln|\mu|)
  + i(r^{+}-r^{-})\sin(\alpha+|k|\ln|\mu|) \right) \ .
\end{equation}
The values of $\t{\ub{e}}_{\omega}(\mu)$ trace out an ellipse on
the complex plane, parameterized by $\ln|\mu|$. The length of
semi-major and semi-minor axis of this ellipse is equal to,
respectively
\begin{subequations}\label{eq:axis}\begin{align}
  r^{+}+r^{-}   &= \sup_{\mu}|\tilde{\ub{e}}_{\omega}|  &
  |r^{+}-r^{-}| &= \inf_{\mu}|\tilde{\ub{e}}_{\omega}| \ ,
    \tag{\ref{eq:axis}}
\end{align}\end{subequations}
whereas the phase $\alpha$ is related to positions of maxima of
$\tilde{\ub{e}}_{\omega}$ as follows
\begin{equation}\label{eq:phase}
  |\ub{e}_{\omega}| = r^{+}+r^{-} \quad \Leftrightarrow \quad
  \alpha+|k|\ln|\mu|\ =\ n\pi\ , \quad  n\in\integer \ .
\end{equation}
The remaining phase $\beta$ is just the phase of
$\tilde{\ub{e}}_{\omega}$ at maximum. The sign of $(r^{+}-r^{-})$
is, on the other hand, determined by direction of the rotation of
the curve as $\mu$ increases.

The method specified above allows us to calculate the
decomposition in $\ub{e}_k$ basis of a function $\ub{e}_{\omega}$
specified in the form of numerical data (i.e. array of values at
sufficiently large domain). The same algorithm can be applied to
identify the WDW limit of any eigenfunction of the LQC operator
$\Theta$. Indeed given an eigenfunction $e_{\omega}(\mu)$
supported on the lattice $\La_{|\epsilon|}$ (or
$\La_{-|\epsilon|}$) one can again define $\tilde{e}_{\omega}$
analogously to $\tilde{\ub{e}}_{\omega}$ and find its (local)
extrema for large $\mu$. (For definiteness, we restrict our
consideration to finding the limit on the positive $\mu$ side,
however this method can be used of course also for the negative
$\mu$ domain.) Next, the positions of extremas and values of
$\tilde{e}_{\omega}$ at them can be used to calculate coefficients
$r^{+}\pm r^{-},\alpha,\beta$ at each extremum independently. If
$\tilde{e}_{\omega} \to \tilde{\ub{e}}_{\omega}$ then these
coefficients form sequences $(\{r^++r^-\}_i, \{r^+-r^-\}_i,
\{\alpha\}_i, \{\beta\}_i)$ which converge to the analogous
coefficients corresponding to $\tilde{\ub{e}}_{\omega}$ as $\mu
\to \infty$. Finding the WDW limit of $\tilde{e}_{\omega}$ reduces
then to finding the limit of $(\{r^++r^-\}_i, \{r^+-r^-\}_i,
\{\alpha\}_i, \{\beta\}_i)$.

In actual numerical work the following method was used:
\begin{itemize}
  \item After given eigenfunction $e_{\omega}$ was calculated using
    \eqref{eq:eigen} the positions of extrema $\{\mu\}_i$ were found.
  \item Around the extrema the function $e_{\omega}$ supported on
    $\La_{|\epsilon|}$ (or $\La_{-|\epsilon|}$) was extended to
    neighborhoods of $\{\mu\}_i$ via polynomial interpolation. Then
    the positions and values of extrema were recalculated with use of
    this extension. This allowed us to construct sequences converging to
    the WDW limit much more quickly than the ones constructed in the first
    step. The motivation for this construction is the expectation that
    for sufficiently large $\mu$ the values of $e_{\omega}$ at
    $\La_{|\epsilon|}$ should be good estimates of its WDW limit (being
    regular function defined on entire $\re^+$).
  \item Extrema found in previous step were next used to calculate
    sequences $(\{r^++r^-\}_i, \{r^+-r^-\}_i, \{\alpha\}_i,
    \{\beta\}_i)$ in a way analogous to that specified by Eqs.
    (\ref{eq:axis})and (\ref{eq:phase}) and in the description below them.
  \item  Finally the limits of coefficients at ${1}/{\mu} \to 0$
    were calculated by polynomial extrapolation.
\end{itemize}

\section{An alternate physical Hilbert space}
\label{a3}

In this Appendix we will construct a physical Hilbert space $\Hp'$
in LQC which is qualitatively different from the spaces
$\Hp^\epsilon$ constructed in section \ref{s4.2}. In its features,
it interpolates between these and the Hilbert space $\Hpwdw$ of
section \ref{s3.2}. For completeness, we will first explain why a
new representation of the algebra of Dirac observables can arise,
then summarize the results and finally compare and contrast them
with those obtained in sections \ref{s3.2} and \ref{s4.2}. The
first part is somewhat technical but we have organized the
presentation such that the readers can go directly to the summary
without loss of continuity.

Let us begin by recalling the situation for a general system with
a single constraint $C$. In the refined version of Dirac
quantization \cite{almmt}, one introduces an auxiliary Hilbert
space $\Ha$, and represents the constraint by a self-adjoint
operator $\hat{C}$ on it. The technically difficult task is to
chooses a dense sub-space $\Phi$ of $\Ha$ such that for all $f,g
\in \Phi$,
\be (\Psi_f| := \int_{-\infty}^\infty d\lambda\,\, \langle
e^{-i\lambda\hat{C}}\, f|\ee
is a well-defined element of $\Phi^\star$, such that the action
$(\Psi_f|g\rangle$ of $(\Psi_f|\in \Phi^\star$ on $|g\rangle\in
\Phi$ yields a Hermitian scalar product on the space of solutions
$(\Psi_f|$ to the quantum constraint (see, e.g.
\cite{dm,almmt,abc}). Results in section \ref{s4.2} were obtained
using $L^2(\R_{\rm Bohr}, B(\mu)\,\dd\mu_{\rm Bohr})\otimes
L^2(\R,\dd\phi)$ for $\Ha$, and the space of rapidly decreasing
functions $f(\mu, \phi)$ in this $\Ha$ for $\Phi$. 
In the LQC literature, $\Phi$ is sometimes called $\cyl$ and
$\Phi^\star$ is taken to be its algebraic dual, denoted by
$\cyl^\star$.

The construction given above is rather general. For the model
under consideration, we can extend this construction by using an
entirely different subspace of $\Phi^\star$ for the auxiliary
Hilbert space. This is possible because by duality the action of
$\hat{C}$ can be extended to all of $\Phi^\star$. Let us set
$\Ha^\prime := L^2(\R^2, B(\mu)\dd\mu\dd\phi)$. This is a subspace
of $\Phi^\star$ because each $\Psi \in \Ha^\prime$ defines a
linear map from $\Phi$ to $\Comp$:
\be (\Psi|f\rangle := \sum_\mu \int_{-\infty}^\infty \dd\phi
B(\mu) \bar\Psi(\mu,\phi)\, f(\mu, \phi) \quad\quad \forall\Psi
\in \Ha^\prime \ee
where the sum over $\mu$ converges because $f \in \Phi$ has
support only on a countable number of points on the $\mu$-axis and
a rapid fall-off. The dual action of $\hat{C}$ on $\Ha^\prime$ can
now be calculated: Since
\ba (\Psi|\hat{C}f\rangle \,=\, \sum_\mu \int_{-\infty}^\infty
\dd\phi B(\mu) \bar\Psi\,  \bigg[\f{\p^2 f}{\p\phi^2} &+&
[B(\mu)]^{-1}\, (C^+(\mu) f(\mu+4\mu_o, \phi)\nonumber\\
&+& C^o(\mu) f(\mu, \phi) + C^{-}(\mu) f(\mu-4\mu_o))\bigg]\ea
it follows from the definitions of $C^\pm(\mu)$ that
\ba \left(\hat{C}\Psi\right)(\mu,\phi ) &=& \f{\p^2
\Psi}{\p\phi^2} + [B(\mu)]^{-1}\, \left(C^+(\mu) \Psi(\mu+4\mu_o,
\phi) + C^o(\mu) \Psi(\mu, \phi) + C^{-}(\mu) \Psi(\mu-4\mu_o)\right)
\nonumber\\
&\equiv& \bigg[\f{\p^2 f}{\p\phi^2} + \Theta \bigg] \Psi(\mu,\phi)\ea
It is straightforward to verify that $\hat\Theta$ and
 $\hat{C}$ are
 self-adjoint on
$\Ha^\prime$. Therefore, we can carry out group averaging on
$\Ha'$ and obtain a new physical Hilbert space. As in the \WDW
theory of section \ref{s3.2}, there are two superselected sectors.
We will work with the positive frequency sector and denote it by
$\Hp^\prime$. The Dirac observables $\hat{p}_\phi$ and
$\widehat{|\mu|_{\phi_o}}$ on $\Phi$ act by duality on
$\Ha^\prime$ and descend naturally to $\Hp^\prime$.

The final results can be summarized as follows. The new physical
Hilbert space $\Hp^\prime$ is the space of functions
$\Psi(\mu,\phi)$ satisfying the positive frequency equation:
$-i\p_\phi \Psi = \sqrt{\Theta} \Psi$, with finite norm:
\be \label{ip3}||\Psi||^{\prime\, 2}_{\rm phy} =
\int_{\phi=\phi_o} d\mu B(\mu) |\Psi(\mu,\phi)|^2 \ee
and the action of the Dirac observables is the standard one:
\be \label{dirac5}\widehat{|\mu|_{\phi_o}}\, \Psi(\mu,\phi) =
e^{i\sqrt{{\Theta}}(\phi-\phi_o)}\,|\mu| \,\Psi(\mu,\phi_o),\quad
{\rm and} \quad \hat{p}_\phi \Psi(\mu,\phi) = - i\hbar\, \f{\p
\Psi(\mu,\phi)}{\p\phi} \, . \ee

Note that the final physical theory is different from both the
\WDW theory of \ref{s3.2} and the `standard' LQC theory of
\ref{s4.2}. Since the inner product (\ref{ip3}) involves an
integral rather than a sum, states $\Psi(\mu,\phi)$ now have
support on continuous intervals of the $\mu$-axis as in the \WDW
theory, rather than on a countable number of points as in LQC.
However, the states satisfy the LQC type positive frequency
equation $-i\p_\phi \Psi = \sqrt{\Theta} \Psi$, where the operator
on the right is the square-root of a positive, self-adjoint \emph{difference} operator
rather than of a differential operator, and the measure
determining the inner product also involves $B(\mu)$ from LQC
rather than $\ub{B}(\mu)$ from the \WDW theory. Thus, the
dynamical operator is the same as in LQC. \emph{In particular, as
in section \ref{s5}, the quantum states exhibit a big bounce.}
However, since typical states are continuous on the $\mu$-axis,
the spectrum of the Dirac observable $\widehat{|\mu|_{\phi_o}}$
is now continuous. In essence, the states $\Psi(\mu,\phi)$ have
support on \emph{2-d continuous} regions of the $\mu-\phi$ plane
as in the \WDW theory but their dynamics is dictated by a
\emph{difference} operator as in LQC. When the cosmological
constant is non-zero, the analog of this physical Hilbert space
appears to be a natural home to analyze the role of the Kodama
state in quantum cosmology \cite{kodama}.

In the literature on `polymer representations', non-relativistic
quantum mechanics of point particles and the quantum theory of a
Maxwell theory have been discussed in some detail\cite{afw,almax}.
In the first case, the standard Schr\"odinger Hilbert space
$L^2(\R, \dd x)$, and in the second case, the standard Fock space
turned out to be subspaces of $\cyl^\star$ which were especially
helpful for semi-classical analysis. The present auxiliary Hilbert
space $\Ha' \subset \Phi^\star$ is completely analogous to these.
Therefore the resulting $\Hp^\prime$ may be more useful for
semi-classical considerations. Indeed, since it does not refer to
any $\epsilon$, no coarse graining is required to carry out the
semi-classical analysis. Therefore the analog of this construction
may well be useful in more general contexts in full LQG.

\end{appendix}

\end{document}